\documentclass[12pt]{article}
\pdfoutput=1
\usepackage[nosort]{cite}
\usepackage[normalem]{ulem}

\usepackage{epsfig}
\usepackage{amsfonts}
\usepackage{amscd}
\usepackage{latexsym}
\usepackage{amsmath,amssymb}
\usepackage{verbatim}
\usepackage{setspace}
\usepackage[dvipsnames]{xcolor}
\usepackage{fancyhdr}
\usepackage{cite}
\usepackage{hyperref}
\usepackage{slashed}
\usepackage{multirow}
   \usepackage{draft}
\usepackage{hyperref}
\usepackage{graphicx,color,subcaption}
\usepackage{cite}
\usepackage{skak}
\usepackage{bbm}
\usepackage[english]{babel}
\usepackage{amsthm}
\usepackage{soul}
\usepackage{upgreek}
\usepackage[scr]{rsfso}

\usepackage{tikz}
\usepackage{tikz-feynman}
\usetikzlibrary{3d}
\usetikzlibrary{positioning}
\usetikzlibrary{decorations.markings,decorations.pathmorphing,arrows.meta}

\numberwithin{equation}{section}

\begin{document}

\begin{titlepage}

\hfill MIT-CTP/6024

\title{Lattice chiral symmetry from bosons in 3+1d}

\author{Zhiyao Lu$^\text{V}$, Sahand Seifnashri$^\text{A}$, Shu-Heng Shao$^\text{V}$} 

\begin{center}

${}^\text{V}$\textit{Center for Theoretical Physics - a Leinweber Institute,\\ 
Massachusetts Institute of Technology,\\ 
77 Massachusetts Ave., Cambridge, MA 02139 USA}

\vspace{2pt}
${}^\text{A}$\textit{School of Natural Sciences, Institute for Advanced Study,\\ 
1 Einstein Drive, Princeton, NJ 08540, USA}

\end{center}

\vspace{5pt}

\abstract{We present a solvable  Hamiltonian that realizes an exact lattice chiral U(1)$_\text{V}\times \text{U(1)}_\text{A}$ symmetry. 
Nielsen-Ninomiya-type no-go theorems are evaded by using lattice bosons rather than fermions. 
The continuum limit is a compact boson field theory with an axion-like coupling. 
The U(1)$_\text{V}$ symmetry shifts the scalar, while U(1)$_\text{A}$ acts on local operators associated with short axion strings and is transmuted into a higher-form symmetry in the continuum limit. We demonstrate the chiral anomaly by showing that the lattice theta angle is shifted by an axial rotation when U(1)$_\text{V}$ is gauged. 
Gauging either U(1)$_\text{V}$ or U(1)$_\text{A}$ leads to lattice non-invertible and 2-group symmetries, respectively, matching the continuum picture.

}

\end{titlepage}

\tableofcontents

\section{Introduction}

Chiral symmetries are central to high-energy physics, from confinement and chiral Lagrangians to spontaneous symmetry breaking. A longstanding challenge is to realize such symmetries exactly in lattice models \cite{Kaplan:2009yg}. 
Notably, the Nielsen-Ninomiya theorem \cite{Nielsen:1980rz,Nielsen:1981xu,Nielsen:1981hk,Friedan:1982nk} rules out a naive lattice realization of chiral symmetries with lattice fermions. 
More recently, this no-go theorem has been generalized to more general lattice systems beyond free fermion models \cite{Fidkowski:2023sif,Kapustin:2024rrm,Liu:2026atf}.

If the no-go theorems are, at their core, theorems about lattice fermions, then perhaps the right strategy is to change the microscopic building blocks. 
More precisely, these theorems only apply to lattice systems whose local Hilbert space is finite-dimensional, such as the one obtained from quantizing fermion fields. 
Indeed, chiral symmetries in 1+1d can be realized both in Euclidean \cite{Gross:1990ub,Sulejmanpasic:2019ytl,Gorantla:2021svj,DeMarco:2023hoh,Berkowitz:2023pnz} and Hamiltonian \cite{Cheng:2022sgb,Fazza:2022fss,Thorngren:2026ydw,Seifnashri:2026ema} lattice systems of continuous bosonic variables with an infinite-dimensional local Hilbert space. 
However, it was not clear how to extend this construction to realize 
the ordinary chiral U(1)$_\text{V}\times \text{U(1)}_\text{A}$ symmetry and its anomaly in 3+1d lattice systems, since there is no natural notion of bosonization.

 Recent advances in generalized symmetries \cite{Gaiotto:2014kfa,McGreevy:2022oyu,Cordova:2022ruw,Schafer-Nameki:2023jdn,Brennan:2023mmt,Bhardwaj:2023kri,Shao:2023gho,Costa:2024wks,shao2025noninvertible,Kaidi:2026urc}  have reopened this question and motivated new mechanisms to bypass standard obstructions. 
Motivated by the non-invertible chiral symmetry in quantum electrodynamics (QED) \cite{Choi:2022jqy,Cordova:2022ieu}, the authors of \cite{Fidkowski:2025rsq} constructed exact chiral U(1)$_\text{V}\times \text{U(1)}_\text{A}$ symmetry operators in a 3+1d lattice Hilbert space. 
Subsequently, anomaly cancellation of this lattice chiral symmetry was demonstrated in \cite{Thorngren:2026ydw}. 
The no-go theorems are evaded by using continuous lattice bosons, rather than fermions.
 
Now that there is a lattice chiral symmetry operator, what theory actually realizes this symmetry? 
Furthermore, what is chiral about a boson in 3+1d? 
In this paper, we present an exactly solvable Hamiltonian that realizes this lattice chiral U(1)$_\text{V}\times \text{U(1)}_\text{A}$ symmetry in a different but related setting. 
Our lattice model falls into the class of the modified Villain models of \cite{Sulejmanpasic:2019ytl,Gorantla:2021svj,Cheng:2022sgb,Fazza:2022fss}, but with one important constraint relaxed (Section \ref{sec:Hamiltonian}). 
The vector symmetry shifts the lattice boson, while the axial symmetry is generated by the following (Section \ref{sec:U1VA}):
\ie\label{introQA}
Q_\text{A}  = \int w\cup dw
=
\sum_{i,j,k=x,y,z}
\epsilon_{ijk}
\sum_{\vec r}
w_{i}(\vec r)
\left[
w_{ j}(\vec r+\hat i)
- w_{j} (\vec r+\hat i+\hat k)
\right]
\fe
where $w_i(\vec r)$ is an integer lattice field on the link starting at site $\vec r$ and pointing in the $+i$ direction. 
The continuum limit of our lattice model is a compact boson field theory of $\phi$, coupled to the U(1)$_\text{V}\times \text{U(1)}_\text{A}$ background gauge fields  as (Section \ref{sec:matching})
\ie
{f^2\over2} (\partial_\mu \phi - A^\text{V}_\mu)^2
+{i\over 16\pi^2}
\epsilon^{\mu\nu\rho\sigma}\,
\phi\,   F^\text{V}_{\mu\nu}
F^\text{A} _{\rho\sigma}\,.
\fe
This coupling is similar to that of the pion in the chiral Lagrangian or that of an axion, explaining the chiral nature of the symmetry in this bosonic theory.

Anomalies are commonly associated with  \textit{fermions} in the \textit{continuum}. This intuition is rooted in the original calculation by Adler \cite{Adler:1969gk}, Bell, and Jackiw \cite{Bell:1969ts} (ABJ), where the anomaly arises from the divergence of fermions running in the loop diagrams. 
In contrast, our lattice chiral U(1)$_\text{V}\times \text{U(1)}_\text{A}$   symmetry is realized by \textit{bosons} on a \textit{lattice}, and yet it has an exact lattice chiral anomaly. 
In the continuum limit, this anomaly corresponds to the one captured by the triangle diagram of U(1)$_\text{A}\text{-}$U(1)$_\text{V}\text{-}\text{U(1)}_\text{V}$, with no additional anomalies present.  
We demonstrate the lattice anomaly by showing that the U(1)$_\text{A}$ symmetry is broken when U(1)$_\text{V}$ is gauged (Section \ref{sec:anomaly}). Furthermore, a U(1)$_\text{A}$ axial rotation induces a lattice $\theta$-angle for the U(1)$_\text{V}$ gauged theory (Section \ref{sec:theta}). 

While the axial U(1)$_\text{A}$ symmetry generated by \eqref{introQA} acts faithfully on the lattice, it does not act on the local operators in the continuum limit. 
More specifically, one can think of the lattice local operators carrying the axial charge as certain short open axion strings. 
In continuum field theory, such open strings are not invariant under the gauge transformation of the 2-form gauge field $b_{\mu\nu}$ dual to the compact boson $\phi$. 
However, gauge invariance is only imposed energetically in our lattice models, and the open strings are well-defined local operators that create states with a large energy penalty. 
This is similar to Kitaev's toric code \cite{Kitaev:1997wr}, which can be viewed as a $\mathbb{Z}_2$ gauge theory with Gauss's law imposed energetically.  The lattice axial U(1)$_\text{A}$ symmetry acts effectively as a 2-form winding global symmetry in the continuum limit, a phenomenon known as symmetry transmutation \cite{Seiberg:2025bqy}. The microscopic chiral anomaly is matched by an anomaly involving this higher-form global symmetry, similar to the examples discussed in \cite{Komargodski:2018odf,Cordova:2019bsd, Delmastro:2022pfo, Brennan:2022tyl, Antinucci:2024bcm, Dumitrescu:2024jko, Brennan:2025acl}.

\begin{figure}
    \centering
    \scalebox{.9}{
    \begin{tikzpicture}
    \tikzset{
      fermionline/.style={thick},
      fermionarrow/.style={
        postaction={decorate},
        decoration={
          markings,
          mark=at position 0.6 with {\arrow[scale=1.2]{Stealth}}
        }
      },
      myarrow/.style={
    postaction={decorate},
    decoration={
      markings,
      mark=at position 1 with {\arrow[scale=1.2]{Stealth}}
    }
  }
    }
    \begin{feynman}
    
    \vertex (a) at (0,1);
    \vertex (b) at (-0.87,-0.5);
    \vertex (c) at (0.87,-0.5);
    
    \vertex (p1) at (0,2.5);
    \vertex (p2) at (-2.16,-1.25);
    \vertex (p3) at (2.16,-1.25);
    
    \diagram*{
        (a) -- [photon,thick] (p1),
        (b) -- [photon,thick] (p2),
        (c) -- [photon,thick] (p3),
    };
    \draw[fermionline, fermionarrow] (a) -- (b);
    \draw[fermionline, fermionarrow] (b) -- (c);
    \draw[fermionline, fermionarrow] (c) -- (a);

    \end{feynman}
    \draw[thick,myarrow] (-2.5,0) -- (-5.5,0) node[midway, above] {gauge U$(1)_\text{A}$};
    \draw[thick,myarrow] (2.5,0) -- (5.5,0) node[midway, above] {gauge U$(1)_\text{V}$};

    \node[right] at (p1) { U$(1)_\text{A}$};
    \node[below] at (p2) { U$(1)_\text{V}$};
    \node[below] at (p3) { U$(1)_\text{V}$};

    \node at (-7.3,0) {\large 2-group};
    \node[align=center] at (7.3,0) {\large non-invertible\\ \large symmetry};

    \node[align=center] at (7.3,1.2) {\small lattice: \cite{Fidkowski:2025rsq}, Section \ref{sec:noninvertible}};
    \node[align=center] at (7.3,-1.2) {\small continuum: \cite{Choi:2022jqy,Cordova:2022ieu}};

     \node[align=center] at (-7.3,1.2) {\small lattice:  Section \ref{sec:2group}};
    \node[align=center] at (-7.3,-1.2) {\small continuum: \cite{Cordova:2018cvg}};
    
    \end{tikzpicture}
    }

    \caption{Consider a theory with a mixed 't Hooft anomaly between the U$(1)_\text{V}$ and U$(1)_\text{A}$ global symmetries captured by the triangle diagram. Gauging U$(1)_\text{V}$ turns U$(1)_\text{A}$ into a non-invertible symmetry, while gauging U$(1)_\text{A}$ results in a 2-group symmetry. 
     See \cite{Cordova:2022ruw} for a review.
    }
    \label{fig:AVVanomaly}
\end{figure}
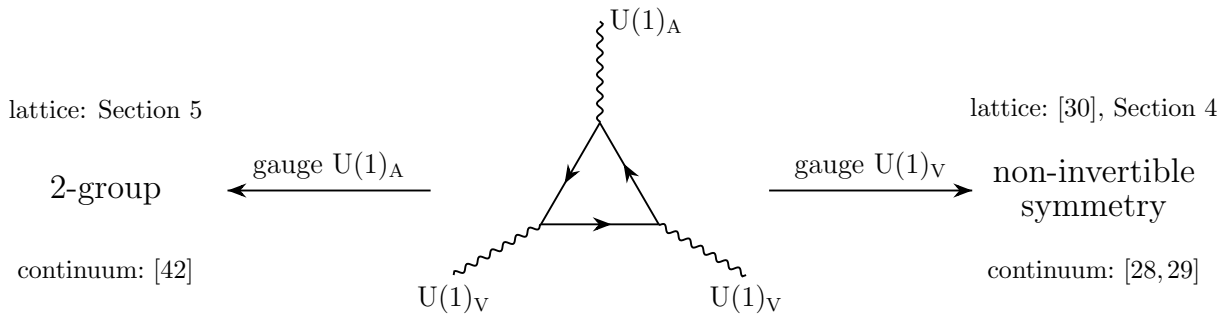

Generalized symmetries provide a more refined check of this anomaly. In continuum field theory, gauging an ordinary symmetry that participates in an 't Hooft anomaly often produces generalized global symmetries \cite{Tachikawa:2017gyf}. In the present lattice model, gauging U(1)$_\text{V}$ turns U(1)$_\text{A}$ into a non-invertible symmetry, while gauging U(1)$_\text{A}$ yields a 2-group symmetry with a lattice Green-Schwarz term. These are in direct parallel with the corresponding continuum constructions \cite{Cordova:2018cvg,Choi:2022jqy,Cordova:2022ieu} (Figure \ref{fig:AVVanomaly}). 
These generalized symmetries therefore provide further evidence for the lattice chiral anomaly. Related non-invertible and higher-group phenomena in modified Villain models have been explored in \cite{Choi:2021kmx,Honda:2024sdz,Honda:2024xmk,Jacobson:2024muj,Honda:2024kvf}.

This paper is organized as follows. Section \ref{sec:model} introduces the lattice model (Sections \ref{sec:Hilbert} and \ref{sec:Hamiltonian}), its chiral symmetry (Section \ref{sec:U1VA}), and anomaly (Section \ref{sec:anomaly}). 
We solve the model exactly by presenting its spectrum in Section \ref{sec:spectrum}. 
We then discuss its continuum limit in Section \ref{sec:contlimit}, which is a compact boson field theory. 
In Section \ref{sec:matching}, we discuss how the lattice chiral anomaly is matched by the axion coupling of the field theory via symmetry transmutation. 
Section \ref{sec:Yukawa} discusses a continuum field theory locally sharing the same symmetries and anomalies as our lattice model.
% in the same universality class as our lattice model. 

In Section \ref{sec:noninvertible} we discuss the lattice model with  U(1)$_\text{V}$ gauged. 
We demonstrate the chiral anomaly in Section \ref{sec:theta} by showing that an axial rotation shifts the lattice $\theta$-angle. 
Section \ref{sec:subnoninv} reviews the lattice non-invertible axial symmetry of \cite{Fidkowski:2025rsq}. 
Section \ref{sec:abelianHiggs} discusses the continuum picture of the gauged lattice system in terms of the Abelian Higgs model. 
In parallel, Section \ref{sec:2group} presents the lattice model with U(1)$_\text{A}$ gauged, and we demonstrate the lattice 2-group symmetry by identifying the Green-Schwarz term.

Appendix \ref{app:cochain} reviews chains, cochains, and cup products on a hypercubic lattice. 
In Appendix \ref{app:more}, we discuss an exact lattice duality and a winding symmetry of our lattice Hamiltonian. Appendix \ref{app:gauging} discusses the gauging of a continuous global symmetry on the lattice. Finally, we review the ABJ anomaly in QED in Appendix \ref{app:QED}.

\section{The lattice model}\label{sec:model}

\subsection{The Hilbert space}\label{sec:Hilbert}

Let space be a 3-dimensional cubic lattice $M_3$  and let time be continuous. 
On every site $s$, we begin with an $\mathbb{R}$-valued operator $\phi_s$  and its conjugate variable $p_s$.\footnote{Sometimes we use $\vec r$ instead of $s$ to denote a site on the lattice to make the lattice expression look closer to the continuum.} 
On every link $\ell$ there is a $\mathbb{Z}$-valued operator $w_\ell$ and its conjugate variable $b_\ell$.  
They obey the commutation relations:
\ie\label{commutators}
[\phi_s, p _{s'} ] = i  \delta_{s,s'}
\,,~~~ 
[w_\ell , b_{\ell'}]  =  -i\delta_{\ell,\ell'}\,.
\fe

The operator $w_\ell$, known as the Villain gauge field, is introduced to make the real scalar field $\phi_s$ compact. 
It gauges a $\mathbb{Z}$ symmetry that shifts the scalar field as $\phi_s \sim \phi_s+2\pi$. 
More specifically, we impose the $\mathbb{Z}$ gauge invariance:
\ie\label{Zgauge}
\phi_s \sim \phi_s +2\pi m_s\,,~~~~
w_\ell \sim w_\ell -(dm)_\ell\,,
\fe
where $m_s\in\mathbb{Z}$. 
Here $d$ is the lattice exterior derivative and $(dm)_\ell = m_{s_1} - m_{s_2}$ where $s_1$ ($s_2$) is the head (tail) of the oriented link $\ell$; see Appendix \ref{app:cochain}. 
The $\mathbb{Z}$ gauge invariance makes $\phi_s$ a compact boson field effectively valued in $\mathbb{R}/2\pi\mathbb{Z}\simeq U(1)$. 
In the Hamiltonian formalism, this gauge invariance is implemented by the following Gauss law operators
\ie\label{gauss1}
&\exp \left(
2\pi i p_s - i (\delta b)_s
\right)=1\,,~~~~\forall~s\,.
\fe
Here $\delta$ is the lattice divergence and $(\delta b)_s = \sum_{\ell\ni s} b_\ell$, where the sum is over every link whose tail is the site $s$; see Figure \ref{fig:coboundary}. 
Gauge-invariant operators, such as $(d\phi)_\ell+2\pi w_\ell$, are those which commute with the Gauss law operators.

Furthermore, since the Villain gauge field $w_\ell$ is integer-valued, we have another operator constraint:
\ie\label{gauss2}
&\exp\left(2\pi i w_\ell \right)=1\,,~~~~\forall ~\ell \,.
\fe
It implies that $b_\ell$ is not a gauge-invariant operator, but $e^{i b_\ell}$ is.

The Hilbert space of this lattice model is constructed by quantizing the fields $\phi_s,p_s,w_\ell,b_\ell$ following the canonical commutation relations in \eqref{commutators}. 
Each local Hilbert space is infinite-dimensional from quantizing the continuous variables, such as $\phi_s$.  
We further impose the two Gauss law constraints \eqref{gauss1} and \eqref{gauss2} strictly on the Hilbert space to project to the gauge-invariant states. 
Therefore, the Hilbert space is not a tensor product of local Hilbert spaces.

We can view $\phi_s, p_s$ as elements in $C^0(M_3,\mathbb{R})$ and write them as $\phi^{(0)}, p^{(0)}$ in the cochain notations. Similarly, we write $w^{(1)}\in C^{1}(M_3, \mathbb{Z}), b^{(1)}\in C^{1} (M_3, \mathbb{R}/2\pi \mathbb{Z})$, which can also be viewed as living on the dual plaquettes.

\subsection{Villain Hamiltonian}\label{sec:Hamiltonian}

A concrete quadratic Hamiltonian on this Hilbert space is 
\ie\label{Hamiltonian}
H=  
{1\over 2\beta}\sum_s p_s^2
+{\beta\over2}
\sum_\ell \left( ( d\phi)_\ell +2\pi w_\ell\right)^2
+{\lambda\over2} \sum_p [(dw)_p ]^2
\,.
\fe
The first two terms are the usual kinetic terms for a boson $\phi_s$, coupled to the Villain integer gauge field $w_\ell$. 
Here $(dw)_p$ is the (oriented) sum of $w_{\ell}$ around the plaquette $p$ (see Figure \ref{fig:boundary}). 
As a consistency check, this Hamiltonian commutes with the two Gauss law constraints \eqref{gauss1} and \eqref{gauss2}.

This Hamiltonian is a 3+1d generalization of the 2+1d model introduced in \cite{Fazza:2022fss}. (See also \cite{Cheng:2022sgb,Seifnashri:2026ema} for its 1+1d counterpart.) 
However, there is one crucial distinction: here we do \textit{not} impose $(dw)_p=0$. 
This condition is a flatness condition for the Villain gauge field $w_\ell$, which physically means that the vortices are strictly suppressed. 

This condition $dw=0$ can also be interpreted as a Gauss law constraint after a duality transformation discussed in Appendix \ref{app:duality}. 
In the continuum, this duality maps the compact boson field $\phi$ to a 2-form gauge field $b_{\mu\nu}$ via $\partial_\mu\phi \sim \epsilon_{\mu\nu\rho\sigma}\partial^\nu b^{\rho\sigma}$. See \cite{Witten:2026twr} for a recent review of this duality in the continuum. 
On the lattice, the dual 2-form gauge field arises from $b_\ell$ on the links, which are equivalent to the dual plaquettes.
Hence, $b_\ell$ is naturally a 2-form field on the dual lattice, and $w_\ell$ is its conjugate electric field. 
The flatness constraint $dw=0$ is the Gauss law implementing a gauge transformation which in the continuum corresponds to
\ie\label{2formgauge}
b_{\mu\nu} \sim b_{\mu\nu} +\partial_\mu \Lambda_\nu - \partial_\nu \Lambda_\mu\,.
\fe
However, we do not impose this gauge invariance strictly on the lattice. 
Rather, there are states with nonzero $dw$, but they are energetically penalized by the $\lambda$ term in \eqref{Hamiltonian}. These states will become important as we discuss the lattice chiral symmetry below. 
For these states to be present we need to keep $\lambda$ finite, but we also cannot set it to zero: as we will show in Section~\ref{sec:contlimit}, the model with $\lambda = 0$ is sick because of an extensive ground-state degeneracy arising from local excitations with non-zero $dw$.

\subsection{U(1)$_\text{V}\times$U(1)$_\text{A}$ chiral symmetry operators}\label{sec:U1VA}

The Hamiltonian \eqref{Hamiltonian} has an obvious U(1)$_\text{V}$ global symmetry which shifts the boson field by a constant:
\ie
e^{i \alpha Q_\text{V} } \phi_s e^{- i \alpha Q_\text{V}} = \phi_s + \alpha\,,
\fe
where the vector charge $Q_\text{V}$ operator is
\ie
Q_\text{V} = \sum_s p_s 
=\int_{\tilde M_3} \star p^{(0)} \,,
\fe
where in the last expression we have introduced the local charge density $q_\text{V}=\star p^{(0)}$ written  in the cochain language. 
Here $\star$ is the lattice Hodge dual operator that maps a $q$-cochain on the original lattice $M_3$ to a $(3-q)$-cochain on the dual lattice $\tilde M_3$. 
Equivalently, the local, gauge-invariant operators charged under U(1)$_\text{V}$ are $e^{i \phi_s}$, i.e., $[Q_\text{V} , e^{i \phi_s} ]  = e^{i\phi_s}$.

The charge density $q_\text{V}= \star p^{(0)}$ is gauge-invariant, i.e., it commutes with the Gauss law constraints \eqref{gauss1} and \eqref{gauss2}. 
However, $q_\text{V}$ is not quantized, i.e., $q_\text{V}\notin\mathbb{Z}$. 
We can introduce another vector charge density 
\ie
\tilde q_\text{V} =  \star\left( p^{(0)} -{(\delta b)^{(0)}\over 2\pi}\right)
\fe
which gives the same total charge $Q_\text{V}=  \int_{\tilde M_3} \tilde q_\text{V}$ by using $\delta = \star d \star$. 
Thanks to the first Gauss law constraint \eqref{gauss1}, $\tilde q_\text{V}$ is quantized. However, it does not commute with the second Gauss law constraint \eqref{gauss2} and is therefore not gauge-invariant. 
Nonetheless, the total charge $Q_\text{V}$ is both gauge-invariant and quantized, and hence generates a (compact) U(1)$_\text{V}$ global symmetry.

Next, motivated by \cite{Fidkowski:2025rsq,Thorngren:2026ydw},  we define the axial charge as,
\ie\label{QA}
&Q_\text{A}  = \int_{M_3} w^{(1)} \cup dw^{(1)}
=
\sum_{i,j,k=x,y,z}
\epsilon_{ijk}
\sum_{\vec r}
w_{i}(\vec r)
\left[
w_{ j}(\vec r+\hat i)
- w_{j} (\vec r+\hat i+\hat k)
\right]\,,
\fe
where $w_i(\vec r)$ is an equivalent expression for $w_\ell$ on the link starting at site $\vec r$ pointing in the $+i$ direction. 
This operator takes the schematic form of a Chern-Simons term $\epsilon_{ijk} w_i \partial_k w_j$.   
Here $\cup$ is the cup product reviewed in Appendix \ref{app:cochain}, particularly in Figure \ref{fig:cupproduct3}. 
It can be roughly thought of as the lattice counterpart of the wedge product in the continuum. 
This axial charge commutes with the Hamiltonian because the latter is independent of $b_\ell$, the conjugate field of $w_\ell$.

The axial charge density $\tilde q_\text{A}= w^{(1)}\cup dw^{(1)}$ is quantized, but is not gauge-invariant under \eqref{Zgauge}. 
We can define another axial charge density
\ie \label{local.a.charge}
 q_\text{A} = \left( w^{(1)}  + {d\phi^{(0)}\over 2\pi}\right)\cup
 dw^{(1)}\,,
\fe
which is gauge-invariant under \eqref{Zgauge}, but not quantized. 
Nonetheless, the total charge $Q_\text{A}$ is both gauge-invariant and quantized, and hence generates a (compact) U(1)$_\text{A}$ global symmetry.

\begin{figure}[h]
    \centering

\begin{tikzpicture}[
    scale=1.2,
    x={(1cm,0cm)},
    z={(0cm,1cm)},
    y={({0.5cm*cos(45)},{0.5cm*sin(45)})},
    >=stealth,
    midarrow/.style={
        postaction={
            decorate,
            decoration={
                markings,
                mark=at position 0.56 with {\arrow[scale=1]{Stealth}}
            }
        }
    }
]

\coordinate (A) at (0,0,0);        
\coordinate (B) at (2,0,0);

\coordinate (C1) at (2,0,0);
\coordinate (C2) at (2,2,0);
\coordinate (C3) at (2,0,2);
\coordinate (C4) at (2,2,2);

\coordinate (D1) at (0,0,0);
\coordinate (D2) at (0,-2,0);
\coordinate (D3) at (0,0,-2);
\coordinate (D4) at (0,-2,-2);

\draw[thick,midarrow] (A) -- (B);
\node at (1,0,-0.3) { $e^{i b_\ell}$};

\draw[fill=blue, fill opacity=0.15] (C1) -- (C2) -- (C4) -- (C3) -- cycle;
\draw[->,thick] 
    [canvas is yz plane at x=2]
    (1+0.383,1+0.321) arc[start angle=40, end angle=330, radius=0.5];
\node at (2.1,1,1.1) {$ dw$};
\draw[fill=blue, fill opacity=0.15] (D1) -- (D2) -- (D4) -- (D3) -- cycle;
\draw[->,thick] 
    [canvas is yz plane at x=0]
    (-1+0.383,-1+0.321) arc[start angle=40, end angle=330, radius=0.5];
\node at (-0.15,-.2,-1.2) {$ dw$};

\end{tikzpicture}
\caption{The axial charge of the short string $e^{ib_\ell}$ on a link $\ell$ is determined by the values of $dw$ on the two adjacent plaquettes $\mathfrak{t}(\ell)$ and  $\mathfrak{t}^{-1} (\ell)$, whose centers are displaced from the link center by half-lattice translations $\pm (\frac 12,\frac 12,\frac12)$.}
\label{fig:QA_act_on_b}
\end{figure}
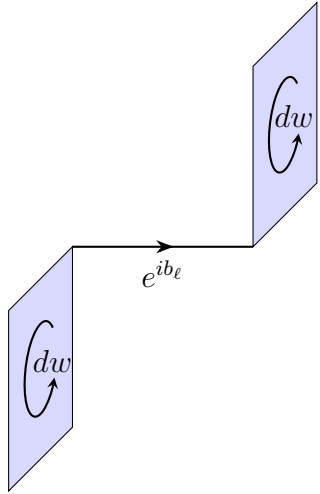

The axial U(1)$_\text{A}$ symmetry acts on the local operator $e^{ib_\ell}$ at  link $\ell$ as (Figure \ref{fig:QA_act_on_b}):\footnote{This equation can be written in terms of the cup product as
\ie
[Q_\text{A}, e^{i b_\ell} ] 
= 
e^{i b_\ell}
\int_{M_3} \left( 
\mathbf{l}^{(1)} \cup dw^{(1)}
+dw^{(1)}\cup \mathbf{l}^{(1)} \right)
\fe
where $\mathbf{l}^{(1)}$ is a 1-cochain that takes value $1$ on the link $\ell$ and 0 otherwise.
}
\ie\label{eq:QAactonb}
[ Q_\text{A} ,  e^{ib_\ell } ]
= \Big(\,
(dw)_{\mathfrak{t}(\ell) }
+(dw)_{\mathfrak{t}^{-1}(\ell)}
\,
\Big)\,
e^{ib_\ell}
\,,
\fe
where $\mathfrak{t}(\ell)$ is the plaquette whose center is separated from the center of the link $\ell$ by a half lattice translation $(\frac12,\frac12,\frac12)$ (see Figure \ref{fig:t}). 
More explicitly, the above can be written as
\ie
[Q_\text{A} ,  e^{i  b_i(\vec r)}]
=\sum_{j,k}\epsilon_{ijk}
\Big[\,
w_j(\vec r+\hat i)
-w_j(\vec r+\hat i +\hat k)
+w_j(\vec r -\hat j -\hat k)
-w_j(\vec r-\hat j)
\,\Big]\,e^{i b(\vec r)_i} \,.
\fe
Physically, $e^{ib_\ell}$ corresponds to an infinitesimally short axion string. 
The axial symmetry acts as a control gate, with the axial charge determined by  the local vortices $dw$ on the adjacent plaquettes. 
An example of a local operator carrying a fixed charge $q$ under $Q_\text{A}$ is
\ie\label{qQA}
\mathcal{O}_\ell = e^{ib_\ell}  \, \delta_{(dw)_{\mathfrak{t}(\ell)} + (dw)_{\mathfrak{t}^{-1}(\ell)},q} = e^{i b_\ell} \int_0^{2\pi} \frac{\mathrm{d}\theta}{2\pi} e^{i\theta
\left( 
(dw)_{\mathfrak{t}(\ell)} +(dw)_{\mathfrak{t}^{-1}(\ell)}-q
\right)
} ~,
\fe
The last factor is a projection operator to the subspace where $(dw)_{\mathfrak{t}(\ell)} + (dw)_{\mathfrak{t}^{-1}(\ell)} = q$.

The continuum counterpart $e^{i b_{\mu\nu}}$ of our short string is not a local operator because it is not invariant under the 2-form gauge transformation \eqref{2formgauge}. 
In our lattice model, the 2-form gauge invariance is not imposed strictly as discussed in Section \ref{sec:Hamiltonian}, so $e^{ib_\ell}$ is an allowed local operator on the lattice. 
If we had imposed $dw=0$ strictly, this would lead to a trivial axial charge $Q_\text{A}$, consistent with the fact that the charged local operators $e^{ib_\ell}$ are no longer gauge-invariant in that limit. 
We will discuss the continuum limit of our axial charge more in Section \ref{sec:matching}.

Note that $Q_\text{V}$ and $Q_\text{A}$ commute with each other, so the total group is U(1)$_\text{V}\times \text{U(1)}_\text{A}$. 
This is unlike the situation in \cite{Chatterjee:2024gje,Gioia:2025bhl,Xu:2025tdr,Gioia:2025xqj,Misumi:2025yjf} where the chiral symmetry groups are modified in lattice systems with a finite-dimensional local  Hilbert space. 
Our model also has a U(1)$_\text{W}^{(2)}$ winding 2-form global symmetry whose charge is $Q_\text{W}^{(2)}=\int_{\gamma_1}w^{(1)}$, discussed in Appendix \ref{app:winding}.\footnote{We denote a global symmetry or its charge with a superscript $(q)$ to indicate that it is a $q$-form global symmetry \cite{Gaiotto:2014kfa}. Such symmetries are generated by conserved operators of codimension $q$ in space (equivalently, codimension $q+1$ in spacetime). For ordinary, $0$-form global symmetries, we typically omit the superscript $(0)$. } 
However, we will not impose this symmetry. 
For instance, we allow ourselves to add
\ie\label{breakW}
    \sum_\ell \cos (b_\ell) 
    \, \delta_{(dw)_{\mathfrak{t}(\ell)} +(dw)_{\mathfrak{t}^{-1}(\ell)},0},
\fe
to the Hamiltonian to break U(1)$_\text{W}^{(2)}$ while preserving  U(1)$_\text{V}\times \text{U(1)}_\text{A}$.

This lattice axial charge $Q_{\text A}$ is a direct generalization of the constructions in 
\cite{Fidkowski:2025rsq,Thorngren:2026ydw}, which themselves were inspired by earlier work 
\cite{DeMarco:2021erp,DeMarco:2021dii,DeMarco:2023hoh} on Hall conductance. 
The authors of \cite{Fidkowski:2025rsq,Thorngren:2026ydw} work in a tensor-product Hilbert space, 
where the axial charge takes the form
\[
\int \lceil d\phi \rfloor \cup d\lceil d\phi \rfloor .
\]
Here $\lceil x \rfloor$ denotes the integer closest to $x$. 
In contrast, we reintroduce the Villain gauge field $w_\ell$ in place of $\lceil d\phi \rfloor$. 
Let us compare the two approaches. 
Our formulation avoids the apparent discontinuity associated with the function $\lceil x \rfloor$. 
Moreover, it makes explicit the local operators—such as $e^{i b_\ell}$—on which the axial symmetry acts. 
The trade-off is that the resulting Hilbert space no longer factorizes into a tensor product of local Hilbert spaces. 
Nevertheless, even in this setting, we can still gauge various global symmetries that are free of 't~Hooft anomalies, as we will show in later sections and review in Appendix \ref{app:gauging}.

\subsection{$\cal C,P,T$ symmetries}

Let $\mathcal{P}$ be the spatial parity operator, acting just by reversing orientations and reflecting positions.\footnote {$Q_\text{A}$ does not have a simple commutation relation with spatial reflections. 
By applying spatial reflection in the three directions, we find a total of 4 different lattice axial charges.} 
It acts on the vector and axial charges as 
\ie
\mathcal{P} Q_\text{V} \mathcal{P}^{-1} = Q_\text{V},~~~~
\mathcal{P} Q_\text{A} \mathcal{P}^{-1} = -Q_\text{A}.
\fe 
We also define an anti-unitary time-reversal operator $\mathcal{T}$ that acts as 
\ie
&\mathcal{T} \phi_s \mathcal{T}^{-1} = -\phi_s,~~
\mathcal{T} p_s \mathcal{T}^{-1} = p_s,\\
&\mathcal{T} w_\ell \mathcal{T}^{-1} = -w_\ell,~~
\mathcal{T} b_\ell \mathcal{T}^{-1} = b_\ell\,.
\fe
This time-reversal operator acts on the charges as
\ie
\mathcal{T} Q_\text{V} \mathcal{T}^{-1} = Q_\text{V},~~~~
\mathcal{T} Q_\text{A} \mathcal{T}^{-1} = Q_\text{A}.
\fe
There is also a unitary internal $\mathbb{Z}_2$  symmetry generated by $\mathcal{C}$:
\ie
&\mathcal{C} \phi_s \mathcal{C}^{-1} = -\phi_s,~~
\mathcal{C} p_s \mathcal{C}^{-1} = - p_s,\\
&\mathcal{C} w_\ell \mathcal{C}^{-1} = -w_\ell,~~
\mathcal{C} b_\ell \mathcal{C}^{-1} = -b_\ell\,.
\fe
It acts on the vector charge as a charge conjugation, but leaves the axial charge invariant:
\ie
\mathcal{C} Q_\text{V} \mathcal{C}^{-1} = -Q_\text{V},~~~~
\mathcal{C} Q_\text{A} \mathcal{C}^{-1} = Q_\text{A}.
\fe

The anti-unitary operator $\mathcal{CPT}$ acts on the charges as
\ie
(\mathcal{CPT}) Q_\text{V} (\mathcal{CPT})^{-1} = -Q_\text{V}\,,~~~~
(\mathcal{CPT}) Q_\text{A} (\mathcal{CPT})^{-1} = -Q_\text{A}\,.
\fe
See \cite{2023JMP....64i1901O,Kobayashi:2024bts,Li:2024dpq,Sopenko:2025tbc,Seiberg:2025zqx,Seiberg:2026icc} for recent discussions of $\mathcal{CPT}$ symmetries in Hamiltonian lattice models.

\subsection{Chiral anomaly}\label{sec:anomaly}

We have identified the lattice U(1)$_\text{V}\times \text{U(1)}_\text{A}$ symmetry in Section \ref{sec:U1VA}. 
In this subsection, we provide a quick check for the 't Hooft anomaly of the form U(1)$_\text{A}\text{-}\text{U(1)}_\text{V}\text{-}\text{U(1)}_\text{V}$ on the lattice. 
In the continuum, this anomaly corresponds to the following 4+1d topological action:
\ie\label{AVV}
{i\over 4\pi^2} \int A_\text{V}^{(1)} \wedge dA_\text{V}^{(1)} \wedge dA_\text{A}^{(1)}\,,
\fe
where $A_\text{V}^{(1)}$ and $A_\text{A}^{(1)}$ are the 1-form background gauge fields for U(1)$_\text{V}$ and for U(1)$_\text{A}$, respectively.  
More detailed checks of this anomaly will be given in Sections \ref{sec:noninvertible} and \ref{sec:2group} using the lattice $\theta$-angle and generalized symmetries. 
Since $Q_\text{V}$ and $Q_\text{A}$ locally have the same 4+1-dimensional anomaly-inflow term and transform in the same way under the standard $\mathcal{C}$, $\mathcal{P}$, and $\mathcal{T}$ symmetries of a free massless Dirac fermion in the continuum, we refer to them as the vector and axial symmetries, respectively.

To gauge U(1)$_\text{V}$ on the lattice, we introduce the gauge field $A_\ell$ and its conjugate electric field $E_\ell$ on every link, obeying the commutation relation:
\ie
[A_\ell , E_{\ell'} ] = i \delta_{\ell , \ell'}\,.
\fe
Furthermore, we impose the U(1)$_\text{V}$ Gauss law  
\ie\label{U1Vgauss}
(\delta E)_s = p_s\,,
\fe
which is the lattice version of $\nabla \cdot \vec E = q_\text{V}$. 
This Gauss law constraint implements the standard gauge transformation
\ie\label{U1Vgauge}
A_\ell \sim A_\ell +(d\alpha)_\ell\,,~~~~~\phi_s \sim \phi_s + \alpha_s\,,
\fe
where $\alpha_s\in \mathbb{R}$ is the gauge parameter associated with site $s$.

To ensure that the gauge group is a compact U(1)$_\text{V}$ (rather than $\mathbb{R}$), we need to further impose a constraint on the electric field. 
In the absence of the matter field, this constraint would simply require the electric field to be integer-valued, i.e.,  $\exp(2\pi i E_\ell ) = 1$. 
In our case,  the gauge field is coupled to the matter field $\phi_s$ and its Villain field $w_\ell$. It follows that this constraint needs to be modified to
\ie\label{ZVgauss}
\exp(2\pi i E_\ell - i b_\ell) = 1\,.
\fe
The intuitive reason is that both the U(1)$_\text{V}$ gauge transformation \eqref{U1Vgauge} and the Villain gauge transformation \eqref{Zgauge} shift $\phi_s$, so the corresponding conjugate fields $E_\ell$ and $b_\ell$ are subject to a correlated constraint. 
This constraint imposes a $\mathbb{Z}$ gauge transformation:
\ie\label{ZVgauge}
A_\ell \sim A_\ell +2\pi m_\ell \,,~~~~w_\ell \sim w_\ell +m_\ell\,,
\fe
where $m_\ell\in \mathbb{Z}$. 
As a consistency check, the two Gauss law constraints \eqref{U1Vgauss} and \eqref{ZVgauss} together imply the correct constraint $\exp(2\pi i p_s -i  (\delta b)_s)=1$ in \eqref{gauss1} for the Villain gauge transformation. 
See Appendix \ref{app:novillain} for more details.

What happens to the axial charge now that we have gauged U(1)$_\text{V}$?
The axial charge $Q_\text{A} = \int w^{(1)}\cup dw^{(1)}$  is quantized (i.e., $Q_\text{A}\in \mathbb{Z}$) since $w_\ell \in \mathbb{Z}$, but it is not invariant under the $\mathbb{Z}$ gauge transformation in \eqref{ZVgauge}. 
We can attempt to covariantize $Q_\text{A}$ to   find another axial charge
\ie
\widehat Q_\text{A} =  \int_{M_3} \left( w^{(1)} + {(d\phi)^{(1)} -A^{(1)}\over 2\pi } \right) 
\cup d\left( w^{(1)} + {(d\phi)^{(1)} -A^{(1)}\over 2\pi } \right) \,.
\fe
While the Villain field $w_\ell\in \mathbb{Z}$ is integer, $\phi_s,A_\ell$ are not quantized. 
Therefore, the covariant charge $\widehat Q_\text{A}$ is gauge-invariant but not quantized. 
Either way, we do not have a quantized and gauge-invariant axial charge that generates a (compact) U(1)$_\text{A}$ global symmetry after U(1)$_\text{V}$ is gauged. 
This is the lattice manifestation of the chiral anomaly between U(1)$_\text{V}$ and U(1)$_\text{A}$.

The U(1)$_\text{V}$ gauged Hamiltonian is
\ie\label{gaugedH}
H&=  
{1\over 2\gamma} 
\sum_\ell E_\ell^2
+{\gamma\over2}\sum_p \cos(  (dA)_p)
\\
+&
{1\over 2\beta}\sum_s p_s^2
+{\beta\over2}
\sum_\ell \left( (d\phi)_\ell+ 2\pi w_\ell  - A_\ell \right)^2
+{\lambda\over2} \sum_p \left(dw - {dA\over 2\pi} \right)_p ^2
\,.
\fe
We see that the quantized axial charge $Q_\text{A}$ commutes with this Hamiltonian, but it is not gauge-invariant. 
In contrast, the unquantized axial charge $\widehat Q_\text{A}$ is gauge-invariant, but does not commute with the Hamiltonian. 
The properties of the two axial charges are summarized in Table \ref{table:ABJ} and are parallel to their counterparts in QED, reviewed in Appendix \ref{app:QED}.

 \begin{table}
    \centering
    \begin{tabular}{|c|c|c|}
\hline  &&\\
\text{axial charges}&~~~~$Q_\text{A} = \int w\cup dw$~~~~
&~~ ~~$\widehat Q_\text{A} = \int \left( w+ {d\phi -A\over 2\pi } \right) 
\cup d\left( w + {d\phi-A\over 2\pi } \right)$~~~~ \\
&&\\
\hline ~~\text{quantized?}~~ &$\checkmark$& $\times$ \\
\hline ~~\text{gauge-invariant?}~~&  $\times$ &$\checkmark$\\
\hline~~\text{conserved?} ~~&$\checkmark$ &$\times$ \\
\hline \end{tabular}
\caption{After we gauge U(1)$_\text{V}$, there is no gauge-invariant, conserved, and quantized axial charge that generates U(1)$_\text{A}$.}\label{table:ABJ}
\end{table}

\subsection{The spectrum}\label{sec:spectrum}

Here we solve the Hamiltonian \eqref{Hamiltonian} and find its spectrum. For convenience, we copy the Hamiltonian here:
\ie\label{Hamiltonian2}
H=  
{1\over 2\beta}\sum_s p_s^2
+{\beta\over2}
\sum_\ell \left( ( d\phi)_\ell +2\pi w_\ell\right)^2
+{\lambda\over2} \sum_p [(dw)_p ]^2
\,.
\fe
Note that the Hamiltonian is independent of $b_\ell$, so we work in a diagonal basis for $\{w_\ell\}$ and diagonalize $H$.

To simplify the Hamiltonian, it is useful to expand $\phi_s$ around its classical solution in terms of $\{w_\ell\}$ that minimizes the Hamiltonian. Namely, we write
\ie \label{exp.cl}
    \phi = \phi_{\mathrm{cl}}[w] + \varphi \,,
\fe
where $\phi_{\mathrm{cl}}$ minimizes $\sum_\ell [(d\phi +2\pi w)_\ell]^2$ and thus satisfies
\ie
    \delta(d \phi_\mathrm{cl} + 2\pi w) = 0 \,.
\fe
The equation above determines $\phi_{\mathrm{cl}}$ up to an overall constant as a function of $\{w_\ell\}$. In momentum space, the explicit solution is given by
\ie
    (\phi_{\mathrm{cl}})_{\vec{k}} = 2\pi \sum_{i=x,y,z}\frac{1-e^{2\pi i k_i / L}}{\omega_{\vec k}^2} w_{\vec{k},i} ~, 
\fe
for $\vec{k} \neq 0$ where
\ie
    \omega_{\vec k} &= 2\sqrt{\sum_{i=x,y,z} \sin^2\left(\frac{\pi k_i}{L}\right)}\,,
\fe
and
\ie
    w_{\vec k,i} = \frac{1}{\sqrt{L^3}} \sum_{\vec r} e^{2\pi i \vec{k} \cdot \vec r \over L} w_{i}(\vec r) ~,  \qquad \phi_{\vec k} = \frac{1}{\sqrt{L^3}} \sum_{\vec r} e^{2\pi i \vec{k} \cdot \vec r \over L} \phi(\vec r) \,.
\fe

In the following, we work with variables $(p,\varphi)$ and $(w+\frac{d\phi_\mathrm{cl}}{2\pi},b)$ instead of $(p,\phi)$ and $(w,b)$. The advantage is the reduction in gauge redundancies associated with the $\bZ$ gauge transformations in \eqref{Zgauge}. In particular, $w+\frac{d\phi_\mathrm{cl}}{2\pi}$ is gauge invariant, and there is a single residual $\bZ$ gauge transformation that shifts every $\varphi_s$ by $2\pi$ at the same time:
\ie
    \varphi_s \sim  \varphi_s + 2\pi \,,
\fe
associated with the constraint $\exp( 2\pi i \sum_s p_s) = e^{2\pi i Q_{\mathrm{V}}} = 1$. As a result, the variables $(p,\varphi)$ and $(w+\frac{d\phi_\mathrm{cl}}{2\pi},b)$ are decoupled, and their associated Hilbert spaces factorize as a tensor product.

Using the new variables, the Hamiltonian simplifies into
\ie
H=  
{1\over 2\beta}\sum_s p_s^2
+{\beta\over2}
\sum_\ell [ (d\varphi)_\ell ]^2 +{\beta\over2}
\sum_\ell \left[ ( d\phi_\mathrm{cl} + 2\pi w)_\ell\right]^2
+{\lambda\over2} \sum_p [(dw)_p ]^2
\,.
\fe
The first two terms commute with the second two terms, thus each can be diagonalized simultaneously. We write $H = H[\varphi,p]+ H[w]$, and represent $H[\varphi,p]$ in momentum space:
\ie\label{HH}
    H[\varphi,p] &= \sum_{\vec{k} \neq 0} \omega_{\vec k} \left( a_{\vec k}^\dagger a_{\vec k} + \frac{1}{2} \right) + \frac{1}{2\beta L^3}(Q_{\mathrm{V}})^2 \,, \\
    H[w] &= {\beta\over2} \min_\phi \left\{
\sum_\ell \left[ ( d\phi + 2\pi w)_\ell\right]^2 \right\}
+{\lambda\over2} \sum_p [(dw)_p ]^2 \,,
\fe
where
\ie
    a_{\vec k} &= \frac{1}{\sqrt{2\beta \omega_{\vec k}}} p_{\vec{k}} - i \sqrt{\frac{\beta\omega_{\vec k}}{2}} \varphi_{\vec{k}} = \frac{1}{\sqrt{2 \beta L^3 \omega_{\vec k}}} \sum_{\vec r} e^{2\pi i \vec{k} \cdot \vec r \over L} \Big( p(\vec r) - i \beta\omega_{\vec k} \varphi(\vec r) \Big) \,.
\fe
The oscillators $a_{\vec k}$ satisfy the standard commutation relation $[ a_{\vec k},  a_{\vec k}^\dagger] = 1$.

We parametrize the Hilbert space by $\ket{N_{\vec k}, Q_{\mathrm{V}}, w_\ell}$, where $N_{\vec k} = a_{\vec k}^\dagger a_{\vec k}$ is the occupation number for the oscillators with $\vec k \neq 0$, and the wavefunction only depends on the gauge orbit of $\{w_\ell\}$. Namely, $\ket{N_{\vec k}, Q_{\mathrm{V}}, w_\ell} = \ket{N_{\vec k}, Q_{\mathrm{V}}, w_\ell + (d m)_\ell}$ for $m_s \in \bZ$. The energy is given by
\ie \label{energy}
    E = \sum_{\vec{k} \neq 0} \omega_{\vec k} ( N_{\vec k} + \frac{1}{2} ) + \frac{1}{2\beta L^3}(Q_{\mathrm{V}})^2 + {\beta\over2} \min_\phi \left\{
\sum_\ell \left[ ( d\phi + 2\pi w)_\ell\right]^2 \right\}
+{\lambda\over2} \sum_p [(dw)_p ]^2 \,.
\fe
The Hilbert space factorizes into $L^3-1$ oscillators $a_{\vec k \neq0}$, one rotor degree of freedom $Q_{\text{V}}$, and $2L^3+1$ rotors associated with the gauge orbits of $\{w_\ell\}$. To see that the dimension of gauge orbits of $\{w_\ell\}$ is $2L^3+1$, note that there are $L^3-1$ gauge constraints, where the missing gauge constraint is $e^{2\pi i Q_{\text{V}}}=1$, which does not act on $\{ w_\ell \}$.

\section{Continuum picture}

\subsection{Continuum limit}\label{sec:contlimit}

We will show that the continuum limit of our lattice Hamiltonian \eqref{Hamiltonian} with $\lambda>0$ is described by
\ie\label{freeboson}
    \mathcal{L} = \frac{f^2}{2} (\partial_\mu \phi)^2 \,,
\fe
where $\phi(x) \sim \phi(x)+2\pi$ is a compact scalar field. This theory contains oscillators, a vector mode (also known as the momentum mode), and three winding modes.

More specifically, starting from the dimensionless lattice Hamiltonian $H$, we introduce a lattice spacing $a$ and define a dimensionful Hamiltonian $H/a$, which will be compared to the continuum Hamiltonian $H_{\text{continuum}}$. The continuum limit is obtained by sending $a \to 0$ and $L \to \infty$ while keeping the physical size of the spatial torus, $R = La$ and  $f$  fixed. This requires scaling $\beta\sim a^2$ as we discuss below. 
We can subsequently take the large $R$ and fixed $f$ limit to study the scaling of each state.

The continuum limit is to be contrasted with the thermodynamic limit. 
In the latter case we do not introduce a lattice spacing and keep the Hamiltonian dimensionless. 
We take $L\to \infty$ while keeping $\beta$ finite, and analyze the scaling in $L$ for each state. 

We will see that the two limits agree for the $\lambda>0$ model. 
The $\lambda=0$ lattice model, on the other hand, has infinite ground state degeneracy in the continuum limit, and does not correspond to a quantum field theory with finite parameters.

\subsubsection*{Oscillators:}

In the continuum limit $L = \frac{R}{a} \gg 1$ with finite $k_i$, the oscillators in the lattice Hamiltonian \eqref{HH} (divided by $a$) have energy:
\ie
    \frac{1}{a}\omega_{\vec k} = \frac{2}{a}\sqrt{\sum_{i=x,y,z} \sin^2\left(\frac{\pi k_i}{L}\right)} \approx \sqrt{\sum_i \left(\frac{2\pi k_i}{R}\right)^2} \,,
\fe
which matches the dispersion relation for the continuum theory \eqref{freeboson} defined on the 3-torus with size $R = La$.

Since the dispersion relation is independent of $\beta$, the theory is gapless for all values of $\beta$, similar to the other modified Villain lattice models \cite{Sulejmanpasic:2019ytl,Gorantla:2021svj,Cheng:2022sgb,Fazza:2022fss,Seifnashri:2026ema}. 
Indeed, the presence of the lattice chiral anomaly is expected to forbid a gapped phase.  
This is to be contrasted with the standard XY model, where there is no anomaly and a phase transition occurs as we tune the coefficients in the kinetic term for the scalar field.

\subsubsection*{Vector mode:}
To relate $\beta$ to $f$, let us compute the energy of the vector modes in the continuum. Consider a spatially-constant solution $\phi = \phi_0(t)$ in the continuum. The continuum Lagrangian for such a configuration is $L = \frac{f^2 R^3}{2} (\dot{\phi_0})^2$, where $R^3$ is the volume of the spatial 3-torus. Thus, the continuum Hamiltonian is
\ie
    H_{\text{continuum}} = \Pi \dot{\phi_0} - L = \frac{1}{2f^2 R^3} \Pi^2\,,
\fe
where $\Pi$ is the conjugate variable to the periodic scalar $\phi$. Hence, $\Pi \in \bZ$ is quantized and is equal to the vector charge $Q_{\mathrm{V}}$. Comparing this with the lattice answer \eqref{energy} and the fact that the dimensionful Hamiltonian is $H/a$, we find $\frac{1}{2\beta L^3 a} = \frac{1}{2f^2 R^3}$, which implies
\ie
    \beta = f^2 a^2 \,.
\fe

Next, we move on to the winding modes. 
It is useful to write the energy $H[w]$ for the winding modes of the lattice model in \eqref{HH} in momentum space:
\ie \label{winding.energy}
    H[w] &= 2\pi^2\beta L \sum_{i=x,y,z} (Q_{\mathrm{W}}^i)^2 + \frac{\beta}{2} \sum_{i,\vec k \neq 0} \left| 2\pi w_{\vec k, i} + \varepsilon_{\vec k, i}^* \sum_j \frac{-2\pi \varepsilon_{\vec k, j}}{\omega_{\vec k}^2} w_{\vec k, j}  \right|^2 +{\lambda\over2} \sum_p [(dw)_p ]^2  \\
    &= 2\pi^2\beta L \sum_{i=x,y,z} (Q_{\mathrm{W}}^i)^2 + \frac12 \sum_{\vec{k} \neq 0} \left( \frac{\beta (2\pi)^2}{\omega_{\vec k}^2} + \lambda \right) \sum_{ij =xy,yz,zx}\left|(dw)_{\vec{k},ij}\right|^2
\fe
where
\ie \label{global.winding.charge}
    Q^i_{\mathrm{W}} &= \frac{1}{L^2} \sum_{\vec r} w_{i}(\vec r) \,, \qquad \text{and} \qquad \varepsilon_{\vec k, i} = e^{2\pi i k_i / L} - 1 \,.
\fe
Thus, we see that $H[w]$ has contributions from the global winding modes $Q_{\mathrm{W}}^i$, and local winding modes $(dw)_p$. As we will see, the configurations with non-zero local winding modes ($dw\neq0$) are suppressed in the continuum limit, and the finite energy configurations correspond to those with $dw=0$.

\subsubsection*{Local winding modes:}

First, for $\lambda=0$, we show that the local winding modes associated with $(dw)_p \neq 0$ have energy of order $a$. In the continuum limit, the energy of the local winding modes vanishes, leading to an extensive degeneracy. Thus, the model does not have a good continuum limit for $\lambda=0$.

To see this, consider a configuration of $w$, where $w_\ell = 1$ for a single link $\ell = \ell_0$ and $w_\ell=0$ for the rest $\ell \neq \ell_0$. The energy, measured by the dimensionful Hamiltonian $H/a$, of this configuration is
\ie
    \frac{\beta}{2a} \sum_{ij,\vec{k} \neq 0}  \frac{ (2\pi)^2}{\omega_{\vec k}^2}|(dw)_{\vec{k},i,j}|^2 = \frac{2\pi^2 \beta}{a} \sum_{\vec k \neq 0} \left( w_{-\vec k, i} w_{\vec k, i} - \frac{|\varepsilon_{\vec k, i} w_{\vec k, i}|^2}{\omega_{\vec k}} \right) \leq \frac{2\pi^2 \beta}{a} \sum_{\vec r,i} (w_{i}(\vec r))^2 = \frac{2\pi^2 \beta}{a} \,.
\fe
This energy is at most ${2\pi^2 \beta}/{a} = 2 \pi^2 f^2 a$ and hence vanishes in the continuum limit. 

To find a sensible theory in the continuum limit, we must choose $\lambda > 0$ to lift the energy of the local winding modes. In that case, local winding modes receive an energy $4{\lambda}/{a}$ from the $\lambda \sum_p [(dw)_p]^2$ term. All local winding modes become infinitely massive in the continuum limit, and only those configurations with $dw=0$ survive. This is consistent with the continuum theory, where there is no configuration with $dw\neq0$, and there are only global winding modes associated with non-contractible 1-cycles of the spatial manifold.

The $\lambda=0$ model is somewhat reminiscent of the exotic field theories in \cite{Seiberg:2020bhn,Seiberg:2020wsg,Seiberg:2020cxy,Gorantla:2020xap}. 
In particular, the thermodynamic limit and continuum limit are different \cite{Gorantla:2021bda}. 
In the continuum limit, where $L\to \infty$ with $\beta =\mathcal{O}( 1/L^2)$, the local winding modes are lighter than all other modes. 
On the other hand, in the thermodynamic limit where $L\to \infty$ with $\beta$ held fixed, the local winding modes ($H=\mathcal{O}(1)$) are much heavier than the vector modes $(H= \mathcal{O}(1/L^3)$) and the oscillators $(H=\mathcal{O}(1/L)$).

\subsubsection*{Global winding modes:}

Given $dw=0$, the global winding charges $Q_{\mathrm{W}}^i$, defined in \eqref{global.winding.charge}, are quantized as integers. They become the winding charges in the continuum limit. We now compute the energy of such winding modes in the continuum theory and compare it with the lattice answer.

Consider the continuum configuration $\phi = \frac{2\pi nx }{R}$, which has winding charge $Q_{\mathrm{W}}^x = n$ and $Q_{\mathrm{W}}^{y}=Q_{\mathrm{W}}^{z}=0$. The energy of this configuration is 
\ie
    H_{\text{continuum}} = \frac{f^2R^3}{2} \left(\partial_x \frac{2\pi n x}{R} \right)^2 = 2\pi^2 f^2 R n^2 \,,
\fe
which matches the winding contribution to the dimensionful lattice Hamiltonian \eqref{winding.energy}
\ie
\frac{H}{a}= \frac{2\pi^2 \beta L}{a} n^2
\fe
 upon setting $R=La$ and $\beta=f^2a^2$. 

We conclude that, for any $\lambda>0$, the continuum limit of our lattice Hamiltonian \eqref{Hamiltonian2} gives the free boson field theory in \eqref{freeboson}.

\subsection{Symmetry transmutation and the axion coupling}\label{sec:matching}

We have shown in the previous subsection that the continuum limit of our lattice Hamiltonian \eqref{Hamiltonian} is a free compact boson field theory ${\cal L}= {f^2\over 2}(\partial_\mu\phi)^2$ in \eqref{freeboson}. 
It has an ordinary (0-form) U(1)$_\text{V}$ symmetry that shifts $\phi$ by a constant, which is spontaneously broken. 
The vector current is given by $j^\text{V}_\mu=if^2 \partial_\mu\phi$. 
There is also a 2-form winding U(1)$_\text{W}^{(2)}$ symmetry, whose current is $j^\text{W}_{\mu\nu\rho} = {1\over 2\pi} \epsilon_{\mu\nu\rho\sigma}\partial^\sigma\phi$. 
The two global symmetries have a mixed 't Hooft anomaly captured by the following 4+1d topological action:
\ie\label{IRanomaly}
{i\over 2\pi}\int A_\text{V}^{(1)}\wedge dC^{(3)}
\fe
where $A_\text{V}^{(1)}$ and $C^{(3)}$ are the 1-form and 3-form background gauge fields for U(1)$_\text{V}\times \text{U(1)}_\text{W}^{(2)}$. 
Both symmetries arise from exact symmetries of the microscopic lattice Hamiltonian \eqref{Hamiltonian}; however, we do not impose the lattice U(1)$_\text{W}^{(2)}$ symmetry (see Appendix \ref{app:winding}).

While the U(1)$_\text{A}$ axial symmetry generated by $Q_\text{A}$ in \eqref{QA} acts faithfully in the UV lattice model, it does not act faithfully in this IR field theory. 
The axial current $j^\text{A}_\mu$ is therefore a redundant operator.\footnote{An operator is called \emph{redundant} if its correlation functions vanish at separated points, although it may produce contact terms when operator insertions coincide. See, e.g., \cite{Closset:2012vg,Closset:2012vp,Cordova:2018cvg} for further discussions.}  
Indeed, the short strings, on which U(1)$_\text{A}$ acts,  correspond to excitations with large energy for nonzero $\lambda$ and therefore decouple from the low-energy dynamics. 
Instead, the lattice axial symmetry acts effectively as a 2-form winding symmetry. 
This is an example of the phenomenon known as symmetry transmutation \cite{Seiberg:2025bqy}. More generally, a theory with a faithful higher-form symmetry can be coupled to background gauge fields of lower-form symmetries through symmetry fractionalization (see, e.g.,  \cite{Barkeshli:2014cna}).

Next, we discuss anomaly matching from our lattice model to this field theory. 
The microscopic anomaly ${i\over 4\pi^2} \int A_\text{V}^{(1)} \wedge dA_\text{V}^{(1)} \wedge dA_\text{A}^{(1)}$ in \eqref{AVV} between U(1)$_\text{V}$ and U(1)$_\text{A}$ is matched by the IR anomaly in \eqref{IRanomaly} with the gauge field $C^{(3)}$ for the 2-form global symmetry chosen to be
\ie
C^{(3)} = {1\over 2\pi} A_\text{V}^{(1)}\wedge dA_\text{A}^{(1)}\,.
\fe
The IR Lagrangian coupled to the U(1)$_\text{V}\times \text{U(1)}_\text{A}$ background gauge fields is
\ie\label{IRQFT}
{f^2\over2} (d\phi^{(0)} - A_\text{V}^{(1)})\wedge \star 
 (d\phi^{(0)} - A_\text{V}^{(1)})
 - {i\over 4\pi^2} d\phi^{(0)} \wedge A_\text{V}^{(1)} \wedge dA_\text{A}^{(1)}\,
 \,.
\fe
Under a gauge transformation
\ie
\phi^{(0)}\sim \phi^{(0)}+ \lambda_\text{V}^{(0)}\,,~~~A_\text{V}^{(1)}\sim A_\text{V}^{(1)}+d\lambda_\text{V}^{(0)}\,,~~~A_\text{A}^{(1)}\sim A_\text{A}^{(1)}+d\lambda^{(0)}_\text{A}
\fe
this effective action transforms by a term
\ie
-{i\over 4\pi^2} \int d\lambda_\text{V}^{(0)} \wedge A_\text{V}^{(1)} \wedge dA_\text{A}^{(1)}
\fe
which is canceled by inflow from the UV anomaly \eqref{AVV}.

In quantum field theory,  chiral anomalies imply the presence of certain non-analytic contributions to the current three-point correlation functions $\langle j^\text{A}_\mu(x)j^\text{V}_\nu(y)j^\text{V}_\rho(z)\rangle$ in momentum space \cite{Frishman:1980dq,Coleman:1982yg,Cordova:2018cvg}. 

Taking functional derivatives of \eqref{IRQFT} with respect to $A_\text{A}^{(1)}$ and $A_\text{V}^{(1)}$, we obtain the current three-point function
\ie
\langle j^\text{A}_\mu(x) j^\text{V}_\nu(y) j^\text{V}_\rho(z) \rangle
=
\frac{f^2}{4\pi^2}\epsilon_{\mu\nu\alpha\beta}
\partial^\beta \delta^{(4)}(x-y)
\langle \partial^\alpha\phi (y)
\partial_\rho \phi(z)
\rangle
+(y\leftrightarrow z, \nu\leftrightarrow \rho) \,.
\fe
This is a \textit{partial} contact term in two of the three points; it is not a complete contact term supported only when all three points coincide. 
Such a partial contact term cannot be removed by local counterterms and contributes to the non-analytic structure in momentum space responsible for the chiral anomaly.

\subsection{Yukawa field theory}\label{sec:Yukawa}

The lattice axial U(1)$_\text{A}$ symmetry acts faithfully in the UV lattice model  \eqref{Hamiltonian}, but it acts effectively as a higher-form symmetry in the IR field theory \eqref{IRQFT}. 
Here, we discuss a UV quantum field theory that has the same symmetry structure and flows to the same IR field theory. 
This model was discussed in \cite{Cordova:2018cvg}.

The UV field theory contains 4 massless Weyl fermions $\psi_a^{(I)}$ and a complex scalar field $\Phi$. Here $I=1,\cdots,4$ is the flavor index and $a$ is the left-handed, two-component spinor index. 
The fermions and the scalar are coupled through the following Yukawa coupling:
\ie\label{yukawa}
g \Phi^\dagger  \epsilon^{ab} \psi_a^{(1)} \psi_b^{(3)}
+g \Phi  \epsilon^{ab} \psi_a^{(2)} \psi_b^{(4)}+\text{h.c.}
\fe
This UV field theory has an ordinary, faithful U(1)$_\text{V}\times \text{U(1)}_\text{A}$ global symmetry with the following charge assignment:
\ie\label{Yukawacharge}
\left.\begin{array}{|c|c|c|c|c|c|}
\hline  &~~\psi^{(1)}_a ~~& ~~\psi^{(2)}_a~~&~~\psi^{(3)}_a~~&~~\psi^{(4)}_a ~~ &~~~~\Phi~~~~\\
\hline ~~Q_\text{V}~~ &+1& -1&0&0 &+1\\
\hline ~~Q_\text{A}~~&  +1 &+1&-1&-1 &0\\
\hline \end{array}\right.
\fe
The fermion charges are chosen so that the only 't Hooft anomaly is the one between U(1)$_\text{A}\text{-}\text{U(1)}_\text{V}\text{-}\text{U(1)}_\text{V}$ \cite{Cordova:2018cvg,Thorngren:2026ydw}. 
In particular, there is no mixed gravitational anomaly and no self anomaly for U(1)$_\text{A}$. 
Therefore, this is a quantum field theory locally sharing the same global symmetry and 't Hooft anomaly as our lattice model.
% Therefore, this is a quantum field theory in the same universality class as our lattice Hamiltonian \eqref{Hamiltonian}: they share the same global symmetry and 't Hooft anomaly.
\footnote{Strictly speaking, this field theory \eqref{yukawa} is a spin/fermionic quantum field theory which depends on the choice of the spin structure for spacetime, while our lattice model \eqref{Hamiltonian} is non-spin/bosonic. This distinction is not important for the rest of the discussion since the 't Hooft anomaly \eqref{AVV} is compatible with non-spin theories. The lattice Hamiltonian \eqref{Hamiltonian} also has a winding 2-form symmetry discussed in Appendix \ref{app:winding}, while this UV field theory \eqref{yukawa} does not. However, we do not impose the lattice winding symmetry and allow ourselves to add local terms like \eqref{breakW} to break it. } 
There, the axial U(1)$_\text{A}$ symmetry acts on the short strings, while here it acts on the fermions.

Having specified the UV field theory, we now turn on a Higgs potential for $\Phi$ so that it acquires a vev. 
All 4 fermions acquire a mass through the Yukawa coupling, and the vector U(1)$_\text{V}$ global symmetry is spontaneously broken. 
The IR field theory is described by a Goldstone boson $\phi\sim \phi+2\pi$, which is the phase of $\Phi$ as $\Phi = \rho  e^{ i\phi}$. 
The 't Hooft anomaly matching argument dictates that the Goldstone boson is coupled to the background gauge fields $A_\text{A}^{(1)},A_\text{V}^{(1)}$ as in \eqref{IRQFT}, the same IR field theory as our lattice model \eqref{Hamiltonian}.  
U(1)$_\text{A}$ does not act faithfully on the low-energy local degrees of freedom and is transmuted into a higher-form symmetry. 
\section{Non-invertible symmetry from gauging U(1)$_\text{V}$}\label{sec:noninvertible}

\subsection{The gauged model}

In Section \ref{sec:anomaly}, we gauged the U$(1)_{\text{V}}$ symmetry by coupling the scalar theory in \eqref{Hamiltonian} to a gauge field $A^{(1)}$ (and its conjugate $E^{(1)}$). While this representation contains the minimal set of operators, it does not preserve the magnetic 1-form symmetry, which will play an important role in our later discussions on the $\theta$-angle and non-invertible symmetries.

In this section, we will gauge U(1)$_\text{V}$ by further introducing a 2-form Villain gauge field $n^{(2)}$ for $A^{(1)}$.
We denote the conjugate operator of $n^{(2)}$ by $\tilde{A}^{(2)}$. They satisfy the following commutation relation,
\ie
\left[\tilde{A}_p,n_{p'}\right]=i\delta_{p,p'}
\fe
$n^{(2)}$ gauges a $\mathbb{Z}$ symmetry which shifts the gauge field as 
$A_\ell\sim A_\ell+2\pi$. Starting from the scalar Hamiltonian \eqref{Hamiltonian}, we gauge the U$(1)_\text{V}$ symmetry and the $\mathbb{Z}$ symmetry following the steps in Appendix \ref{app:villain}, arriving at the following Hamiltonian
\ie\label{HV}
H_\text{V} & = {1\over 2\gamma} \sum_\ell E_\ell^2
+{\gamma\over2}
\sum_p ((dA)_p -2\pi n_p)^2\\
+& {1\over 2\beta} \sum_s p_s^2
+{\beta\over2}
\sum_\ell ( (d\phi)_\ell +2\pi w_\ell -A_\ell)^2
+{\lambda\over2} \sum_p ((dw)_p - n_p)^2 \,.
\fe
We also obtain the following Gauss law constraints
\ie\label{eq:Gausslaw_gaugeV}
&\exp(2\pi i w_\ell) = 1\,,\\
&\exp(2\pi i n_p)=1\,,\\
&(\delta E)_s  = p_s\,,\\
&\exp (2\pi i E_\ell -i b_\ell - i (\delta \tilde A)_\ell)=1\,.
\fe
We impose the flatness condition $dn^{(2)}=0$ to suppress 
monopoles of the U$(1)_\text{V}$ gauge field. The first two equations in \eqref{eq:Gausslaw_gaugeV} imply that $w^{(1)}$ and $n^{(2)}$ are $\mathbb{Z}$-valued. The third Gauss law 
is the lattice analog of $\nabla \cdot \vec E = q_\text{V}$, 
which implements the U$(1)_\text{V}$ gauge transformation,
\ie\label{eq:AgaugeU1}
A^{(1)} \sim A^{(1)} + d\alpha^{(0)}\,,~~~~\phi^{(0)} \sim \phi^{(0)}  + \alpha^{(0)}\,.
\fe
where $\alpha_s\in \mathbb{R}$. The fourth Gauss law enforces the $\mathbb{Z}$ gauge invariance under
\ie\label{eq:AgaugeZ}
A^{(1)} \sim A^{(1)} + 2\pi k^{(1)}\,,~~~
w^{(1)} \sim w^{(1)}  +k^{(1)}\,,~~~
n^{(2)} \sim n^{(2)} + d k^{(1)}\,,
\fe
where $k_\ell\in\mathbb{Z}$. The earlier Gauss law \eqref{gauss1}, which is associated with the $\mathbb{Z}$ gauge transformation of the scalar field $\phi^{(0)}$, need not be included explicitly, since it is implied by the third and fourth equations in \eqref{eq:Gausslaw_gaugeV}.

The advantage of this alternative way of gauging U(1)$_\text{V}$ is that the magnetic 1-form global symmetry U(1)$_m^{(1)}$ is manifest and is generated by the following operator,
\ie
Q_m = \int_{\Sigma_2} (n^{(2)} - dw^{(1)})\,.
\fe
where $\Sigma_2$ is a 2-cycle. 
This corresponds to the magnetic flux through $\Sigma_2$ in the continuum. 
This operator is gauge-invariant because it commutes with all the Gauss law constraints. It commutes with the Hamiltonian and is therefore conserved; it is also topological following the flatness condition $dn^{(2)}=0$. Note that the $dw^{(1)}$ term vanishes upon integration; it is included so that the charge density $n^{(2)}-dw^{(1)}$ is manifestly gauge invariant. 
Note that if $\lambda$ is sent to infinity, the magnetic 1-form symmetry charge $Q_m^{(1)}$ trivializes, so does the axial charge $Q_\text{A}$ for the scalar Hamiltonian.

\subsection{Axial charges and the lattice $\theta$-angle}\label{sec:theta}

After gauging U$(1)_\text{V}$, the original axial charge \eqref{QA} is no longer gauge invariant. We may seek alternative definitions of the axial charge $Q_\text{A}$. For this purpose, it is convenient to define the following Chern-Simons operator,
\ie
\text{CS}[x^{(1)} ]=  \int_{M_3} (x^{(1)}\cup dx^{(1)} - n^{(2)}\cup x^{(1)} - x^{(1)}\cup n^{(2)})\,.
\fe
Here $x^{(1)}$ is a general 1-cochain. 
Later we will specialize $x^{(1)}$ to either $A^{(1)}/2\pi$ or $w^{(1)}$. When $x^{(1)}= A^{(1)}/2\pi$, the operator $\text{CS}[A^{(1)}/2\pi ]$ agrees with the Chern-Simons action introduced in \cite{Jacobson:2023cmr} using $dn^{(2)}=0$, which we imposed by hand. 
See also \cite{Chen:2019mjw,DeMarco:2019pqv,Han:2021wsx,Han:2022cnr,Jacobson:2024hov,Xu:2024hyo,Ikeda:2026lyl,Xu:2026ygx} for recent related works.

Now, we define the following axial charges,
\ie
&\widehat{\mathbf{Q}}_\text{A}  = \text{CS}[w^{(1)}] - \text{CS}\left[{A^{(1)}\over2\pi}\right],\\
&\mathbf{Q}_\text{A}  =\text{CS}[w^{(1)}] \,.
\fe
See \cite{Fidkowski:2025rsq} for the corresponding current operators. 
Both charges reduce to $\int w^{(1)}\cup dw^{(1)}$ when the U$(1)_\text{V}$ gauge field and the associated Villain gauge field are turned off, and therefore provide natural covariantizations of the original axial charge. Both $\mathbf{Q}_\text{A}$ and $\widehat{\mathbf{Q}}_\text{A}$ are invariant under the U$(1)_\text{V}$ gauge transformation \eqref{eq:AgaugeU1}. The charge $\widehat{\mathbf{Q}}_\text{A}$ is also invariant under the $\mathbb{Z}$ gauge transformation \eqref{eq:AgaugeZ}, but it is neither quantized nor  conserved. By contrast, $\mathbf{Q}_\text{A}$ is quantized and conserved but not $\mathbb{Z}$ gauge invariant. The relation
\ie
\mathbf{Q}_\text{A} = \widehat{\mathbf{Q}}_\text{A} + \text{CS} 
\left[ {A^{(1)}
\over2\pi}\right]
\fe
is the lattice counterpart of the continuum relation $\star j_\text{A}  =\star \hat j_\text{A}  - {1\over 8\pi^2 }AdA$ in QED from \eqref{j} and \eqref{jhat}, 
where $ j_\text{A}$ is conserved but not gauge invariant, while $\hat j_\text{A}$ is gauge invariant but not conserved. 

In the continuum, the axial anomaly is reflected in a shift of the $\theta$-angle under an axial rotation; this is also the case on the lattice. We implement this by performing a unitary transformation by $\exp\left(i\frac{\theta}{2}\mathbf{Q}_\text{A}\right)$ on the Hamiltonian and Gauss law constraints. 
Here we use $\theta/2$  because the $\theta$-angle is $4\pi$-periodic in bosonic systems  \cite{Witten:1995gf}.
Under this transformation, the Hamiltonian remains invariant, but the fourth Gauss law in \eqref{eq:Gausslaw_gaugeV} is mapped to
\ie\label{eq:thetaGauss}
\exp\left( 2\pi i E_\ell  - ib_\ell - i \left(\delta \tilde A\right)_\ell\right)
=\exp\left(-i{\theta\over 2}\left( n_{\mathfrak{t}(\ell)} +n_{\mathfrak{t}^{-1}(\ell)}\right)\right)\,.
\fe
Here $\mathfrak{t}$ is a lattice translation by $\left({1\over 2},{1\over 2},{1\over 2}\right)$, which sends a link to a plaquette (Figure \ref{fig:t}). 
Equation \eqref{eq:thetaGauss} is the lattice counterpart of a $\theta$-term in the continuum. 
Indeed, integrating the equation above over a 2-cycle $\Sigma_2$ leads to the following quantization condition,
\ie
\int_{\Sigma_2}\left(\star_{\mathfrak{t}} E^{(1)}+\frac{\theta}{4\pi}n^{(2)}+\frac{\theta}{4\pi}\star_{\mathfrak{t}}^2n^{(2)}\right)=\int_{\Sigma_2}\left(\star_{\mathfrak{t}} E^{(1)}+\frac{\theta}{2\pi}n^{(2)}\right)\in \mathbb{Z}\,.
\fe
The operator $\star_{\mathfrak{t}}$ is defined on cochains by $(\star_{\mathfrak{t}}n)_{\ell}=n_{\mathfrak{t}(\ell)}$ (see Appendix \ref{app:cochain}). The operator $\star_\mathfrak{t}^2$ acts on $n^{(2)}$ as a lattice translation by $(1,1,1)$. Consequently, the integrals of $n^{(2)}$ and $\star_\mathfrak{t}^2n^{(2)}$ over a 2-cycle $\Sigma_2$ are equal, thanks to the flatness condition $dn^{(2)}=0$. Since $E^{(1)}$ and $n^{(2)}$ correspond to the electric field and magnetic field, respectively, the above condition reduces in the continuum limit to
\ie\label{eq:witteneffect}
    q_e+\frac{\theta}{2\pi}q_m\in \mathbb{Z}\,,
\fe
where $q_e$ and $q_m$ denote the electric charge and magnetic charge, respectively. This is precisely the Witten effect, in which the electric charge is shifted by the magnetic charge in the presence of a non-zero $\theta$\cite{Witten:1979ey}. The Euclidean version of this formulation of the lattice $\theta$-angle was studied in \cite{Sulejmanpasic:2019ytl,Anosova:2022cjm,Aoki:2026pvq}. See also \cite{Seiberg:1984id} for earlier work.

The $\theta$-angle can also be described in a way that is closer to the continuum. We conjugate the system by a unitary operator 
\ie
\exp\left(i\frac{\theta}{4\pi}
\sum_\ell A_\ell\left(n_{\mathfrak{t}(\ell)}+n_{\mathfrak{t}^{-1}(\ell)}-\frac{1}{2\pi}(dA)_{\mathfrak{t}(\ell)}\right)\right)\,.
\fe
The Gauss law constraints are restored to the original ones \eqref{eq:Gausslaw_gaugeV}, while the Hamiltonian \eqref{HV} is changed to the following,
\ie\label{eq:thetaHamiltonian}
H_{\text{V},\theta} & = {1\over 2\gamma} \sum_\ell \left(E_\ell+\frac{\theta}{8\pi^2}\left[(dA-2\pi n)_{\mathfrak{t}(\ell)}+(dA-2\pi n)_{\mathfrak{t}^{-1}(\ell)}\right]\right)^2
+{\gamma\over2}
\sum_p ((dA)_p -2\pi n_p)^2\\
+& {1\over 2\beta} \sum_s p_s^2
+{\beta\over2}
\sum_\ell ( (d\phi)_\ell +2\pi w_\ell -A_\ell)^2
+{\lambda\over2} \sum_p ((dw)_p - n_p)^2 \,.
\fe
We now show that the expression above agrees with the Hamiltonian of a continuum U$(1)$ gauge theory. Since the $\theta$-angle does not affect the matter sector, it suffices to compare with the pure U$(1)$ gauge theory in the continuum. The corresponding (Lorentzian) Lagrangian density is given by
\ie
\mathcal{L}=-\frac{1}{4e^2} F_{\mu\nu}F^{\mu\nu}-\frac{\theta}{32\pi^2}\epsilon^{\mu\nu\rho\sigma}F_{\mu\nu}F_{\rho\sigma}\,.
\fe
Denoting $\Pi_i={\partial\mathcal{L}\over \partial \dot{A}_i}$ as the conjugate momentum of $A_i$,  the Hamiltonian density is\footnote{
On the lattice, we denote the conjugate momentum of the gauge field by $E_\ell$, which in the continuum corresponds to a linear combination of the electric and magnetic fields in the presence of the $\theta$-angle.} 
\ie
\mathcal{H}=\Pi_i\dot{A}_i-\mathcal{L}=\frac{e^2}{2}\left(\Pi_i+\frac{\theta}{4\pi^2}\epsilon^{ijk}\partial_j A_k\right)^2+\frac{1}{4e^2}F_{ij}F_{ij}\,.
\fe
This agrees with \eqref{eq:thetaHamiltonian} under the identifications $\Pi_i\sim E_\ell$ with $\ell$ a link in the $i$ direction, and  $F_{ij}\sim (dA-2\pi n)_p$ with $p$ a plaquette on the $ij$-plane.

\subsection{Non-invertible symmetry}\label{sec:subnoninv}

In the continuum QED with a massless electron, it was shown in \cite{Choi:2022jqy,Cordova:2022ieu} that the classical U(1)$_\text{A}$ symmetry is not entirely broken by the ABJ anomaly; rather, an axial rotation by angle $2\pi p/N$ with $p/N\in \mathbb{Q}/\mathbb{Z}$ becomes a non-invertible global symmetry. 
This motivated the authors of \cite{Fidkowski:2025rsq} to find a lattice construction of this non-invertible symmetry, which led them to discover the lattice axial charge $Q_\text{A}$. 
In this subsection, we reformulate these non-invertible symmetry operators in terms of the Villain gauge fields within our framework.

Consider the $\mathbb{Z}_N^{(1)}$ subgroup of the U$(1)_m^{(1)}$ magnetic 1-form symmetry, and restrict to the subspace that is invariant under this symmetry by imposing the following condition,
\ie\label{eq:ZNsubspace}
\exp\left({2\pi i\over N} \int_{\Sigma_2} (n^{(2)} - dw^{(1)})\right)=1\,,
\fe
for all 2-cycles $\Sigma_2$. 
This implies that in this subspace, the 2-cochain $n^{(2)} - dw^{(1)}$ is exact modulo $N$:
\ie
    n^{(2)} - dw^{(1)}=d a^{(1)}~\text{mod}~N\,,
\fe
where $a^{(1)}$ is a gauge-invariant, integer-valued 1-cochain. 
In the subspace satisfying \eqref{eq:ZNsubspace}, there exists a gauge-invariant $\mathbb{Z}_N^{(0)}$ 0-form symmetry operator, given by
\ie\label{eq:noninv}
\exp\left(
{2\pi  i \over N}
\int_{M_3} 
a^{(1)} \cup da^{(1)}
\right)\,.
\fe
This operator generates a $\mathbb{Z}_N$ subgroup of U$(1)_\text{A}$, as it reduces to $e^{{2\pi i\over N}Q_\text{A}}$ when the gauge field is turned off, with $Q_\text{A}$ defined in \eqref{QA}. This operator is only defined on the subspace satisfying \eqref{eq:ZNsubspace}.  
To extend it to the entire Hilbert space, it must be multiplied with a projection operator onto this subspace:
\ie\label{eq:condensation}
\mathcal{C} = {1\over N}\sum_{\Sigma_2\in H_2 (M_3,\mathbb{Z}_N)}
\exp\left(
{2\pi i \over N}\int_{\Sigma_2} (n^{(2)} -dw^{(1)})
\right)
\fe
This operator is known as the condensation operator in \cite{Roumpedakis:2022aik,Choi:2022zal}. 
The resulting operator, which is the product of \eqref{eq:noninv} and \eqref{eq:condensation}, thus defines a non-invertible symmetry operator.

\subsection{Anomaly-free  non-invertible symmetries in the Abelian Higgs model}\label{sec:abelianHiggs}

We discuss the continuum picture of this non-invertible symmetry and its 't Hooft anomaly. 

We start with the UV continuum field theory of Section \ref{sec:Yukawa}.
This field theory has 4 massless left-handed Weyl fermions coupled to a complex scalar $\Phi$ via the Yukawa coupling in \eqref{yukawa}. 
It has a U(1)$_\text{V}\times \text{U(1)}_\text{A}$ global symmetry, with  charge assignments given in \eqref{Yukawacharge} and 
a U(1)$_\text{A}\text{-}\text{U(1)}_\text{V}\text{-}\text{U(1)}_\text{V}$ 't Hooft anomaly.

Next, we gauge U(1)$_\text{V}$ to find the scalar-fermion-QED with a Yukawa coupling. 
\ie
{\cal L}& = 
i \psi^{\dagger(1)}\bar \sigma^\mu(\partial_\mu - i A_\mu)\psi^{(1)}+
i \psi^{\dagger(2)}\bar \sigma^\mu(\partial_\mu +i A_\mu)\psi^{(2)}
+\sum_{I=3,4}i \psi^{\dagger(I)}\bar \sigma^\mu\partial_\mu\psi^{(I)}\\
&+\frac 12 |(\partial_\mu-i  A_\mu)\Phi|^2 -V(\Phi)
-{1\over 4e^2}F_{\mu\nu}F^{\mu\nu}
+\left(
g \Phi^\dagger  \epsilon^{ab} \psi_a^{(1)} \psi_b^{(3)}
+g \Phi  \epsilon^{ab} \psi_a^{(2)} \psi_b^{(4)}+\text{h.c.}\right)\,.
\fe
The construction of \cite{Choi:2022jqy,Cordova:2022ieu} implies that the  U(1)$_\text{A}$ symmetry, which acts on the fermions with charges $(+1,+1,-1,-1)$, is not broken by the anomaly, but becomes a non-invertible global symmetry which acts on the fermion fields.

We now turn on a Higgs potential $V(\Phi)$ for $\Phi=\rho e^{i\phi}$ so that it acquires a vev. 
All fermions become massive, and the IR limit is described by the Abelian Higgs model: 
\ie\label{abelianHiggs}
 {\cal L}_\text{V} = {f^2\over 2}(\partial_\mu \phi - A_\mu)^2-{1\over 4e^2} F_{\mu\nu}F^{\mu\nu}  \,,~~~\phi\sim \phi+2\pi\,,
\fe
which is trivially gapped. 
One can straightforwardly verify that this is also the continuum limit of our gauged lattice Hamiltonian $H_\text{V}$. 
In the IR Abelian Higgs model, the non-invertible global symmetry becomes non-faithful since it does not act on $\phi$.
It has no 't Hooft anomaly, in the sense that it is compatible with a trivially gapped phase.\footnote{The non-invertible chiral symmetry in QED  \cite{Choi:2022jqy,Cordova:2022ieu} is of a different nature and may still have an 't Hooft anomaly because it arises from an axial U(1)$_\text{A}$ symmetry with a self 't Hooft anomaly before gauging U(1)$_\text{V}$. In contrast, the non-invertible symmetry considered here arises from the field theory in Section \ref{sec:Yukawa}, where the U(1)$_\text{A}$ global symmetry has no self anomaly. 
See \cite{Choi:2021kmx,Choi:2022zal,Choi:2022rfe,Kaidi:2022cpf,Apte:2022xtu,Cordova:2023bja,Antinucci:2023ezl,Gorantla:2024ocs,Putrov:2026csz} for discussions of 't Hooft anomalies of related non-invertible symmetries in 3+1d.}

\section{2-group from gauging U(1)$_\text{A}$}\label{sec:2group}

\subsection{The gauged model}

In this subsection, we gauge the U$(1)_\text{A}$ symmetry of our lattice model \eqref{Hamiltonian}, following the steps outlined in Appendix \ref{app:villain}. We introduce a U$(1)_\text{A}$ gauge field $A^{(1)}$ with conjugate operator $E^{(1)}$ on every link, together with the integer Villain gauge field $n^{(2)}$ and its conjugate operator $\tilde{A}^{(2)}$ on every plaquette. These operators satisfy the following commutation relations,
\ie
[A_\ell,E_{\ell'}]=i\delta_{\ell,\ell'},\quad [\tilde{A}_p,n_{p'}]=i\delta_{p,p'}\,.
\fe
The Hamiltonian of the gauged theory is
\ie\label{eq:AgaugeHamiltonian}
H_\text{A}&=  
{1\over 2\gamma} 
\sum_\ell E_\ell  ^2
+{\gamma\over2}\sum_p ((dA)_p - 2\pi n_p)^2
\\
+&
{1\over 2\beta}\sum_s \left(p_s +\frac{1}{2\pi} (dw \cup A)_{\mathfrak{t}(s)} \right)^2
+{\beta\over2}
\sum_\ell \left( (d\phi)_\ell +2\pi w_\ell\right)^2
+{\lambda\over2} \sum_p (dw)_p^2
\,,
\fe
with the following Gauss law constraints,
\ie\label{eq:AgaugeGauss}
&\exp(2\pi i w_\ell) = 1\,,\\
&\exp(2\pi i n_p) = 1\,,\\
&(\delta E)_s  = -\left(
\left( w +{ d\phi \over 2\pi}
\right)
\cup 
d w \right)_{\mathfrak{t}^{-1}(s)}\,, \\
&\exp(2\pi i p_s - i (\delta b)_s)=1\,,\\
&\exp(2\pi i E_\ell + i(\phi \cup dw)_{\mathfrak{t}^{-1}(\ell)} - i (\delta \tilde{A})_\ell ) = 1\,.
\fe
The first two constraints just mean that $w_\ell, n_p$ are integers. 
The third one is the lattice analog of $\nabla\cdot \vec E =- q_\text{A}$, which implements the following U$(1)_\text{A}$ gauge transformation,
\ie\label{eq:Agaugetransform1}
b_\ell\sim b_\ell&
+\int \left(\mathbf{l}\cup dw\cup \alpha+dw\cup \mathbf{l}\cup \alpha+\left(w+\frac{d\phi}{2\pi}\right)\cup \mathbf{l}\cup d\alpha\right)\\
&p_s\sim p_s+\frac{1}{2\pi}(dw\cup d\alpha)_{\mathfrak{t}(s)}, ~~~~~~ A_\ell\sim A_\ell-(d\alpha)_{\ell}
\fe
where $\alpha^{(0)}$ is an $\mathbb{R}$ valued 0-cochain, $\mathbf{l}^{(1)}$ is the 1-cochain that is 1 on $\ell$ and 0 otherwise. Indeed, when $d\alpha$ is set to zero, the gauge transformation above reduces to the global U$(1)_\text{A}$ action \eqref{eq:QAactonb}. The fourth and fifth Gauss law constraints implement the $\mathbb{Z}$ gauge transformations for the matter fields \eqref{Zgauge} and for the gauge fields, respectively. The $\mathbb{Z}$ gauge transformation of the gauge fields takes the following form,
\ie  \label{eq:Agaugetransform2}
A_\ell\sim A_\ell+2\pi k_\ell,\ n_p\sim n_p+(dk)_p,\ p_s\sim p_s-\left(d w\cup k\right)_{\mathfrak{t}(s)},\
b_\ell\sim b_\ell
+\int\phi\cup d \mathbf{l}\cup k\,,
\fe
where $k^{(1)}$ is a $\mathbb{Z}$-valued 1-cochain. We also impose the flatness condition $dn^{(2)}=0$ for the Villain gauge field.

\subsection{2-group}

As shown in \cite{Cordova:2018cvg} in the continuum, the U$(1)_\text{V}$ symmetry becomes part of a 2-group symmetry after gauging U$(1)_\text{A}$. In this subsection, we demonstrate that the same structure arises in our lattice model. 
We begin by reviewing the 2-group symmetry involving a U(1)$^{(0)}$ 0-form symmetry and a U(1)$^{(1)}$ 1-form symmetry in 3+1d continuum field theory. 
We denote their associated currents by $j_\mu(x)$ and $j_{\mu\nu}(x)$, respectively. 
They obey the conservation equations
\ie
\partial^\mu j_\mu(x) = 0 \,,~~~~~
\partial^\mu j_{\mu\nu}(x)=0\,.
\fe
The hallmark of the 2-group symmetry is encoded in the contact term of the following Euclidean OPE:
\ie\label{2groupOPE}
\partial^\mu j_\mu (x) j_\nu(0)
= {\widehat{\kappa} \over 2\pi}  
\partial^\lambda\delta^{(4)} (x) j_{\nu\lambda}(0)\,,
\fe
where $\widehat\kappa$ is an integer known as the 2-group structure constant. 
The Lorentzian counterpart of the above equation is a nontrivial equal-time commutator reminiscent of the Schwinger term in 1+1d: 
\ie\label{ETC}
[j_t(t,\vec x) , j_t(0)]
= i {\widehat{\kappa}\over 2\pi} 
\partial_i \delta^{(3)} (\vec x)
\, j_{ti }(0)\,.
\fe

We now return to the U$(1)_\text{A}$ gauge theory described by \eqref{eq:AgaugeHamiltonian} and \eqref{eq:AgaugeGauss}. 
There is a magnetic U(1)$^{(1)}_m$ 1-form symmetry generated by
\ie\label{eq:AgaugeQm}
Q_m^{(1)} =  \int_{\Sigma_2}  q_m= \int_{\Sigma_2} \left(n^{(2)}-\frac{dA^{(1)}}{2\pi}\right)
\fe
This operator is quantized, gauge-invariant, and conserved. It is also topological because of the flatness condition $dn=0$.

The original vector charge $Q_\text{V}=\int_{M_3} \star_t \left(p - {\delta b\over 2\pi}\right)$ is no longer gauge-invariant. 
Rather, there is a new U$(1)_{\text{V}'}^{(0)}$ (0-form) symmetry generated by the following charge, 
\ie\label{eq:AgaugeQV}
Q^{(0)}_{\text{V}'}=\int_{M_3}\tilde{q}_{\text{V}'}=\int_{{M}_3}\star_{\mathfrak{t}} \left(p^{(0)}
-{\delta b^{(1)}\over 2\pi}
+\star_{\mathfrak{t}}\left(w^{(1)}\cup n^{(2)}\right)\right)
\fe
This operator commutes with the Hamiltonian and is quantized. 
We show that it commutes with the Gauss law constraints in \eqref{eq:AgaugeGauss} in Appendix \ref{app:commutator}. 
The expression above makes it obvious that the charge is quantized, but the charge density $\tilde{q}_{\text{V}'}$ is not gauge invariant. We can alternatively write $Q^{(0)}_{\text{V}'}$ as
\ie
Q^{(0)}_{\text{V}'}=\int_{M_3}q_{\text{V}'}=\int_{{M}_3}\star_{\mathfrak{t}} \left(p^{(0)}
+\star_{\mathfrak{t}}\left(dw^{(1)}\cup{A^{(1)}\over 2\pi}\right)
+\star_{\mathfrak{t}}\left[\left(w+{d\phi^{(0)}\over 2\pi}\right)\cup \left(n^{(2)}-{dA^{(1)}\over 2\pi}\right)\right]
\right)
\fe
The charge density $q_{\text{V}'}$ is obviously gauge-invariant, as it is constructed from terms appearing in the Hamiltonian \eqref{eq:AgaugeHamiltonian}. It is also straightforward to verify that $\star_{\mathfrak{t}}q_{\text{V}'}$ and $\star_{\mathfrak{t}}\tilde{q}_{\text{V}'}$ differ by a total derivative.

The 2-group structure is manifested in the commutator of the gauge-invariant charge densities,
\ie\label{eq:latticeETC}
[ j^{\text{V}'}(\vec r_1),j^{\text{V}'}(\vec r_2)]={i\over 2\pi}\sum_{i=x,y,z}
\Big[\, 
\delta_{\vec r_1+\hat{i},\vec r_2} \, j^m_i(\vec r_1)-\delta_{\vec r_1,\vec r_2+\hat{i}}\, j^m_i(\vec r_2)
\,\Big]\,.
\fe
Here we have defined $j^{\text{V}'}=\star_{\mathfrak{t}^{-1}}q_{\text{V}'}$, $j^m=\star_{\mathfrak{t}}q_m$ to simplify the expression. The equation above is the lattice counterpart of \eqref{ETC} with $\widehat{\kappa}=1$, demonstrating the 2-group structure that arises after gauging U$(1)_\text{A}$.

\subsection{Lattice Green-Schwarz term}

Another way to characterize the 2-group symmetry is through the Green-Schwarz term \cite{Green:1984sg} in the background gauge transformations \cite{Cordova:2018cvg}. 
Consider a continuum theory with a U$(1)_{\text{V}'}^{(0)}$ 0-form symmetry and a U$(1)_m^{(1)}$ 1-form symmetry that together form a 2-group. We couple the theory to a background 1-form gauge field $A_{\text{V}'}^{(1)}$ and a 2-form gauge field  $B_m^{(2)}$, associated with U$(1)_{\text{V}'}^{(0)}$ and U$(1)_m^{(1)}$, respectively. The theory is invariant under the following gauge transformation,
\ie\label{eq:2groupGScontinuum}
A_{\text{V}'}^{(1)}\sim A_{\text{V}'}^{(1)}+d\lambda^{(0)},\quad B_m^{(2)}\sim  B_m^{(2)}+d\Lambda^{(1)}+\frac{\widehat{\kappa}}{2\pi}\lambda^{(0)}  F_{\text{V}'}^{(2)}
\fe
The last term is known as the Green-Schwarz term, which encodes the 2-group structure through the mixing of the U$(1)_{\text{V}'}^{(0)}$ and U$(1)_m^{(1)}$ gauge transformations.

In our lattice model described by \eqref{eq:AgaugeHamiltonian} and \eqref{eq:AgaugeGauss}, there is likewise a Green-Schwarz term that signals the presence of a 2-group structure. The U$(1)_{\text{V}'}^{(0)}$ 0-form symmetry and a U$(1)_m^{(1)}$ 1-form symmetry are generated by \eqref{eq:AgaugeQV} and \eqref{eq:AgaugeQm}, respectively. Following Appendix \ref{app:novillain}, we couple the system to background gauge fields $\mathcal{A}^{(1)}$ and $\mathcal{B}^{(2)}$. The Hamiltonian remains unchanged from \eqref{eq:AgaugeHamiltonian}, while the Gauss law constraints are changed to the following,
\ie\label{eq:2groupGauss1}
&\exp(2\pi i w_\ell) = \exp(i\mathcal{A}_\ell)\,,\\
&\exp(2\pi i n_p) = 1\,,\\
&\exp\left(2\pi i p_s - i \left(\delta b\right)_s\right)=\exp\left(i\int \mathcal{A}^{(1)}\cup \mathbf{s}^{(0)}\cup n^{(2)}\right)\,,\\
&(\delta E)_s  = -\left(
\left( w +{ d\phi \over 2\pi}
\right)
\cup 
dw \right)_{\mathfrak{t}^{-1}(s)}\,, \\
&\exp(2\pi i E_\ell + i(\phi \cup dw)_{\mathfrak{t}^{-1}(\ell)} - i (\delta \tilde{A})_\ell ) = \exp\left(-i  \mathcal{B}_{\mathfrak{t}^{-1}(\ell)}+i\left(\mathcal{A}\cup w\right)_{\mathfrak{t}^{-1}(\ell)}\right)\,.
\fe
Here $\mathbf{s}^{(0)}$ is a 0-chain that is 1 on site $s$ and zero otherwise. 

To find the gauge transformations associated with $\mathcal{A}^{(1)}$ and $\mathcal{B}^{(2)}$, we promote the two background gauge fields to operators and introduce their conjugate momenta, $\mathcal{E}^{(1)}$ and $\mathcal{H}^{(2)}$, satisfying the commutation relations,
\ie
[\mathcal{A}_\ell,\mathcal{E}_{\ell'}]=i\delta_{\ell,\ell'},\quad [\mathcal{B}_p,\mathcal{H}_{p'}]=i\delta_{p,p'}
\fe
We have the following Gauss law constraints associated with U$(1)_\text{V}^{(0)}$ and U$(1)_m^{(1)}$:
\ie\label{eq:2groupGauss2}
&(\delta \mathcal{H})_\ell+  n_{\mathfrak{t}(\ell)}=0\,,\\
&\left(\delta \mathcal{E}\right)_s+ p_s - { \left(\delta b\right)_s\over2\pi}+ (w\cup n)_{\mathfrak{t}(s)} - \frac{1}{2\pi}\int\mathcal{A}^{(1)}\cup d\mathbf{s}^{(0)}\cup\star_{\mathfrak{t}^{-1}}\mathcal{H}^{(2)} =0\,.
\fe 
The last term in the second Gauss law involving $\mathcal{H}^{(2)}$ signals a 2-group structure. To see this more clearly, we derive the gauge transformations on $\mathcal{A}^{(1)}$ and $\mathcal{B}^{(2)}$ generated by \eqref{eq:2groupGauss2}.
\ie
&\mathcal{A}^{(1)}\sim \mathcal{A}^{(1)}+d\lambda^{(0)},\quad \\
&\mathcal{B}^{(2)}\sim\mathcal{B}^{(2)}+d\left(\Lambda^{(1)}-\frac{1}{2\pi}\mathcal{A}^{(1)}\cup\lambda^{(0)}\right)+\frac{1}{2\pi}d\mathcal{A}^{(1)}\cup\lambda^{(0)}\,.
\fe
We are free to absorb ${1\over 2\pi}\mathcal{A}^{(1)}\cup\lambda^{(0)}$ into $\Lambda^{(1)}$, after which the gauge transformation above agrees with \eqref{eq:2groupGScontinuum} with $\widehat{\kappa}=1$, indicating a nontrivial 2-group symmetry on the lattice.

\subsection{2-group from a trivial 3-group with an anomaly}

In this subsection, we discuss the continuum limit of the U$(1)_\text{A}$ gauge theory and its 2-group symmetry.

Our lattice model locally shares the same symmetries and anomalies as the continuum field theory of \cite{Cordova:2018cvg}, which we review in Section \ref{sec:Yukawa}. 
% Our lattice model is in the same universality class as the continuum field theory of \cite{Cordova:2018cvg}, which we review in Section \ref{sec:Yukawa}. 
This field theory has 4 massless left-handed Weyl fermions coupled to a complex scalar $\Phi$ via the Yukawa coupling. It has a U$(1)_\text{V}\times$U$(1)_\text{A}$ global symmetry with charge assignment given in \eqref{Yukawacharge}, and the only 't Hooft anomaly is the one between U$(1)_\text{A}-$U$(1)_\text{V}-$U$(1)_\text{V}$.

We gauge the U$(1)_\text{A}$ symmetry to find the following Lagrangian density, 
\ie
{\cal L}& = 
\sum_{I=1,2}i \psi^{\dagger(I)}\bar \sigma^\mu(\partial_\mu - i A_\mu)\psi^{(I)}+\sum_{I=3,4}i \psi^{\dagger(I)}\bar \sigma^\mu(\partial_\mu + i A_\mu)\psi^{(I)}\\
&+\frac 12 |\partial_\mu\Phi|^2 -V(\Phi)
-{1\over 4e^2}F_{\mu\nu}F^{\mu\nu}
+\left(g \Phi^\dagger  \epsilon^{ab} \psi_a^{(1)} \psi_b^{(3)}
+g \Phi  \epsilon^{ab} \psi_a^{(2)} \psi_b^{(4)}+\text{h.c.}\right)\,.
\fe
Following the discussion in \cite{Cordova:2018cvg}, this theory has a U$(1)_{\text{V}'}^{(0)}$ 0-form symmetry and a U$(1)_m^{(1)}$ 1-form symmetry which form a 2-group.

We turn on a Higgs potential $V(\Phi)$ so that $\Phi$ acquires a vev. All fermions become massive, and the IR theory is described by a free compact scalar theory and a decoupled free Maxwell theory.
\ie\label{eq:IRLA}
\mathcal{L}_\text{A}=\frac{f^2}{2}(\partial_\mu\phi)^2-\frac{1}{4e^2}F_{\mu\nu}F^{\mu\nu},\ ~~~\phi\sim\phi+2\pi\,.
\fe
One can verify that this theory is also the continuum limit of the Hamiltonian theory described by \eqref{eq:AgaugeHamiltonian} and \eqref{eq:AgaugeGauss}. 
As discussed in Section \ref{sec:contlimit}, in the continuum limit with $\lambda>0$, we effectively have $dw\to 0$, and \eqref{eq:AgaugeHamiltonian} becomes decoupled. 
Nonetheless, the 2-group structure remains nontrivial, as can be seen from \eqref{eq:latticeETC}.

To understand this, we first note that the decoupled theory \eqref{eq:IRLA} has a trivial 3-group which is the direct product U$(1)_\text{W}^{(2)}\times$U$(1)_m^{(1)}\times$U$(1)^{(0)}_\text{V}$. Here U$(1)_\text{W}^{(2)}$ is the 2-form winding symmetry discussed in Section \ref{sec:matching}. U$(1)_m^{(1)}$ is the magnetic 1-form symmetry and U$(1)^{(0)}_\text{V}$ is the 0-form symmetry that shifts $\phi$ by a constant. Note that U$(1)^{(0)}_\text{V}$ differs from the continuum limit of the U$(1)^{(0)}_{\text{V}'}$ symmetry discussed earlier in this section. Indeed, in addition to the terms that shift $\phi$, \eqref{eq:AgaugeQV} contains an extra contribution $w^{(1)}\cup n^{(2)}$, which is the cup product of the charge densities of U$(1)_\text{W}^{(2)}$ and U$(1)_m^{(1)}$. 
In the continuum, this can be schematically understood by defining a current for a new U(1)$^{(0)}_{\text{V}'}$ symmetry as $J_{\text{V}'}^{(3)} =J^{(3)}_\text{V} + J_m^{(2)}\wedge J_\text{W}^{(1)}$ \cite{Brauner:2020rtz,Heidenreich:2020pkc}, where we have defined $J=\star j$ to simplify this equation. 
The mixed anomaly between U(1)$_\text{V}^{(0)}$ and U(1)$_\text{W}^{(2)}$ in \eqref{IRanomaly} then induces a 2-group structure between U(1)$_{\text{V}'}^{(0)}$ and U(1)$_m^{(1)}$. 
This is how a nontrivial 2-group is embedded into a trivial 3-group with an 't Hooft anomaly.

\section{Discussions}

In this paper we present an exactly solvable lattice Hamiltonian \eqref{Hamiltonian} preserving the U(1)$_\text{V}\times \text{U(1)}_\text{A}$ chiral symmetry. 
For any $\beta,\lambda>0$, our model is in a phase where U(1)$_\text{V}$ is spontaneously broken. 
It would be interesting to explore other possible phases by adding symmetric deformations such as  \eqref{breakW}.

't~Hooft anomalies of continuous chiral global symmetries in continuum field theory imply that the IR theory must be gapless~\cite{tHooft:1979rat,Frishman:1980dq,Coleman:1982yg}.
For example, the IR theory may be described by a strongly interacting conformal field theory, or by gapless excitations such as Goldstone bosons or massless fermions.
This follows because the anomaly induces non-analytic contributions to current correlation functions, which cannot be reproduced by any gapped theory. 
In particular, such anomalies cannot be matched by a topological quantum field theory. 
It is therefore natural to ask whether our lattice $\mathrm{U}(1)_\text{V} \times \mathrm{U}(1)_\text{A}$ symmetry similarly enforces gapless phases.
Lattice global symmetries that enforce gaplessness have recently been discussed in ~\cite{Chatterjee:2024gje,Pace:2024oys,Xu:2025hfs,Pace:2025rfu,Onogi:2025xir,Kim:2025tzc,Ando:2026ffy,Misumi:2026ckr}.

Conversely, an anomaly-free global symmetry is expected to be compatible with a trivially gapped phase, in the spirit of symmetric mass generation (see \cite{Wang:2022ucy} for a review). 
If we take multiple copies of the Hamiltonian~\eqref{Hamiltonian}, labeled by an index~$\alpha$, we can find such a symmetry by taking a linear combination of the vector and axial charges $Q = \sum_\alpha \bigl( n_\text{V}^\alpha Q_\text{V}^\alpha + n_\text{A}^\alpha Q_\text{A}^\alpha \bigr)$
where the integers satisfy $\sum_\alpha n_\text{A}^\alpha (n_\text{V}^\alpha)^2 = 0$.  
It would be interesting to find a gapping Hamiltonian preserving this anomaly-free U(1)$_Q$ symmetry generated by $Q$.

Anomaly-free chiral global symmetries also lead to new lattice chiral gauge theories \cite{Thorngren:2026ydw}.
Following the steps in Appendix~\ref{app:gauging}, we can gauge $\mathrm{U}(1)_Q$ to find a lattice chiral gauge theory whose continuum limit is
\begin{equation}
\sum_\alpha \frac{f_\alpha^2}{2}
\bigl(\partial_\mu \phi^\alpha - n_\text{V}^\alpha A_\mu \bigr)^2
+ \frac{i}{16\pi^2} \epsilon^{\mu\nu\rho\sigma}
\left( \sum_\alpha 
n_\text{V}^\alpha n_\text{A}^\alpha \,\phi^\alpha \right)\,
F_{\mu\nu} F_{\rho\sigma}\,.
\end{equation}
In this field theory, certain linear combinations of the scalar fields behave as axions (typically parity-odd), while others behave as Stückelberg or Higgs fields (typically parity-even).
Consequently, there is no (manifest) time-reversal or parity symmetry. 
It would be exciting to explicitly construct such lattice chiral gauge theories and relate them to phenomenological Peccei--Quinn models~\cite{Peccei:1977hh}.

Finally, this bosonic lattice model can perhaps be used as a building block to help construct chiral symmetries and anomalies in lattice models involving fermions. 
It would also be interesting to consider the Euclidean version of this bosonic model. 
We leave these directions for future investigations.

\section*{Acknowledgements}

We thank Tom Banks, Meng Cheng, Yichul Choi, Ross Dempsey, Lukasz Fidkowski, Po-Shen Hsin, Igor Klebanov, Elijah Lew-Smith, Gregory Moore,  Salvatore Pace, Nathan Seiberg, Wilbur Shirley, Nikita Sopenko, and Ryan Thorngren for helpful discussions. 
We thank Aleksey Cherman, Thomas Dumitrescu, Theo Jacobson, and Tin Sulejmanpasic for useful comments on a draft. 
S.S. and S.H.S. were supported in part by the Simons Collaboration on Ultra-Quantum Matter, which is a grant from the Simons Foundation (651444, NS, SHS). 
S.S. also gratefully
acknowledges support from the Ambrose Monell Foundation at the Institute for Advanced
Study. 
S.H.S. is grateful to the Institute for Advanced Study for its hospitality during the final stage of this project. 
The authors of this paper are ordered alphabetically.

\appendix

\section{Chains and cochains on a hypercubic lattice}\label{app:cochain}

In this appendix, we review chains, cochains, and cup products on a hypercubic lattice. Our conventions largely follow \cite{Sulejmanpasic:2019ytl}. 
See, for example, \cite{Chen:2021ppt,Jacobson:2023cmr} for other helpful references. 

\subsection{Chains}

Consider a $d$-dimensional infinite hypercubic lattice $\Lambda$, the sites are labeled by $\vec{r}=(r_1,r_2,\cdots,r_d)$ where $r_i\in\mathbb{Z}$. We define an $n$-cell $c_n(\vec{r})_{i_1\cdots i_n}$ with $1\le i_1<\cdots<i_n\le d$ as follows,
\ie
c_n(\vec{r})_{i_1\cdots i_n}=\{\vec{r}\}\cup & \{\vec{r}+\hat{i}_a,1\le a\le n\}\\
&\cup \{\vec{r}+\hat{i}_a+\hat{i}_b,1\le a<b\le n\}\cup\cdots\cup \{\vec{r}+\hat{i}_1+\cdots+\hat{i}_n\}
\fe
where $\hat{i}$ denotes the unit vector in the direction $i$. For each natural number $n$, we define a formal sum of such $n$-cells. We also include $0$ as the identity element. We extend the definition above to a more general ordering of the subscripts by,
\ie
c_n(\vec{r})_{i_1\cdots i_a\cdots i_b\cdots i_n}=-c_n(\vec{r})_{i_1\cdots i_b\cdots i_a\cdots i_n}
\fe
Here, the minus sign is understood as the inverse under the formal sum. For a given $n$, all $n$-cells together with $0$ generate an Abelian group $C_n(\Lambda,\mathbb{Z})$. A generic element in this group can be written as a formal sum of $c_n$ with integer coefficients.
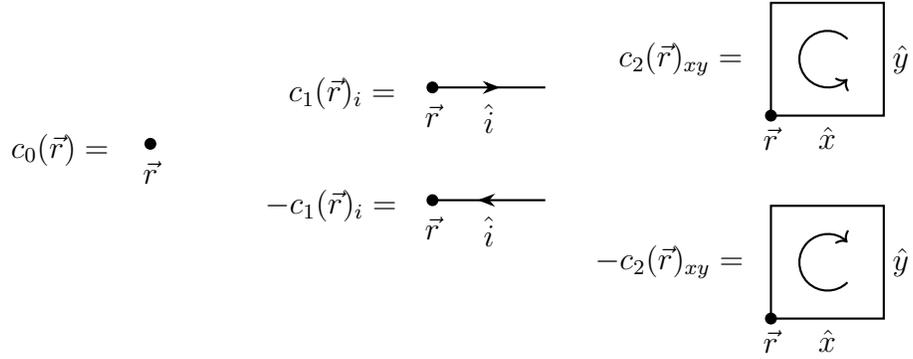
\begin{figure}[h]
    \centering
    \begin{tikzpicture}[scale=1.5,
    midarrow/.style={
        postaction={decorate},
        decoration={
            markings,
            mark=at position 0.6 with {\arrow[scale=1]{Stealth}}
        }
    }]

\node at (-0.8,0) {$c_0(\vec{r})=$};
\fill (0,0.05) circle (1.5pt);
\node at (0,-0.2) {$\vec{r}$};

\begin{scope}[xshift=2.5cm,yshift=0.5cm]
    \node at (-0.8,0) {$c_1(\vec{r})_i=$};
    \draw[thick,midarrow] (0,0.05) -- (1,0.05);
    \fill (0,0.05) circle (1.5pt);
    \node at (0,-0.2) {$\vec{r}$};
    \node[below] at (0.5,0) {$\hat{i}$};
\end{scope}
\begin{scope}[xshift=2.5cm,yshift=-0.5cm]
    \node at (-0.9,0) {$-c_1(\vec{r})_i=$};
    \draw[thick,midarrow] (1,0.05) -- (0,0.05);
    \fill (0,0.05) circle (1.5pt);
    \node at (0,-0.2) {$\vec{r}$};
    \node[below] at (0.5,0) {$\hat{i}$};
\end{scope}

\begin{scope}[xshift=5.5cm,yshift=0.3cm]
    \node at (-0.8,0.5) {$c_2(\vec{r})_{xy}=$};
    \draw[thick] (0,0) -- (1,0) -- (1,1) -- (0,1) -- cycle;
   
    \draw[->,thick] (0.5+0.25/1.4,0.5+0.25/1.4) arc (45:315:0.25);
    \fill (0,0) circle (1.5pt);
    \node at (0,-0.2) {$\vec{r}$};
    \node[below] at (0.5,0) {$\hat{x}$};
    \node at (1.15,0.5) {$\hat{y}$};
\end{scope}
\begin{scope}[xshift=5.5cm,yshift=-1.5cm]
    \node at (-0.9,0.5) {$-c_2(\vec{r})_{xy}=$};
    \draw[thick] (0,0) -- (1,0) -- (1,1) -- (0,1) -- cycle;
    \draw[->,thick] (0.5+0.25/1.4,0.5-0.25/1.4) arc (315:45:0.25);
    \fill (0,0) circle (1.5pt);
    \node at (0,-0.2) {$\vec{r}$};
    \node[below] at (0.5,0) {$\hat{x}$};
    \node at (1.15,0.5) {$\hat{y}$};
\end{scope}

\end{tikzpicture}
    \caption{Sites, links and plaquettes as 0-, 1-, and 2-cells.}
    \label{fig:sellp}
\end{figure}

As concrete examples, a site at position $\vec{r}$ is represented by $c_0(\vec{r})$. A link originating at $\vec{r}$ and pointing in the $\hat{i}$ direction is described by $c_1(\vec{r})_i$, while a link with the opposite orientation is given by $-c_1(\vec{r})_{i}$. A plaquette lying in the $xy$ plane is denoted by $c_2(\vec{r})_{xy}$, and the plaquette with the opposite orientation is given by $c_2(\vec{r})_{yx}=-c_2(\vec{r})_{xy}$. See Figure~\ref{fig:sellp}. In the main text, we denote sites, links, plaquettes, and cubes by $s$, $\ell$, $p$, and $c$, respectively, to simplify the notation. 

\begin{figure}[h]
    \centering
    \begin{tikzpicture}[scale=1.5,
    midarrow/.style={
        postaction={decorate},
        decoration={
            markings,
            mark=at position 0.6 with {\arrow[scale=1]{Stealth}}
        }
    }]

\begin{scope}
    \draw[thick,midarrow,red] (0,-0.5) -- (0,0.5);
    \node at (0.3,0) {$=$};
\end{scope}

\begin{scope}[xshift=0.6cm]
    \draw (0,-0.5) -- (0,0.5);
    \fill[blue] (0,0.5) circle (1.5pt);
    \node at (0.3,0) {$-$};
\end{scope}
\begin{scope}[xshift=1.2cm]
    \draw (0,-0.5) -- (0,0.5);
    \fill[blue] (0,-0.5) circle (1.5pt);
\end{scope}

\begin{scope}[xshift=2.5cm,yshift=-0.5cm]
    \draw[thick] (0,0) -- (1,0) -- (1,1) -- (0,1) -- cycle;
    \fill[red,opacity=0.4] (0,0) -- (1,0) -- (1,1) -- (0,1) -- cycle;
    \draw[->,thick] (0.5+0.3/1.4,0.5+0.3/1.4) arc (45:315:0.3);
    \node at (1.3,0.5) {$=$};
\end{scope}
\begin{scope}[xshift=4.1cm,yshift=-0.5cm]
    \draw[thick,midarrow,blue] (0,0) -- (1,0);
    \draw[thick,midarrow,blue] (1,0) -- (1,1);
    \draw[thick,midarrow,blue] (1,1) -- (0,1);
    \draw[thick,midarrow,blue] (0,1) -- (0,0);
\end{scope}
\end{tikzpicture}

~
~
~

\begin{tikzpicture}[scale=1.4,x={(1cm,0cm)},
    y={(0.6cm,0.4cm)},
    z={(0cm,1cm)}]

\begin{scope}
\coordinate (O) at (0,0,0);
\coordinate (A) at (1,0,0);
\coordinate (B) at (1,1,0);
\coordinate (C) at (0,1,0);
\coordinate (D) at (0,0,1);
\coordinate (E) at (1,0,1);
\coordinate (F) at (1,1,1);
\coordinate (G) at (0,1,1);

\node at (1.9,0,0.6) {$=$};

\fill[red,opacity=0.4] (O) -- (A) -- (E) -- (D) -- cycle;
\fill[red,opacity=0.4] (A) -- (B) -- (F) -- (E) -- cycle;
\fill[red,opacity=0.4] (D) -- (E) -- (F) -- (G) -- cycle;

\draw[thick] (O) -- (A) -- (E) -- (D) -- cycle;
\draw[thick] (D) -- (G);
\draw[thick] (G) -- (F);
\draw[thick] (E) -- (F);
\draw[thick] (A) -- (B);
\draw[thick] (B) -- (F);
\draw (O) -- (C);
\draw (C) -- (G);
\draw (C) -- (B);

\end{scope}

\begin{scope}[xshift=2.2cm]
\coordinate (O) at (0,0,0);
\coordinate (A) at (1,0,0);
\coordinate (B) at (1,1,0);
\coordinate (C) at (0,1,0);
\coordinate (D) at (0,0,1);
\coordinate (E) at (1,0,1);
\coordinate (F) at (1,1,1);
\coordinate (G) at (0,1,1);

\node at (2,0,0.6) {$+$};

\fill[blue,opacity=0.4] (O) -- (A) -- (E) -- (D) -- cycle;
\fill[blue,opacity=0.4] (A) -- (B) -- (F) -- (E) -- cycle;
\fill[blue,opacity=0.4] (D) -- (E) -- (F) -- (G) -- cycle;

\draw[white,thick] (O) -- (A) -- (E) -- (D) -- cycle;
\draw[white,thick] (D) -- (G);
\draw[white,thick] (G) -- (F);
\draw[white,thick] (E) -- (F);
\draw[white,thick] (A) -- (B);
\draw[white,thick] (B) -- (F);

\draw[->,thick][canvas is xz plane at y=0](0.5+0.25/1.4,0.5+0.25/1.4) arc (45:315:0.25);

\draw[->,thick][canvas is yz plane at x=1](0.5+0.25/1.4,0.5+0.25/1.4) arc (45:315:0.25);

\draw[->,thick][canvas is xy plane at z=1](0.5+0.25/1.4,0.5+0.25/1.4) arc (45:315:0.25);

\end{scope}

\begin{scope}[xshift=4.6cm]

\coordinate (O) at (0,0,0);
\coordinate (A) at (1,0,0);
\coordinate (B) at (1,1,0);
\coordinate (C) at (0,1,0);
\coordinate (D) at (0,0,1);
\coordinate (E) at (1,0,1);
\coordinate (F) at (1,1,1);
\coordinate (G) at (0,1,1);

\fill[blue!40] (C) -- (B) -- (F) -- (G) -- cycle;
\fill[blue!40] (O) -- (C) -- (G) -- (D) -- cycle;
\fill[blue!40] (O) -- (A) -- (B) -- (C) -- cycle;

\draw[white,thick] (O) -- (A) -- (B) -- (C) -- cycle;
\draw[white,thick] (B) -- (F);
\draw[white,thick] (F) -- (G);
\draw[white,thick] (C) -- (G);
\draw[white,thick] (O) -- (D);
\draw[white,thick] (D) -- (G);

\draw[->,thick][canvas is xz plane at y=1](0.5+0.25/1.4,0.5-0.25/1.4) arc (315:45:0.25);

\draw[->,thick][canvas is yz plane at x=0](0.5+0.25,0.5) arc (360:90:0.25);

\draw[->,thick][canvas is xy plane at z=0](0.5+0.25,0.5) arc (360:90:0.25);

\end{scope}
\end{tikzpicture}
\caption{Illustration of the boundary operator $\partial:C_{n+1}(\Lambda,\mathbb{Z}) \to C_{n}(\Lambda,\mathbb{Z})$ in \eqref{eq:boundary}, as well as the exterior derivative $d:C^n(\Lambda,\mathbb{R}) \to C^{n+1}(\Lambda,\mathbb{R})$ in \eqref{eq:derivative} with $n=0,1,2$. The colors in this figure match the ones in \eqref{eq:boundary} and \eqref{eq:derivative}.}
    \label{fig:boundary}
\end{figure}
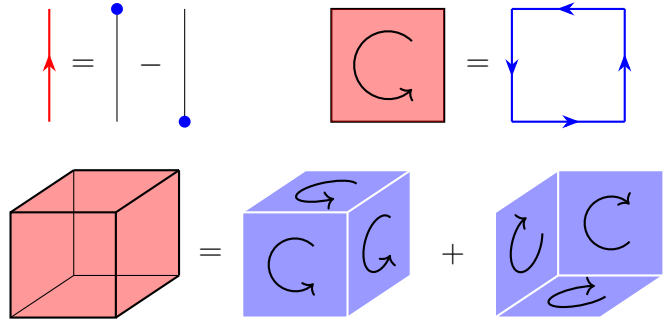

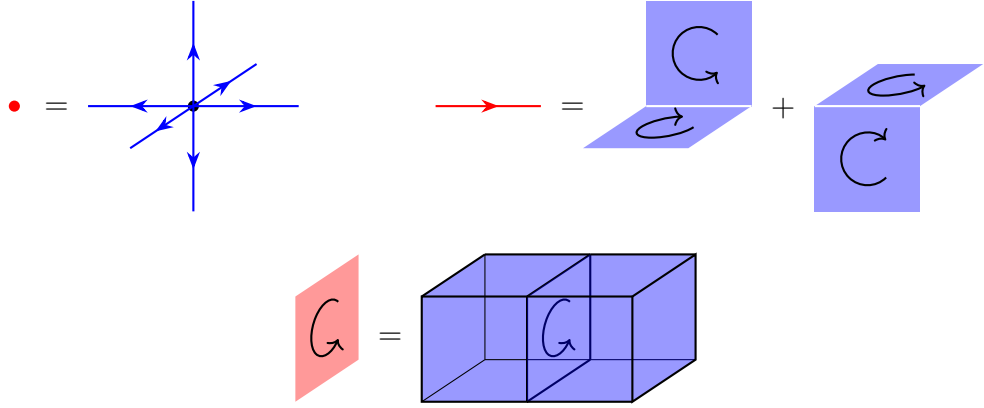
\begin{figure}
    \centering
    \begin{tikzpicture}[scale=1.4,x={(1cm,0cm)},
    y={(0.6cm,0.4cm)},
    z={(0cm,1cm)},midarrow/.style={
        postaction={decorate},
        decoration={
            markings,
            mark=at position 0.6 with {\arrow[scale=1]{Stealth}}
        }
    }]

    \begin{scope}
        \fill[red] (0,0,0) circle (1.5pt);
        \node at (0.4,0,-0.02) {$=$};
    \end{scope}
    
    \begin{scope}[xshift=1.7cm]
        \coordinate (O) at (0,0,0);
        \coordinate (A) at (1,0,0);
        \coordinate (B) at (1,1,0);
        \coordinate (C) at (0,1,0);
        \coordinate (D) at (0,0,1);
        \coordinate (E) at (1,0,1);
        \coordinate (F) at (1,1,1);
        \coordinate (G) at (0,1,1);
        \coordinate (A1) at (-1,0,0);
        \coordinate (C1) at (0,-1,0);
        \coordinate (D1) at (0,0,-1);

        \fill (O) circle (1.5pt);
        \draw[thick,midarrow,blue] (O) -- (D);
        \draw[thick,midarrow,blue] (O) -- (A);
        \draw[thick,midarrow,blue] (O) -- (C);
        \draw[thick,midarrow,blue] (O) -- (D1);
        \draw[thick,midarrow,blue] (O) -- (A1);
        \draw[thick,midarrow,blue] (O) -- (C1);
    \end{scope}

    \begin{scope}[xshift=4cm]
        \node at (1.3,0,-0.02) {$=$};
        \draw[thick,midarrow,red] (0,0,0) -- (1,0,0);
    \end{scope}

    \begin{scope}[xshift=6cm]
        \coordinate (O) at (0,0,0);
        \coordinate (A) at (1,0,0);
        \coordinate (B) at (1,1,0);
        \coordinate (C) at (0,1,0);
        \coordinate (D) at (0,0,1);
        \coordinate (E) at (1,0,1);
        \coordinate (F) at (1,1,1);
        \coordinate (G) at (0,1,1);
        \coordinate (C1) at (0,-1,0);
        \coordinate (B1) at (1,-1,0);
        \coordinate (D1) at (0,0,-1);
        \coordinate (E1) at (1,0,-1);

        \fill[blue!40] (O) -- (A) -- (B1) -- (C1) -- cycle;
        \fill[blue!40] (O) -- (A) -- (E) -- (D) -- cycle;

        \draw[white,thick] (O) -- (A);

        \node at (1.3,0,-0.02) {$+$};

        \draw[->,thick][canvas is xz plane at y=0](0.5+0.25/1.4,0.5+0.25/1.4) arc (45:315:0.25);
        \draw[->,thick][canvas is xy plane at z=0](0.5+0.25,-0.5) arc (360:90:0.25);
    \end{scope}

    \begin{scope}[xshift=7.6cm]
        \coordinate (O) at (0,0,0);
        \coordinate (A) at (1,0,0);
        \coordinate (B) at (1,1,0);
        \coordinate (C) at (0,1,0);
        \coordinate (D) at (0,0,1);
        \coordinate (E) at (1,0,1);
        \coordinate (F) at (1,1,1);
        \coordinate (G) at (0,1,1);
        \coordinate (C1) at (0,-1,0);
        \coordinate (B1) at (1,-1,0);
        \coordinate (D1) at (0,0,-1);
        \coordinate (E1) at (1,0,-1);

        \fill[blue!40] (O) -- (A) -- (B) -- (C) -- cycle;
        \fill[blue!40] (O) -- (A) -- (E1) -- (D1) -- cycle;

        \draw[white,thick] (O) -- (A);

        \draw[->,thick][canvas is xz plane at y=0](0.5+0.25/1.4,-0.5-0.25/1.4) arc (315:45:0.25);
        \draw[->,thick][canvas is xy plane at z=0](0.5,0.5+0.25) arc (90:360:0.25);
    \end{scope}
    
    \end{tikzpicture}

    ~
    ~
    ~

    \begin{tikzpicture}[scale=1.4,x={(1cm,0cm)},
    y={(0.6cm,0.4cm)},
    z={(0cm,1cm)}]

    \begin{scope}
        \coordinate (O) at (0,0,0);
        \coordinate (A) at (1,0,0);
        \coordinate (B) at (1,1,0);
        \coordinate (C) at (0,1,0);
        \coordinate (D) at (0,0,1);
        \coordinate (E) at (1,0,1);
        \coordinate (F) at (1,1,1);
        \coordinate (G) at (0,1,1);

        \node at (0.9,0,0.6) {$=$};

        \fill[red!40] (O) -- (C) -- (G) -- (D) -- cycle;

        \draw[->,thick][canvas is yz plane at x=0](0.5+0.25/1.4,0.5+0.25/1.4) arc (45:315:0.25);
    \end{scope}

    \begin{scope}[xshift=1.2cm]
        \coordinate (O) at (0,0,0);
        \coordinate (A) at (1,0,0);
        \coordinate (B) at (1,1,0);
        \coordinate (C) at (0,1,0);
        \coordinate (D) at (0,0,1);
        \coordinate (E) at (1,0,1);
        \coordinate (F) at (1,1,1);
        \coordinate (G) at (0,1,1);
        \coordinate (E1) at (2,0,1);
        \coordinate (F1) at (2,1,1);
        \coordinate (A1) at (2,0,0);
        \coordinate (B1) at (2,1,0);

        \draw[->,thick][canvas is yz plane at x=1](0.5+0.25/1.4,0.5+0.25/1.4) arc (45:315:0.25);

        \fill[blue,opacity=0.4] (O) -- (A) -- (E) -- (D) -- cycle;
        \fill[blue,opacity=0.4] (D) -- (E) -- (F) -- (G) -- cycle;
        \draw[thick] (O) -- (A) -- (E) -- (D) -- cycle;
        \draw[thick] (D) -- (G);
        \draw[thick] (G) -- (F);
        \draw[thick] (E) -- (F);
        \draw[thick] (A) -- (B);
        \draw[thick] (B) -- (F);
        \draw (O) -- (C);
        \draw (C) -- (G);
        \draw (C) -- (B1);

        \fill[blue,opacity=0.4] (A) -- (A1) -- (E1) -- (E) -- cycle;
        \fill[blue,opacity=0.4] (A1) -- (B1) -- (F1) -- (E1) -- cycle;
        \fill[blue,opacity=0.4] (E) -- (E1) -- (F1) -- (F) -- cycle;
        \draw[thick] (A1) -- (B1) -- (F1) -- (E1) -- cycle;
        \draw[thick] (F) -- (F1);
        \draw[thick] (E) -- (E1);
        \draw[thick] (A) -- (A1);

    \end{scope}

    \end{tikzpicture}

    \caption{Illustration of the coboundary operator $\hat{\partial}: C_{n-1}(\Lambda,\mathbb{Z}) \to C_{n}(\Lambda,\mathbb{Z})$ in \eqref{eq:coboundary}, as well as the divergence operator $\delta: C^n(\Lambda,\mathbb{R}) \to C^{n-1}(\Lambda,\mathbb{R})$ in  \eqref{eq:divergence} with $n=1,2,3$ in a 3-dimensional lattice. The colors in this figure match the ones in \eqref{eq:coboundary} and \eqref{eq:divergence}.}
    \label{fig:coboundary}
\end{figure}

We define a boundary operator $\partial$ that acts on $n$-cells in the following way,
\ie\label{eq:boundary}
\partial {\color{red} c_n(\vec{r})_{i_1\cdots i_n}}={\color{blue}\sum_{k=1}^n(-1)^{k+1}\left[c_{n-1}(\vec{r}+\hat{i}_k)_{i_1\cdots\mathring{i}_k\cdots i_n}-c_{n-1}(\vec{r})_{i_1\cdots\mathring{i}_k\cdots i_n}\right]}\,,
\fe
where $\mathring{i}$ denotes the omission of the index $i$. For 0-cells, we define $\partial c_0(\vec{r})=0$. The operator $\partial$ extends by linearity to a homomorphism $C_n(\Lambda,\mathbb{Z})\rightarrow C_{n-1}(\Lambda,\mathbb{Z})$. See Figure~\ref{fig:boundary} for examples of low-dimensional cells. We also define the coboundary operator $\hat{\partial}$ as follows,
\ie\label{eq:coboundary}
\hat{\partial}{\color{red}c_n(\vec{r})_{i_1\cdots i_n}}={\color{blue}\sum_{j\ne i_1\cdots i_n}\left[c_{n+1}(\vec{r})_{i_1\cdots i_n j}-c_{n+1}(\vec{r}-\hat{j})_{i_1\cdots i_n j}\right]}\,.
\fe
For $d$-cells, we define $\hat{\partial}c_d(\vec{r})_{1\cdots d}=0$. $\hat{\partial}$ extends linearly to a homomorphism $C_n(\Lambda,\mathbb{Z})\rightarrow C_{n+1}(\Lambda,\mathbb{Z})$. See Figure~\ref{fig:coboundary} for examples of low-dimensional cells in a three-dimensional lattice. It is straightforward to verify that $\partial$ and $\hat{\partial}$ are nilpotent. Therefore, the collection of groups $C_n(\Lambda,\mathbb{Z})$, together with the boundary operator $\partial$, form a chain complex. An element $\Sigma_n$ in the Abelian group generated by $n$-cells is called an $n$-chain. $\Sigma_n$ is called a cycle if $\partial \Sigma_n=0$, and a boundary if there exists some $\Sigma'_{n+1}$ such that $\Sigma_n=\partial \Sigma'_{n+1}$.

A dual lattice $\tilde{\Lambda}$ is the hypercubic lattice with sites $\vec{\tilde{r}}$ located at the center of $c_d(\vec{r})_{1\cdots d}$, namely $\vec{\tilde{r}}=\vec{r}+\frac{1}{2}(\hat{1}+\cdots \hat{d})$. We define a $\star$ operator that identifies $n$-cells on $\Lambda$ with $(d-n)$-cells on $\tilde{\Lambda}$,
\ie
&\star c_n(\vec{r})_{i_1\cdots i_n}=\frac{1}{(d-n)!}\sum_{i'_{n+1}\cdots i'_d}\epsilon_{i_1\cdots i_n i'_{n+1}\cdots i'_{d}}\tilde{c}_{d-n}(\vec{\tilde{r}}-\hat{i}'_{n+1}-\cdots-\hat{i}'_{d})_{i'_{n+1}\cdots i'_d}\\
&\star \tilde{c}_n(\vec{\tilde{r}})_{i_1\cdots i_n}=\frac{1}{(d-n)!}\sum_{i'_{n+1}\cdots i'_d}\epsilon_{i_1\cdots i_n i'_{n+1}\cdots i'_{d}}c_{d-n}(\vec{r}+\hat{i}_{1}+\cdots+\hat{i}_{n})_{i'_{n+1}\cdots i'_d}
\fe
It can be extended by linearity to the following homomorphism,
\ie
    \star: C_n(\Lambda,\mathbb{Z})\rightarrow C_{d-n}(\tilde{\Lambda},\mathbb{Z})
\fe
It is straightforward to check that the square of the $\star$ operator is the identity up to a sign, 
\ie
\star^2c_n=(-1)^{n(d-n)}c_n,\ \star^2\tilde{c}_n=(-1)^{n(d-n)}\tilde{c}_n
\fe
It is also useful to note the relation of the $\star$ operator with the boundary and coboundary operators,
\ie\label{eq:starboundary}
\partial\star=\star\hat{\partial},\ \star\partial=(-1)^{d+1}\hat{\partial}\star
\fe

\begin{figure}
    \centering
    \begin{tikzpicture}[scale=1.5,x={(1cm,0cm)},
    y={(0.6cm,0.4cm)},
    z={(0cm,1cm)},
    midarrow/.style={
        postaction={decorate},
        decoration={
            markings,
            mark=at position 0.6 with {\arrow[scale=1]{Stealth}}
        }
    }]
        \begin{scope}
        \coordinate (O) at (0,0,0);
        \coordinate (A) at (1,0,0);
        \coordinate (B) at (1,1,0);
        \coordinate (C) at (0,1,0);
        \coordinate (D) at (0,0,1);
        \coordinate (E) at (1,0,1);
        \coordinate (F) at (1,1,1);
        \coordinate (G) at (0,1,1);

        \draw[thick] (A) -- (B) -- (F) -- (E) -- cycle;
        \draw[thick] (D) -- (G);
        \draw[thick] (G) -- (F);
        \draw[thick] (O) -- (D);
        \draw[thick] (D) -- (E);
        \draw (O) -- (C);
        \draw (C) -- (G);
        \draw (C) -- (B);
        \draw[thick, midarrow, red] (O) -- (A);
        \fill[blue, opacity=0.4] (A) -- (B) -- (F) -- (E) -- cycle;
        \draw[->,thick][canvas is yz plane at x=1](0.5+0.25/1.4,0.5+0.25/1.4) arc (45:315:0.25);
        \node at (0.5,0,-0.2) {$\color{red}\ell$};
        \node at (1.6,0.5,0.5) {$\color{blue}\mathfrak{t}(\ell)$};
        \end{scope}

        \begin{scope}[xshift=2.8cm]
        \coordinate (O) at (0,0,0);
        \coordinate (A) at (1,0,0);
        \coordinate (B) at (1,1,0);
        \coordinate (C) at (0,1,0);
        \coordinate (D) at (0,0,1);
        \coordinate (E) at (1,0,1);
        \coordinate (F) at (1,1,1);
        \coordinate (G) at (0,1,1);

        \draw[thick] (A) -- (B) -- (F) -- (E) -- cycle;
        \draw[thick] (D) -- (G);
        \draw[thick] (O) -- (A);
        \draw[thick] (O) -- (D);
        \draw[thick] (D) -- (E);
        \draw (O) -- (C);
        \draw (C) -- (G);
        \draw (C) -- (B);
        \draw[thick, midarrow, blue] (G) -- (F);
        \fill[red, opacity=0.4] (O) -- (C) -- (G) -- (D) -- cycle;
        \draw[->,thick][canvas is yz plane at x=0](0.5+0.25/1.4,0.5+0.25/1.4) arc (45:315:0.25);
        \node at (-0.5,0.5,0.5) {$\color{red}p$};
        \node at (0.5,1,1.25) {$\color{blue}\mathfrak{t}(p)$};
        \end{scope}

        \begin{scope}[xshift=5.6cm]
        \coordinate (O) at (0,0,0);
        \coordinate (A) at (1,0,0);
        \coordinate (B) at (1,1,0);
        \coordinate (C) at (0,1,0);
        \coordinate (D) at (0,0,1);
        \coordinate (E) at (1,0,1);
        \coordinate (F) at (1,1,1);
        \coordinate (G) at (0,1,1);

        \fill[blue,opacity=0.4] (O) -- (A) -- (E) -- (D) -- cycle;
        \fill[blue,opacity=0.4] (A) -- (B) -- (F) -- (E) -- cycle;
        \fill[blue,opacity=0.4] (D) -- (E) -- (F) -- (G) -- cycle;

        \draw[thick] (O) -- (A) -- (E) -- (D) -- cycle;
        \draw[thick] (D) -- (G);
        \draw[thick] (G) -- (F);
        \draw[thick] (E) -- (F);
        \draw[thick] (A) -- (B);
        \draw[thick] (B) -- (F);
        \draw (O) -- (C);
        \draw (C) -- (G);
        \draw (C) -- (B);
        \fill[red] (O) circle (2pt);
        \node at (-0.2,0,-0.2) {$\color{red} s$};
        \node at (1.6,0,1.6) {$\color{blue} \mathfrak{t}(s)$};
        \end{scope}

        \begin{scope}[xshift=8.4cm]
        \coordinate (O) at (0,0,0);
        \coordinate (A) at (1,0,0);
        \coordinate (B) at (1,1,0);
        \coordinate (C) at (0,1,0);
        \coordinate (D) at (0,0,1);
        \coordinate (E) at (1,0,1);
        \coordinate (F) at (1,1,1);
        \coordinate (G) at (0,1,1);

        \fill[red,opacity=0.4] (O) -- (A) -- (E) -- (D) -- cycle;
        \fill[red,opacity=0.4] (A) -- (B) -- (F) -- (E) -- cycle;
        \fill[red,opacity=0.4] (D) -- (E) -- (F) -- (G) -- cycle;

        \draw[thick] (O) -- (A) -- (E) -- (D) -- cycle;
        \draw[thick] (D) -- (G);
        \draw[thick] (G) -- (F);
        \draw[thick] (E) -- (F);
        \draw[thick] (A) -- (B);
        \draw[thick] (B) -- (F);
        \draw (O) -- (C);
        \draw (C) -- (G);
        \draw (C) -- (B);
        \fill[blue] (1,1,1) circle (2pt);
        \node at (-0.2,0,-0.2) {$\color{red} c$};
        \node at (1.6,0,1.6) {$\color{blue} \mathfrak{t}(c)$};
        \end{scope}
    \end{tikzpicture}
    \caption{Examples of the half lattice translation $\mathfrak{t}$ in 3 dimensions, where $s,\ell,p,c$ denote a site, link, plaquette and cube, respectively. }
    \label{fig:t}
\end{figure}
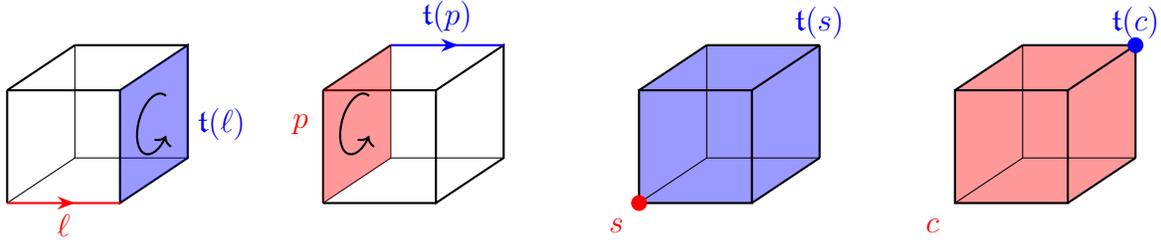

There is another important operator $\mathfrak{t}$, which maps an $n$-cell on $\Lambda$ to a $(d-n)$-cell on the original lattice $\Lambda$. Its action on $n$-cells is given by,
\ie
&\mathfrak{t}\left(c_n(\vec{r})_{i_1\cdots i_n}\right)=\frac{1}{(d-n)!}\sum_{i'_{n+1}\cdots i'_{d}}\epsilon_{i_1 \cdots i_n i'_{n+1}\cdots i'_{d}}  c_{d-n}(\vec{r}+\hat{i}_1+\cdots+\hat{i}_r)_{i'_{n+1}\cdots i'_{d}}\\
&\mathfrak{t}^{-1}\left(c_n(\vec{r})_{i_1\cdots i_n}\right)=\frac{1}{(d-n)!}\sum_{i'_{n+1}\cdots i'_{d}}\epsilon_{i_1 \cdots i_n i'_{n+1}\cdots i'_{d}} c_{d-n}(\vec{r}-\hat{i}'_{n+1}-\cdots-\hat{i}'_d)_{i'_{n+1}\cdots i'_{d}}
\fe
It extends by linearity to the following homomorphism,
\ie
    \mathfrak{t}: C_n(\Lambda,\mathbb{Z})\rightarrow C_{d-n}(\Lambda,\mathbb{Z})
\fe
The operator $\mathfrak{t}$ can be understood schematically as a half-lattice translation, since the centers of $c_n(\vec{r})$ and $\mathfrak{t}\left(c_n(\vec{r})\right)$ are related by a translation by $\left({1\over 2},\cdots,{1\over 2}\right)$, see Figure \ref{fig:t} for examples in 3 dimensions. Although $\mathfrak{t}$ and $\star$ are similar in that both map an $n$-cell to a $(d-n)$-cell, the image of $\mathfrak{t}$ lies in the original lattice $\Lambda$, whereas the image of $\star$ lies in the dual lattice $\tilde\Lambda$. It is useful to note the relation of the $\mathfrak{t}$ operator with the boundary and coboundary operators,
\ie\label{eq:startboundary}
&\mathfrak{t}\partial=(-1)^{d+1}\hat{\partial }\mathfrak{t}\,,~~~~\partial \mathfrak{t}=\mathfrak{t}\hat{\partial}\,,\\
&\mathfrak{t}^{-1}\partial=\hat{\partial }\mathfrak{t}^{-1}\,,~~~~~~~~ \partial \mathfrak{t}^{-1}=(-1)^{d+1}\mathfrak{t}^{-1}\hat{\partial}\,.
\fe

\subsection{Cochains}

An $n$-cochain, denoted by $A^{(n)}$, is a $\mathbb{Z}$-linear map from $n$-chains to real numbers. In particular, for any $n$-cell $c_n$, $A^{(n)}$ assigns a number denoted by $A_{c_n}$, satisfying $A_{-c_n}=-A_{c_n}$. 
We also use the following notations $A_{i_1\cdots i_n}(\vec{r})=A_{c_n(\vec{r})_{i_1\cdots i_n}}$. 
The Abelian group of $n$-cochains is denoted by $C^n(\Lambda,\mathbb{R})$.

In the previous subsection, we introduced several operators that act on chains. These operators can be naturally extended to cochains by linearity. We define the exterior derivative $d$ which maps an $n$-cochain to an $(n+1)$-cochain,
\ie\label{eq:derivative}
{\color{red}(dA)_{c_{n+1}}}=\sum_{c_n\in\partial c_{n+1}}{\color{blue}A_{c_n}}
\fe
Similarly, we define the divergence operator $\delta$ which maps an $n$-cochain to an $(n-1)$-cochain,
\ie\label{eq:divergence}
    {\color{red}(\delta A)_{c_{n-1}}}=\sum_{c_n\in\hat{\partial }c_{n-1}}{\color{blue}A_{c_n}}
\fe
The exterior derivative and divergence operators are both nilpotent, as follows directly from the nilpotence of the boundary and coboundary operators. The two operators also satisfy the following identity, which is particularly useful in calculations,
\ie \label{int.by.part}
\sum_{c_n}(dA)_{c_n}B_{c_n}=(-1)^n\sum_{c_{n-1}}A_{c_{n-1}}(\delta B)_{c_{n-1}}
\fe

The operators $\star$ and $\mathfrak{t}$ also have natural extensions to cochains. For an $n$-cochain $A^{(n)}$ on $\Lambda$, we define a $(d-n)$-cochain $(\star A)^{(d-n)}$ on the dual lattice $\tilde{\Lambda}$ as follows,
\ie
(\star A)_{\tilde{c}_{d-n}}=A_{\star \tilde{c}_{d-n}}
\fe
We also define $(d-n)$-cochains $(\star_\mathfrak{t} A)^{(d-n)}$ and $(\star_{\mathfrak{t}^{-1}} A)^{(d-n)}$ on $\Lambda$ as follows,
\ie
\left(\star_\mathfrak{t}A\right)_{c_{d-n}}=A_{\mathfrak{t}\left(c_{d-n}\right)},\ \left(\star_{\mathfrak{t}^{-1}}A\right)_{c_{d-n}}=A_{\mathfrak{t}^{-1}\left(c_{d-n}\right)}
\fe
It follows that $\star_{\mathfrak{t}^{-1}}=(\star_{\mathfrak{t}})^{-1}$. 

The operators that act on cochains satisfy similar relations as the ones in equations \eqref{eq:starboundary} and \eqref{eq:startboundary}. These relations are summarized in the following equations,
\ie\label{eq:dualwithd}
&d\star=(-1)^{d+1}\star\delta,~~~~~~~\star d=\delta\star\\
&d\ \star_\mathfrak{t}=(-1)^{d+1}\star_\mathfrak{t} \delta,~~~~\star_\mathfrak{t} d=\delta\star_\mathfrak{t}\\
&d\ \star_{\mathfrak{t}^{-1}}=\star_{\mathfrak{t}^{-1}} \delta,~~~~~~~~~~~\star_{\mathfrak{t}^{-1}} d=(-1)^{d+1}\delta\star_{\mathfrak{t}^{-1}}
\fe

The integration of an $n$-cochain on an $n$-chain is defined as follows,
\ie
\int _{\Sigma_n}A^{(n)}=\sum_{c_n\in\Sigma_n}A_{c_n}
\fe
There is a lattice version of the Stokes theorem, which can be proved directly by using the definition \eqref{eq:derivative},
\ie
\int _{\Sigma_{n+1}}dA^{(n)}=\int_{\partial \Sigma_{n+1}}A^{(n)}
\fe
In particular, if $\Sigma_{n+1}$ is a cycle, namely $\partial \Sigma_{n+1}=0$, the integral above vanishes. 

We can similarly define the integration of dual cochains on dual chains, which takes the following form
\ie\label{eq:intstar}
\int _{\tilde{\Sigma}_{d-n}}\star A^{(n)}=\int _{\star \tilde{\Sigma}_{d-n}}A^{(n)}=\sum_{\tilde{c}_{d-n}\in\tilde{\Sigma}_{d-n}}(\star A)_{\tilde{c}_{d-n}}
\fe
A closely related quantity is
\ie\label{eq:intstart}
\int _{\Sigma_{d-n}}\star_{\mathfrak{t}}A^{(n)}=\int _{\mathfrak{t}(\Sigma_{d-n})}A^{(n)}=\sum_{c_{d-n}\in \Sigma_{d-n}}A_{\mathfrak{t}(c_{d-n})}
\fe
Note that \eqref{eq:intstar} integrates an $n$-cochain $A^{(n)}\in C^n(\Lambda,\mathbb{R})$ over a dual $(d-n)$-chain $\tilde{\Sigma}_{d-n}\in C_{d-n}(\tilde{\Lambda},\mathbb{Z})$, while \eqref{eq:intstart} integrates an $n$-cochain $A^{(n)}\in C^n(\Lambda,\mathbb{R})$ over a $(d-n)$-chain $\Sigma_{d-n}\in C_{d-n}(\Lambda,\mathbb{Z})$. Although conceptually different, \eqref{eq:intstar} and \eqref{eq:intstart} are closely related: they produce the same value when $\Sigma_{d-n}=\mathfrak{t}^{-1}(\star \tilde{\Sigma}_{d-n})$. They also have the same continuum interpretation. In the main text, we use \eqref{eq:intstar} to make closer contact with the continuum picture, and \eqref{eq:intstart} for explicit lattice calculations.

\subsection{Cup product}

\begin{figure}[h]
    \centering
    \begin{tikzpicture}[scale=1.5,x={(1cm,0cm)},
    y={(0.6cm,0.4cm)},
    z={(0cm,1cm)},
    midarrow/.style={
        postaction={decorate},
        decoration={
            markings,
            mark=at position 0.6 with {\arrow[scale=1]{Stealth}}
        }
    }]
        \begin{scope}
        \coordinate (O) at (0,0,0);
        \coordinate (A) at (1,0,0);
        \coordinate (B) at (1,1,0);
        \coordinate (C) at (0,1,0);
        \coordinate (D) at (0,0,1);
        \coordinate (E) at (1,0,1);
        \coordinate (F) at (1,1,1);
        \coordinate (G) at (0,1,1);

        \node at (-1,0,0.6) {${\color{red}A^{(1)}}\cup {\color{blue}B^{(2)}}=$};

        \draw[thick] (A) -- (B) -- (F) -- (E) -- cycle;
        \draw[thick] (D) -- (G);
        \draw[thick] (G) -- (F);
        \draw[thick] (O) -- (D);
        \draw[thick] (D) -- (E);
        \draw (O) -- (C);
        \draw (C) -- (G);
        \draw (C) -- (B);
        \draw[thick, midarrow, red] (O) -- (A);
        \fill[blue, opacity=0.4] (A) -- (B) -- (F) -- (E) -- cycle;
        \draw[->,thick][canvas is yz plane at x=1](0.5+0.25/1.4,0.5+0.25/1.4) arc (45:315:0.25);
        \end{scope}

        \begin{scope}[xshift=2.25cm]
        \coordinate (O) at (0,0,0);
        \coordinate (A) at (1,0,0);
        \coordinate (B) at (1,1,0);
        \coordinate (C) at (0,1,0);
        \coordinate (D) at (0,0,1);
        \coordinate (E) at (1,0,1);
        \coordinate (F) at (1,1,1);
        \coordinate (G) at (0,1,1);

        \node at (-0.3,0,0.6) {$+$};

        \draw[thick] (A) -- (B) -- (F) -- (E) -- cycle;
        \draw[thick] (D) -- (G);
        \draw[thick] (G) -- (F);
        \draw[thick] (O) -- (D);
        \draw[thick] (D) -- (E);
        \draw[thick] (O) -- (A);
        \draw (O) -- (C);
        \draw (C) -- (G);
        \draw (C) -- (B);
        \draw[thick, midarrow, red] (O) -- (D);
        \fill[blue, opacity=0.4] (D) -- (E) -- (F) -- (G) -- cycle;
        \draw[->,thick][canvas is xy plane at z=1](0.5+0.25/1.4,0.5+0.25/1.4) arc (45:315:0.25);
        
        \end{scope}

        \begin{scope}[xshift=4.45cm]
        \coordinate (O) at (0,0,0);
        \coordinate (A) at (1,0,0);
        \coordinate (B) at (1,1,0);
        \coordinate (C) at (0,1,0);
        \coordinate (D) at (0,0,1);
        \coordinate (E) at (1,0,1);
        \coordinate (F) at (1,1,1);
        \coordinate (G) at (0,1,1);

        \node at (-0.3,0,0.6) {$+$};

        \draw[thick] (A) -- (B) -- (F) -- (E) -- cycle;
        \draw[thick] (D) -- (G);
        \draw[thick] (G) -- (F);
        \draw[thick] (O) -- (D);
        \draw[thick] (D) -- (E);
        \draw[thick] (O) -- (A);
        \draw (C) -- (G);
        \draw (C) -- (B);
        \draw[thick, midarrow, red] (O) -- (C);
        \fill[blue, opacity=0.4] (C) -- (B) -- (F) -- (G) -- cycle;
        \draw[->,thick][canvas is xz plane at y=1](0.5+0.25/1.4,0.5-0.25/1.4) arc (315:45:0.25);
        
        \end{scope}
    \end{tikzpicture}

    ~

        \begin{tikzpicture}[scale=1.5,x={(1cm,0cm)},
    y={(0.6cm,0.4cm)},
    z={(0cm,1cm)},
    midarrow/.style={
        postaction={decorate},
        decoration={
            markings,
            mark=at position 0.6 with {\arrow[scale=1]{Stealth}}
        }
    }]
        \begin{scope}
        \coordinate (O) at (0,0,0);
        \coordinate (A) at (1,0,0);
        \coordinate (B) at (1,1,0);
        \coordinate (C) at (0,1,0);
        \coordinate (D) at (0,0,1);
        \coordinate (E) at (1,0,1);
        \coordinate (F) at (1,1,1);
        \coordinate (G) at (0,1,1);

        \node at (-1,0,0.6) {${\color{blue}B^{(2)}}\cup {\color{red}A^{(1)}}=$};

        \draw[thick] (A) -- (B) -- (F) -- (E) -- cycle;
        \draw[thick] (D) -- (G);
        \draw[thick] (O) -- (A);
        \draw[thick] (O) -- (D);
        \draw[thick] (D) -- (E);
        \draw (O) -- (C);
        \draw (C) -- (G);
        \draw (C) -- (B);
        \draw[thick, midarrow, red] (G) -- (F);
        \fill[blue, opacity=0.4] (O) -- (C) -- (G) -- (D) -- cycle;
        \draw[->,thick][canvas is yz plane at x=0](0.5+0.25/1.4,0.5+0.25/1.4) arc (45:315:0.25);
        \end{scope}

        \begin{scope}[xshift=2.25cm]
        \coordinate (O) at (0,0,0);
        \coordinate (A) at (1,0,0);
        \coordinate (B) at (1,1,0);
        \coordinate (C) at (0,1,0);
        \coordinate (D) at (0,0,1);
        \coordinate (E) at (1,0,1);
        \coordinate (F) at (1,1,1);
        \coordinate (G) at (0,1,1);

        \node at (-0.3,0,0.6) {$+$};

        \draw[thick] (B) -- (A) -- (E) -- (F);
        \draw[thick] (D) -- (G);
        \draw[thick] (G) -- (F);
        \draw[thick] (O) -- (D);
        \draw[thick] (D) -- (E);
        \draw[thick] (D) -- (O);
        \draw[thick] (O) -- (A);
        \draw (O) -- (C);
        \draw (C) -- (G);
        \draw (C) -- (B);
        \draw[thick, midarrow, red] (B) -- (F);
        \fill[blue, opacity=0.4] (O) -- (A) -- (B) -- (C) -- cycle;
        \draw[->,thick][canvas is xy plane at z=0](0.5+0.25/1.4,0.5+0.25/1.4) arc (45:315:0.25);
        
        \end{scope}

        \begin{scope}[xshift=4.45cm]
        \coordinate (O) at (0,0,0);
        \coordinate (A) at (1,0,0);
        \coordinate (B) at (1,1,0);
        \coordinate (C) at (0,1,0);
        \coordinate (D) at (0,0,1);
        \coordinate (E) at (1,0,1);
        \coordinate (F) at (1,1,1);
        \coordinate (G) at (0,1,1);

        \node at (-0.3,0,0.6) {$+$};

        \draw[thick] (E) -- (A) -- (B) -- (F);
        \draw[thick] (D) -- (G);
        \draw[thick] (G) -- (F);
        \draw[thick] (O) -- (D);
        \draw[thick] (D) -- (E);
        \draw[thick] (O) -- (A);
        \draw[thick] (O) -- (C);
        \draw (C) -- (G);
        \draw (C) -- (B);
        \fill[blue, opacity=0.4] (O) -- (A) -- (E) -- (D) -- cycle;
        \draw[thick, midarrow, red] (E) -- (F);
        \draw[->,thick][canvas is xz plane at y=0](0.5+0.25/1.4,0.5-0.25/1.4) arc (315:45:0.25);
        
        \end{scope}
    \end{tikzpicture}

    ~

    \begin{tikzpicture}[scale=1.5,x={(1cm,0cm)},
    y={(0.6cm,0.4cm)},
    z={(0cm,1cm)}]
        \begin{scope}
        \coordinate (O) at (0,0,0);
        \coordinate (A) at (1,0,0);
        \coordinate (B) at (1,1,0);
        \coordinate (C) at (0,1,0);
        \coordinate (D) at (0,0,1);
        \coordinate (E) at (1,0,1);
        \coordinate (F) at (1,1,1);
        \coordinate (G) at (0,1,1);

        \node at (-1,0,0.6) {${\color{red}A^{(0)}}\cup {\color{blue}B^{(3)}} =$};

        \fill[blue,opacity=0.4] (O) -- (A) -- (E) -- (D) -- cycle;
        \fill[blue,opacity=0.4] (A) -- (B) -- (F) -- (E) -- cycle;
        \fill[blue,opacity=0.4] (D) -- (E) -- (F) -- (G) -- cycle;

        \draw[thick] (O) -- (A) -- (E) -- (D) -- cycle;
        \draw[thick] (D) -- (G);
        \draw[thick] (G) -- (F);
        \draw[thick] (E) -- (F);
        \draw[thick] (A) -- (B);
        \draw[thick] (B) -- (F);
        \draw (O) -- (C);
        \draw (C) -- (G);
        \draw (C) -- (B);
        \fill[red] (0,0,0) circle (2pt);
        \end{scope}

        \begin{scope}[xshift=4cm]
        \coordinate (O) at (0,0,0);
        \coordinate (A) at (1,0,0);
        \coordinate (B) at (1,1,0);
        \coordinate (C) at (0,1,0);
        \coordinate (D) at (0,0,1);
        \coordinate (E) at (1,0,1);
        \coordinate (F) at (1,1,1);
        \coordinate (G) at (0,1,1);

        \node at (-1,0,0.6) {${\color{blue}B^{(3)}}\cup {\color{red}A^{(0)}} =$};

        \fill[blue,opacity=0.4] (O) -- (A) -- (E) -- (D) -- cycle;
        \fill[blue,opacity=0.4] (A) -- (B) -- (F) -- (E) -- cycle;
        \fill[blue,opacity=0.4] (D) -- (E) -- (F) -- (G) -- cycle;

        \draw[thick] (O) -- (A) -- (E) -- (D) -- cycle;
        \draw[thick] (D) -- (G);
        \draw[thick] (G) -- (F);
        \draw[thick] (E) -- (F);
        \draw[thick] (A) -- (B);
        \draw[thick] (B) -- (F);
        \draw (O) -- (C);
        \draw (C) -- (G);
        \draw (C) -- (B);
        \fill[red] (1,1,1) circle (2pt);
        \end{scope}
    \end{tikzpicture}
    \caption{Examples of the cup product in 3 dimensions. }
    \label{fig:cupproduct3}
\end{figure}
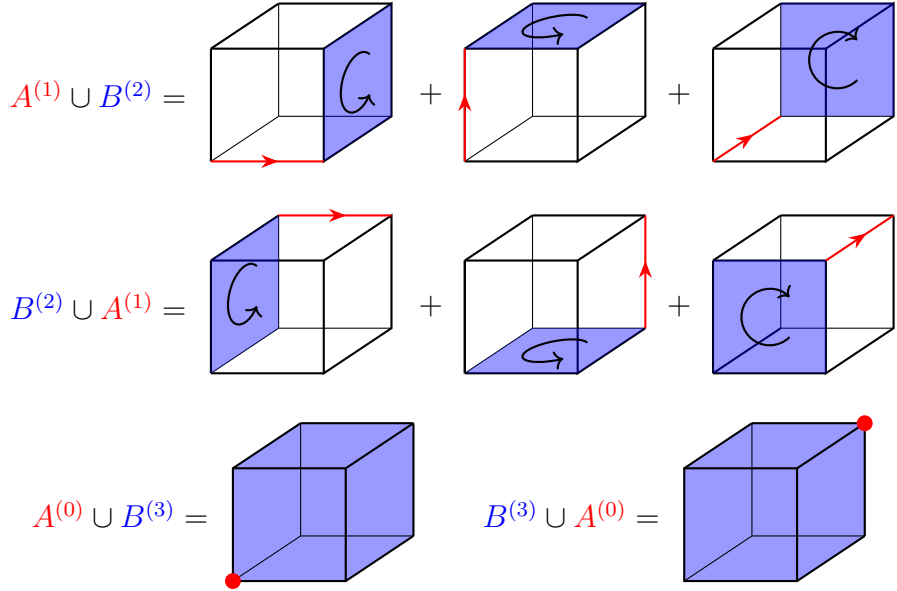

In this subsection, we introduce the cup product of cochains. Given an $n$-cochain $A^{(n)}$ and an $m$-cochain $B^{(m)}$, the cup product can be defined by
\ie
(A\cup B)_{c_{n+m}(\vec{r})_{i_1\cdots i_{n+m}}}=
\frac{1}{n!m!}\sum_{\substack{a_1\cdots a_{n+m}\in\\\{1,\cdots,n+m\}}}
\epsilon_{a_1\cdots a_{n+m}}
A_{c_n(\vec{r})_{i_{a_1}\cdots i_{a_n}}}
B_{c_{m}(\vec{r}+\hat{i}_{a_1}+\cdots+\hat{i}_{a_n})_{i_{a_{n+1}}\cdots i_{a_{n+m}}}}
\fe
Here $\epsilon_{a_1\cdots a_{n+m}}$ is completely antisymmetric with $\epsilon_{1\cdots n+m}=1$. Note that this cup product is associative, but not graded-commutative:
\ie
A^{(n)}\cup B^{(m)}&-(-1)^{nm}B^{(m)}\cup A^{(n)}=\\
&(-1)^{n+m+1}\left[d(A^{(n)}\cup_1 B^{(m)})-dA^{(n)}\cup_1 B^{(m)}-(-1)^nA^{(n)}\cup_1 dB^{(m)}\right]
\fe
where $\cup_1$ is a higher cup product whose definition can be found in \cite{Chen:2021ppt,Jacobson:2023cmr}. See Figure~\ref{fig:cupproduct12} and \ref{fig:cupproduct3} for examples of the cup product involving low-dimensional cells.

\begin{figure}[h]
    \centering
    \begin{tikzpicture}[scale=1.5,
    midarrow/.style={
        postaction={decorate},
        decoration={
            markings,
            mark=at position 0.6 with {\arrow[scale=1]{Stealth}}
        }
    }]
        \begin{scope}
        \node at (-1.1,0) {${\color{red}A^{(0)}}\cup {\color{blue}B^{(1)}}=$};
        \draw[thick,midarrow,blue] (0,0) -- (1,0);
        \fill[red] (0,0) circle (1.5pt);
        \end{scope}
        \begin{scope}[yshift=-0.7cm]
        \node at (-1.1,0) {${\color{blue}B^{(1)}}\cup{\color{red}A^{(0)}}=$};
        \draw[thick,midarrow,blue] (0,0) -- (1,0);
        \fill[red] (1,0) circle (1.5pt);
        \end{scope}

        \begin{scope}[xshift=3.3cm,yshift=-0.3cm]
        \node at (-1.1,0) {${\color{red}A^{(0)}}\cup {\color{blue}B^{(2)}}=$};
        \draw[thick] (0,-0.5) -- (1,-0.5) -- (1,0.5) -- (0,0.5) -- cycle;
        \fill[blue,opacity=0.4] (0,-0.5) -- (1,-0.5) -- (1,0.5) -- (0,0.5) -- cycle;
        \fill[red] (0,-0.5) circle (2pt);
        \draw[->,thick](0.5+0.25/1.4,0.25/1.4) arc (45:315:0.25);
        \end{scope}
        \begin{scope}[xshift=6.5cm,yshift=-0.3cm]
        \node at (-1.1,0) {${\color{blue}B^{(2)}}\cup {\color{red}A^{(0)}}=$};
        \draw[thick] (0,-0.5) -- (1,-0.5) -- (1,0.5) -- (0,0.5) -- cycle;
        \fill[blue,opacity=0.4] (0,-0.5) -- (1,-0.5) -- (1,0.5) -- (0,0.5) -- cycle;
        \fill[red] (1,0.5) circle (2pt);
        \draw[->,thick](0.5+0.25/1.4,0.25/1.4) arc (45:315:0.25);
        \end{scope}
    \end{tikzpicture}
    
    ~
    
    \begin{tikzpicture}[scale=1.5,
    midarrow/.style={
        postaction={decorate},
        decoration={
            markings,
            mark=at position 0.6 with {\arrow[scale=1]{Stealth}}
        }
    }]
        \begin{scope}
        \node at (-1.1,0.5) {${\color{red}A^{(1)}}\cup {\color{blue}B^{(1)}}=$};
        \draw (0,0) -- (1,0) -- (1,1) -- (0,1) -- cycle;
        \draw[thick,midarrow,red] (0,0) -- (1,0);
        \draw[thick,midarrow,blue] (1,0) -- (1,1);
        \node at (1.3,0.45) {$-$};
        \end{scope}
        \begin{scope}[xshift=1.6cm]
        \draw (0,0) -- (1,0) -- (1,1) -- (0,1) -- cycle;
        \draw[thick,midarrow,red] (0,0) -- (0,1);
        \draw[thick,midarrow,blue] (0,1) -- (1,1);
        \end{scope}
    \end{tikzpicture}
    \caption{Examples of the cup product in 1 and 2 dimensions. }
    \label{fig:cupproduct12}
\end{figure}
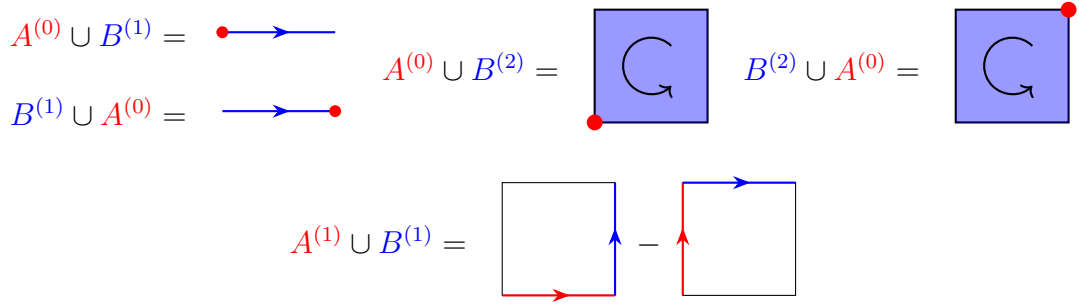

It is useful to study the relations between the cup product and the operators defined earlier. The cup product satisfies the following Leibniz rule, 
\ie
d(A^{(n)}\cup B^{(m)})=d A^{(n)}\cup B^{(m)}+(-1)^nA^{(n)}\cup d B^{(m)}
\fe
Moreover, the operator $\star_{\mathfrak{t}}$ can be expressed in terms of the cup product,
\ie \label{eq:cupandstart}
\star_{\mathfrak{t}}A^{(n)}=\int_{M_d} \mathbf{c}^{(d-n)} \cup A^{(n)},\quad \star_{\mathfrak{t}^{-1}}A^{(n)}=\int_{M_d}  A^{(n)}\cup \mathbf{c}^{(d-n)}
\fe
Here $M_d$ is the entire $d$-dimensional lattice, $\mathbf{c}^{(n)}$ is the $n$-cochain that is 1 on $\mathbf{c}^{(n)}$ and 0 otherwise \cite{Chen:2021xks}. It is also useful to note the following identity
\ie
\int \star_\mathfrak{t}^2 A^{(n)}\cup B^{(d-n)}=\int B^{(d-n)}\cup A^{(n)}
\fe

\subsection{Commutators of cochain operators}\label{app:commutator}

In the main text, we work with cochains that assign to each chain an operator on the Hilbert space, which we refer to as "cochain operators." We now develop techniques for computing their commutators. 

We start with a pair of canonical conjugate cochain operators $X^{(n)}$ and $P^{(n)}$, which satisfy the following commutation relation,
\ie\label{eq:conjugatepair}
[X_{c_n},P_{c'_n}]=i\delta_{c_n,c'_n}
\fe
Here $c_n$ and $c'_n$ are $n$-cells. The delta function on the right-hand side is defined as follows,
\ie
\delta_{c_n,c'_n}=\begin{cases}
    1&c_n=c'_n\\
    -1&c_n=-c'_n\\
    0&\mbox{other cases}
\end{cases}
\fe
Importantly, $\delta_{c_n,c'_n}$ itself can be viewed as an $n$-cochain
\ie
\delta_{c_n,c'_n}=\left(\mathbf{c}^{(n)}\right)_{c'_n}=\left(\mathbf{c}'^{(n)}\right)_{c_n}
\fe
Namely, $\mathbf{c}^{(n)}$ denotes the $n$-cochain that is 1 on $c_n$ and 0 otherwise. 

First, we derive the commutators involving divergence or exterior derivative,
\ie
&[(dX)_{c'_{n+1}},P_{c_n}]=i\left(d\mathbf{c}^{(n)}\right)_{c'_{n+1}}=(-1)^{n+1}i(\delta\mathbf{c}'^{(n+1)})_{c_n},\\
&[X_{c_n},(dP)_{c'_{n+1}}]=i\left(d\mathbf{c}^{(n)}\right)_{c'_{n+1}}=(-1)^{n+1}i(\delta\mathbf{c}'^{(n+1)})_{c_n}\\
&[(\delta X)_{c'_{n-1}},P_{c_n}]=i\left(\delta\mathbf{c}^{(n)}\right)_{c'_{n-1}}=(-1)^{n}i\left(d\mathbf{c}'^{(n-1)}\right)_{c_n}\\
&[X_{c_n},(\delta P)_{c'_{n-1}}]=i\left(\delta\mathbf{c}^{(n)}\right)_{c'_{n-1}}=(-1)^{n}i\left(d\mathbf{c}'^{(n-1)}\right)_{c_n}
\fe
Inspired by this, we generalize \eqref{eq:conjugatepair} to two cochain operators $A^{(n)}$ and $B^{(m)}$ satisfying the following commutation relation,
\ie\label{eq:general-commutator}
\left[A_{c'_n},B_{c_m}\right]=i\left(Q^{(r),c_m}\right)_{c'_n}
\fe
where $Q^{(r),c_m}$ is an $n$-cochain whose definition depends on $c_m$. The action of an exterior derivative or divergence operator on $A^{(n)}$ naturally passes to $Q^{(r),c_m}$.
\ie\label{eq:deltadcommutator}
\left[(dA)_{c'_{n+1}},B_{c_m}\right]=i\left(dQ^{(r),c_m}\right)_{c'_{n+1}},~~~\left[(\delta A)_{c'_{n-1}},B_{c_m}\right]=i\left(\delta Q^{(r),c_m}\right)_{c'_{n-1}}.
\fe
In particular, by taking $A^{(n)}$ to be $X^{(n)}$, $B^{(m)}$ to be $dP^{(n)}$ or $\delta P^{(n)}$, we have
\ie
[(dX)_{c'_{n+1}},(\delta P)_{c_{n-1}}]=[(\delta X)_{c_{n-1}},(d P)_{c'_{n+1}}]=0\,.
\fe
As an example, consider the Hamiltonian model introduced in Section \ref{sec:Hamiltonian}. The last term in the Hamiltonian $[(dw)_p]^2$ is gauge invariant under \eqref{gauss1} because $[(dw)_p,(\delta b)_s]=0$.

Next, we derive commutators involving integration. Starting from the commutator \eqref{eq:general-commutator}, we find the following identity,
\ie\label{eq:intcommutator}
\left[\int_{\Sigma_n} A^{(n)},B_{c_m}\right]=i\int _{\Sigma_n}Q^{(r),c_m}
\fe

Finally, we consider commutators involving the cup product. In addition to \eqref{eq:general-commutator}, we introduce another cochain operator $Y^{(q)}$ satisfying
\ie
\left[Y_{c'_q},B^{(m)}_{c_m}\right]=i\left(Z^{(q),c_m}\right)_{c'_q}
\fe
We have the following identity,
\ie\label{eq:cupcommutator}
\left[\left(A^{(n)}\cup Y^{(q)}\right)_{c'_{n+q}},
B^{(m)}_{c_m}\right]=i \left(Q^{(r),c_m}\cup Y^{(q)}\right)_{c'_{n+q}}+i \left(A^{(n)}\cup Z^{(q),c_m}\right)_{c'_{n+q}}
\fe
In particular, if we set $A^{(n)}$ to be $X^{(n)}$, $B^{(m)}$ to be $P^{(n)}$, $q=d-n$, and $Y^{(q)}$ to commute with $P^{(n)}$, then
\ie
\left[\int_{M_{d}} X^{(n)}\cup Y^{(d-n)},P^{(n)}_{c_n}\right]=i \int_{M_d}\mathbf{c}^{(n)}\cup Y^{(d-n)}=i\left(\star_{\mathfrak{t}} Y^{(d-n)}\right)_{c_n}
\fe

Equations \eqref{eq:deltadcommutator}, \eqref{eq:intcommutator}, and \eqref{eq:cupcommutator} are the basic ingredients for computing more complicated commutators of cochain operators. As a first example, we derive \eqref{eq:QAactonb} by computing the following commutator,
\ie
\left[\int_{M_3}w^{(1)}\cup dw^{(1)},b_\ell\right]&=-i\int_{M_3} \mathbf{l}^{(1)}\cup dw^{(1)}-i\int_{M_3} w^{(1)}\cup d\mathbf{l}^{(1)}\\
&=-i\left((dw)_{\mathfrak{t}(\ell)}+(dw)_{\mathfrak{t}^{-1}(\ell)}\right)
\fe

As a second example, we show that \eqref{eq:AgaugeQV} is invariant under \eqref{eq:Agaugetransform1} and \eqref{eq:Agaugetransform2}. We first note that
\ie
\sum_{\ell\in\hat{\partial}s}(\mathbf{l})_{\ell'}=(\delta \mathbf{l}')_s=-(d\mathbf{s})_{\ell'}
\fe
Under \eqref{eq:Agaugetransform1}, $(\delta b)_s$ is transformed as follows,
\ie
(\delta b)_s\sim &(\delta b)_s-\int_{M_3} \left(d\mathbf{s}\cup dw\cup \alpha+dw\cup d\mathbf{s}\cup \alpha+\left(w+\frac{d\phi}{2\pi}\right)\cup ds\cup d\alpha \right)\\
&=(\delta b)_s+\int_{M_3}s\cup dw\cup d\alpha \\
&=(\delta b)_s+(dw\cup d\alpha)_{\mathfrak{t}(s)}
\fe
In deriving the second line, we used integration by parts. We see that the transformations of $p_s$ and ${(\delta b)_s\over 2\pi}$ under \eqref{eq:Agaugetransform1} cancel. Therefore, $Q_{\text{V}'}^{(0)}$ is invariant under \eqref{eq:Agaugetransform1}. Under \eqref{eq:Agaugetransform2}, $(\delta b)_s$ is transformed as follows,
\ie
(\delta b)_s\sim &(\delta b)_s-\int_{M_3} \phi\cup d^2\mathbf{s}\cup k=(\delta b)_s
\fe
Namely, $(\delta b)_s$ is invariant under \eqref{eq:Agaugetransform2}. The gauge transformation of $Q_{\text{V}'}^{(0)}$ under \eqref{eq:Agaugetransform2} is given by
\ie
Q_{\text{V}'}^{(0)}\sim &Q_{\text{V}'}^{(0)}+\int_{M_3}\star_{\mathfrak{t}}^2\left(-dw\cup k+w\cup dk\right)\\
&=Q_{\text{V}'}^{(0)}-\int_{M_3}d\left[\star_{\mathfrak{t}}^2\left(w\cup k\right)\right]\\
&=Q_{\text{V}'}^{(0)}
\fe
Where in deriving the second line we have used the fact that $d$ commutes with $\star_{\mathfrak{t}}$ in 3 dimensions (see \eqref{eq:dualwithd}). We conclude that $Q_{\text{V}'}^{(0)}$ is invariant under both \eqref{eq:Agaugetransform1} and \eqref{eq:Agaugetransform2}.

As a third example, we present a more involved calculation, in which we verify that the last Gauss law in \eqref{eq:2groupGauss1} commutes with the last Gauss law in \eqref{eq:2groupGauss2}. We verify the vanishing of the following expression,
\ie
&\left[(\phi^{(0)} \cup dw^{(1)})_{\mathfrak{t}^{-1}(\ell)},p_s^{(0)}\right]
-\left[(\delta \tilde{A}^{(2)})_\ell ,(w^{(1)}\cup n^{(2)})_{\mathfrak{t}(s)}\right]-\left[\left(\mathcal{A}^{(1)}\cup w^{(1)}\right)_{\mathfrak{t}^{-1}(\ell)},\left(\delta \mathcal{E}^{(1)}\right)_s\right]\\
&-\frac{1}{2\pi}\left[\mathcal{B}^{(2)}_{\mathfrak{t}^{-1}(\ell)},\int_{M_3}\mathcal{A}^{(1)}\cup d\mathbf{s}^{(0)}\cup\star_{\mathfrak{t}^{-1}}\mathcal{H}^{(2)} \right]+\frac{1}{2\pi}\left[\left(\mathcal{A}^{(1)}\cup w^{(1)}\right)_{\mathfrak{t}^{-1}(\ell)},\left(\delta b^{(1)}\right)_s\right]
\fe
The first line can be simplified as follows,
\ie
i\int_{M_3}\left(\mathbf{s}^{(0)}\cup dw^{(1)}\cup\mathbf{l}^{(1)}-\mathbf{s}^{(0)}\cup w^{(1)}\cup d\mathbf{l}^{(1)}+d\mathbf{s}^{(0)}\cup w^{(1)}\cup\mathbf{l}^{(1)}\right)=0
\fe
where we used the Leibniz rule of the exterior derivative with respect to the cup product. 
The first term in the  second line is $-{i\over 2\pi}\int\mathcal{A}^{(1)}\cup d\mathbf{s}^{(0)}\cup\mathbf{l}^{(1)}$, which cancels with the second term. 
Thus, we verify that the last Gauss law in \eqref{eq:2groupGauss1} indeed commutes with that in \eqref{eq:2groupGauss2}. Other calculations involving cochain operators can be performed in a similar manner.

\section{More on the Villain Hamiltonian}\label{app:more}

\subsection{Duality}\label{app:duality}

In the continuum, a free compact boson field $\phi$ in 3+1d is exactly dual to a 2-form gauge field $b_{\mu\nu}$ via $\partial_\mu\phi \sim \epsilon_{\mu\nu\rho\sigma}\partial^\nu b^{\rho\sigma}$. 
Our lattice Hamiltonian \eqref{Hamiltonian} also realizes this duality exactly, which generalizes the 1+1d lattice T-duality in \cite{Gorantla:2021svj,Cheng:2022sgb,Fazza:2022fss,Seifnashri:2026ema}. 

The duality transformation is implemented by the following unitary transformations,
\ie\label{eq:duality}
    \mathcal{U}:\quad\begin{split}
        \phi_s^{(0)}\mapsto \tilde{\phi}^{(3)}_{\star s}\qquad &p_s^{(0)}\mapsto \left(\tilde{p}^{(3)}+\frac{d\tilde{b}^{(2)}}{2\pi}\right)_{\star s}\\
        b_{\ell}^{(1)}\mapsto \tilde{b}^{(2)}_{\star \ell}\qquad&w^{(1)}_\ell\mapsto \left(\tilde{w}^{(2)}-\frac{\delta \tilde{\phi}^{(3)}}{2\pi}\right)_{\star \ell}
    \end{split}
\fe
where $\star$ is the lattice Hodge dual operator that maps a $p$-chain on the original lattice $M_3$ to a $(3-p)$-chain on the dual lattice $\tilde M_3$. 
Under the duality transformation $\mathcal{U}$, the Villain fields $b^{(1)}$ and $w^{(1)}$ are mapped to the gauge field $\tilde{b}^{(2)}$ and its electric field $\tilde{w}^{(2)}$, while the scalar field $\phi^{(0)}$ and its conjugate $p^{(0)}$ are mapped to new Villain fields $\tilde{\phi}^{(3)}$ and $\tilde{p}^{(3)}$.

The dual theory is described by the following Hamiltonian,
\ie
    \mathcal{U} H\mathcal{U}^{-1}=\frac{1}{2\beta'}\sum_{\tilde{p}}\tilde{w}_{\tilde {p}}^2+
    \frac{\beta'}{2}\sum_{\tilde{c}}((d\tilde{b})_{\tilde{c}}+2\pi\tilde{p}_{\tilde{c}})^2
    +\frac{\lambda}{2}\sum_{\tilde{\ell}}[(\delta \tilde{w})_{\tilde{\ell}}]^2,\quad 
    \beta'=\frac{1}{4\pi^2\beta}
\fe
Using the duality transformation \eqref{eq:duality}, the Gauss law constraints \eqref{gauss1} and \eqref{gauss2} are mapped to the following Gauss law constraints in the dual theory,
\ie
 \exp(2\pi i \tilde{p}_{\tilde{c}})=1,\quad \exp(2\pi i \tilde{w}_{\tilde{p}}-i(\delta\tilde{\phi})_{\tilde{p}})=1
\fe
The first constraint restricts $\tilde{p}$ to take integer valued and identifies $\tilde{\phi}\sim\tilde{\phi}+2\pi$. The second constraint implements the $\mathbb{Z}$ gauge transformation. Importantly, the flatness condition $dw=0$ in the scalar description is mapped under $\mathcal{U}$ to the Gauss law constraint $\delta \tilde{w}=0$ in the dual theory, which generates the following gauge transformation,
\ie
\tilde{b}_{\tilde{p}}\sim \tilde{b}_{\tilde{p}}+(d\Lambda)_{\tilde{p}},\quad \Lambda_{\tilde \ell}\in\mathbb{R} \,.
\fe
This is the lattice counterpart of \eqref{2formgauge}. 
This makes it clear that the dual description is the Villain formulation of a 2-form gauge theory with gauge field $\tilde{b}$, in which the Gauss law $\delta\tilde{w}=0$ is imposed energetically.

\subsection{2-form winding symmetry}\label{app:winding}

Here we discuss a higher-form global symmetry in our lattice Hamiltonian \eqref{Hamiltonian}. 
We do not impose it in our microscopic model, but it plays an important role in the IR field theory.

Given any 1-cycle $\gamma_1$, our Hamiltonian commutes with the following loop operator\footnote{The axial charge $Q_\text{A}$ in \eqref{QA} takes the form of a Chern-Simons term of the charge density $w^{(1)}$ for the 2-form global symmetry. This is reminiscent of the higher gauging in \cite{Roumpedakis:2022aik} and of the Chern-Weil global symmetry in \cite{Brauner:2020rtz,Heidenreich:2020pkc}. In all these cases, one uses a higher-form symmetry to construct an ordinary, 0-form symmetry.} 
\ie
Q_\text{W}^{(2)} = \sum_{\ell \in \gamma_1 }w_\ell
= \int_{\gamma_1} w^{(1)}\,.
\fe
This conserved operator generates a U(1)$_\text{W}^{(2)}$ 2-form global symmetry that counts how many times the compact boson winds around the closed curve $\gamma_1$. 
The charge density $\tilde q_\text{W}=w_\ell$ is quantized but is not gauge-invariant under \eqref{Zgauge}. Instead, we introduce another charge density 
\ie
q_\text{W} = w_\ell + {(d\phi)_\ell \over 2\pi}
\fe
which is gauge-invariant but not quantized. 
Of course, the total charge $Q_\text{W}^{(2)}$ is both quantized and gauge invariant.

The winding charge $Q_\text{W}^{(2)}$  is \emph{non-topological} because it depends on the specific shape of $\gamma_1$. 
If we had imposed $dw = 0$ strictly as a Gauss-law constraint, then $Q_{\mathrm{W}}^{(2)}$ would become topological and depend only on the homology class $[\gamma_1]$. 
See \cite{Seiberg:2019vrp,Qi:2020jrf,Oh:2023bnk,Choi:2024rjm,Gorantla:2024ocs,Liu:2026tcl} for further discussions of the distinction between topological and non-topological higher-form global symmetries.

This symmetry has a mixed 't Hooft anomaly with U(1)$_\text{V}$. 
To see this, we follow the argument in \cite{Thorngren:2026ydw} and perform a $2\pi$ U(1)$_\text{W}^{(2)}$ rotation in a segment $I$ with endpoints $s_1$ and $s_2$, which gives
\ie
\exp\left(2\pi i \int_{I} q_\text{W} \right)
= \exp(i \phi_{s_2} - i \phi_{s_1})\,.
\fe
The endpoints are charged under the U(1)$_\text{V}$ symmetry, signaling the mixed anomaly. 
This is sometimes referred to as ``pumping", which generalizes the notion of spectral flow in 1+1d CFT.

\section{Gauging on the lattice}\label{app:gauging}

Here, we discuss a systematic method of gauging continuous global U(1) symmetries on the lattice. We follow \cite{Seifnashri:2026ema}, and consider U(1) symmetries, where the associated conserved charge is written as a sum of commuting local charges. For gauging more general anomaly-free symmetries on the lattice, see \cite{Seifnashri:2023dpa}.

\subsection{Gauging with Villain fields}\label{app:villain}

We start by gauging a U(1) 0-form symmetry generated by a conserved and quantized charge
\ie \label{app:charge}
    Q = \sum_s q_s \,,
\fe
where the local charges commute with each other, i.e., $[q_{s},q_{s'}]=0$. We assume that the local charges are gauge invariant, but not necessarily quantized. We first couple the theory to a dynamical $\bR$ gauge field $A^{(1)}$ (and its conjugate $E^{(1)}$). Then, we compactify the gauge fields $A^{(1)}$ by gauging the $\bZ$ 1-form symmetry that shifts the real-valued gauge field $A^{(1)}$ by an integer-valued 1-form. We will denote the integer Villain field by $n^{(2)}$, and its conjugate variable by $\tilde{A}^{(2)}$. 

Gauging the $\bR$ 0-form and $\bZ$ 1-form symmetries  each consist of three steps:
\begin{enumerate}
    \item[I.] coupling to background gauge fields
    \item[II.] imposing Gauss's law constraints
    \item[III.] adding kinetic terms for the continuous gauge fields
\end{enumerate}
where the last step is only relevant for gauging continuous symmetries, and not for gauging $\bZ$. Let us start by gauging the $\bR$ symmetry associated with the conserved charge \eqref{app:charge}.

\paragraph{I. Coupling to background gauge fields:} to couple the system to background gauge fields, we conjugate the Hamiltonian and Gauss's laws with the unitary operator
\ie \label{app:bk.gf}
    \exp{\left(-i \sum_s \alpha_s q_s\right)}\,, \qquad \text{where} \quad (d\alpha)_\ell = A_\ell \,.
\fe
Because of the global symmetry, the Hamiltonian and Gauss's constraints commute with the unitary operator above for constant $\alpha_s$, i.e., $\alpha_s = \alpha_0$ for all $s$. Therefore, after conjugation, the system depends only on $d\alpha^{(0)} = A^{(1)}$. Note that we first consider the system on the infinite spatial lattice $\bZ^3$, such that $\alpha^{(0)}$ is determined up to a constant in terms of $A^{(1)}$. The locality of the Hamiltonian and $q_s$ implies that the gauge fields $A_\ell$ couple locally to the system. Thus, we can use the infinite system to put the Hamiltonian on finite lattices.

The idea behind the step above is the following. Consider the background gauge field configuration $A_x(\vec r) = \alpha_0\,\delta_{x,0}$ with trivial $A_y(\vec r)$ and $A_z(\vec r)$. Such a configuration corresponds to inserting a symmetry defect (or twisted boundary condition) along the $yz$-plane at $x=0$. Moreover, such a twisted boundary condition is imposed by applying the symmetry $e^{i\alpha_0 Q}$ on half of the space, which is precisely the unitary operator \eqref{app:bk.gf} for $\alpha(\vec r) = \frac{1+\text{sign}(x)}{2} \alpha_0$, where $d\alpha^{(0)} = A^{(1)}$.

\paragraph{II. Imposing Gauss's law:} to make the gauge fields dynamical, we add the conjugate variables $E_\ell$ (the electric field), which satisfy $[A_\ell, E_{\ell'}] = i \delta_{\ell , \ell'}$. We then impose Gauss's law constraint on sites:
\ie
    (\delta E)_s = q_s \,.
\fe

The constraint above is guaranteed to commute with the system coupled to the gauge field constructed in the previous step. To see this, we note that prior to coupling with $A_\ell$, the original system commutes with the constraint $\delta E =0$. Thus, we only need to show that conjugation by the unitary operator \eqref{app:bk.gf} maps the constraint $\delta E$ to $\delta E - q$. To verify this, choose $\xi_\ell$ such that $\delta \xi = q$, and use the ``integration by part'' formula \eqref{int.by.part} $\sum_s \alpha_s (\delta \xi)_s = - \sum_{\ell} (d\alpha)_\ell \, \xi_\ell$ to write
\ie
    \exp{\left(-i \sum_s \alpha_s q_s\right)} = \exp{\left(i \sum_\ell A_\ell \, \xi_\ell \right)}  \,, \qquad \text{where} \quad \delta \xi = q \,.
\fe
From this expression, it is clear that $E$ is mapped to $E - \xi$ and therefore $\delta E=0$ becomes $\delta E = q$.

\paragraph{III. Adding kinetic terms:} we add the following kinetic term for the gauge fields
\ie
    H_{\text{kinetic}} = \frac{1}{2 \gamma} \sum_\ell E_\ell^2 + \frac{\gamma}{2} \sum_p [(dA)_p]^2 \,.
\fe

To compactify the gauge field $A_\ell$, we gauge the $\bZ$ 1-form symmetry that shifts $A_\ell$ by an integer-valued 1-cochain. Since the charge $Q$ above is quantized, the gauged theory admits a global $\bZ$ 1-form symmetry generated by
\ie
    U\left[k^{(1)}\right]=\exp{\left( i\sum_{\ell} k_\ell j_\ell \right)}\,,  
\fe
for any $k_\ell \in \bZ$ such that $d k^{(1)} = 0$, where $j_\ell$ are local commuting ``charges'': $[j_\ell, j_{\ell'}]=0$.
We can repeat the two steps above to gauge this symmetry:
\begin{enumerate}
    \item \textbf{Coupling to background gauge fields:} We add integer-valued gauge fields $n^{(2)}$ and conjugate the system with
    \ie
    \exp{\left(i \sum_\ell k_\ell j_\ell \right)}\,, \qquad \text{where} \quad (dk)_p = -n_p \,.
    \fe
    \item \textbf{Gauss's laws:} We make the integer-valued gauge field $n^{(2)}$ dynamical by adding its conjugate variable $\tilde{A}^{(2)}$ satisfying $[\tilde{A}_p, n_{p'}]=i \delta_{p,p'}$, and impose the following Gauss's law constraints
    \ie
        \exp\left( 2\pi i n_p \right)=1 \,, \qquad \exp\left(i (\delta \tilde{A})_\ell -i j_\ell \right) = 1 \,.
    \fe
\end{enumerate}

\subsubsection*{Gauging U(1)$_{\mathrm{V}}$}
Let us demonstrate this for the case of gauging $U(1)_{\mathrm{V}}$ of the Hamiltonian \eqref{Hamiltonian}. Take the conserved charge $Q_{\mathrm{V}} = \sum_s p_s$. Conjugating the system with $e^{-i\sum_s \alpha_s p_s}$, and adding a kinetic term for the gauge field, we find the Hamiltonian
\ie
H &= \frac{1}{2 \gamma} \sum_\ell E_\ell^2 + \frac{\gamma}{2} \sum_p [(dA)_p]^2 \\ 
&+ {1\over 2\beta}\sum_s p_s^2
+{\beta\over2}
\sum_\ell \left( ( d\phi)_\ell -A_\ell +2\pi w_\ell\right)^2
+{\lambda\over2} \sum_p [(dw)_p ]^2
\,.
\fe
Note that the unitary transformation maps $\phi^{(0)} \mapsto\phi^{(0)}-\alpha^{(0)} $, and therefore $d\phi^{(0)} \mapsto d\phi^{(0)}-A^{(1)}$. The system is subject to the gauge constraints:
\ie
    &\exp\left(2\pi i w_\ell \right) &= 1 \,, \qquad \exp\left(2\pi i p_s - i(\delta b)_s \right) = 1 \,, \\
    &(\delta E)_s = p_s \,.
\fe
The above system has a $\bZ$ 1-form symmetry generated by $e^{i \sum_\ell k_\ell (2\pi E_\ell - b_\ell)}$ for $d k^{(1)}=0$, which corresponds to $j_\ell = 2\pi E_\ell - b_\ell$. We now gauge this 1-form symmetry. To couple the theory to background gauge fields $n^{(2)}$ for this symmetry, we set $dk^{(1)} = -n^{(2)}$ and conjugate the system with $e^{i\sum_\ell k_\ell (2\pi E_\ell - b_\ell)}$ to find
\ie
H_{\mathrm{V}} &= \frac{1}{2 \gamma} \sum_\ell E_\ell^2 + \frac{\gamma}{2} \sum_p [(dA)_p - 2\pi n_p]^2 \\ 
&+ {1\over 2\beta}\sum_s p_s^2
+{\beta\over2}
\sum_\ell \left( ( d\phi)_\ell -A_\ell +2\pi w_\ell\right)^2
+{\lambda\over2} \sum_p [(dw)_p - n_p]^2
\,.
\fe
We have to impose additional Gauss constraints:
\ie
    \exp\left( 2\pi i n_p \right)=1 \,, \qquad \exp{\left( i(\delta \tilde{A})_\ell - 2\pi i E_\ell + i b_\ell \right)} = 1\,.
\fe
Putting all the constraints together, we find \eqref{eq:Gausslaw_gaugeV}.

\subsubsection*{Gauging U(1)$_{\mathrm{A}}$}

Here, we discuss gauging the U(1)$_\mathrm{A}$ axial symmetry of \eqref{Hamiltonian}. Let us choose the local axial charge to be the gauge invariant charge given in  \eqref{local.a.charge}. Namely,
\ie
    q_s = -\left( \left( w  + {d\phi\over 2\pi}  \right) \cup dw \right)_{\mathfrak{t}^{-1}(s)} \,.
\fe
Conjugating the system with
\ie
    \exp\left( -i \sum_s \alpha_s q_s \right) &= \exp\left( i \sum_s \alpha_s \int  \left( w^{(1)}  + {(d\phi)^{(1)}\over 2\pi}  \right) \cup dw^{(1)} \cup \mathbf{s}^{(0)} \right) \\ &= \exp\left( i \int  \left( w^{(1)}  + {(d\phi)^{(1)}\over 2\pi}  \right) \cup dw^{(1)} \cup \alpha^{(0)} \right) \,,
\fe
and adding a kinetic term for the gauge fields, we find the Hamiltonian
\ie
H &= \frac{1}{2 \gamma} \sum_\ell E_\ell^2 + \frac{\gamma}{2} \sum_p [(dA)_p]^2 \\ 
&+ {1\over 2\beta}\sum_s \left(p_s + \frac{1}{2\pi}(dw \cup A)_{\mathfrak{t}(s)} \right)^2
+{\beta\over2}
\sum_\ell \left( ( d\phi)_\ell +2\pi w_\ell\right)^2
+{\lambda\over2} \sum_p [(dw)_p ]^2
\,,
\fe
where we have used $\int d\phi \cup dw \cup\alpha = -\int \phi \cup dw \cup A$ and that $[\int \phi \cup dw \cup A,p_s] = i(dw \cup A)_{\mathfrak{t}(s)}$. The Gauss law constraints are:
\ie
    &\exp\left(2\pi i w_\ell \right) = 1 \,, \qquad\qquad\qquad\qquad\qquad  \exp\left(2\pi i p_s - i(\delta b)_s \right) = 1 \,, \\
    &(\delta E)_s = -\left( \left( w  + {d\phi\over 2\pi}  \right) \cup dw \right)_{\mathfrak{t}^{-1}(s)} \,.
\fe

The system above has a $\bZ$ 1-form symmetry associated with $j_\ell = 2\pi E_\ell +(\phi \cup dw)_{\mathfrak{t}^{-1}(\ell)}$ acting as \eqref{eq:Agaugetransform2}
\ie \label{unitary.1.z}
    \exp\left(i\sum_\ell j_\ell k_\ell \right) : \quad \begin{aligned} A_\ell &\mapsto A_\ell + 2\pi k_\ell \\ p_s &\mapsto p_s - (dw \cup k)_{\mathfrak{t}(s)}\\
    b_\ell & \mapsto b_\ell +  \int \phi\cup d\mathbf{l}\cup k
    \end{aligned} ~,
\fe
which indeed is gauge-invariant and commutes with the Hamiltonian when $k_\ell \in \bZ$ and $dk^{(1)} =0$. 
To gauge this symmetry, we set $(dk)_p = -n_p$ and act with the unitary operator in \eqref{unitary.1.z} to find the gauged Hamiltonian
\ie
H_{\mathrm{A}} &= \frac{1}{2 \gamma} \sum_\ell E_\ell^2 + \frac{\gamma}{2} \sum_p [(dA)_p - 2\pi n_p]^2 \\ 
&+ {1\over 2\beta}\sum_s \left(p_s + \frac{1}{2\pi}(dw \cup A)_{\mathfrak{t}(s)} \right)^2
+{\beta\over2}
\sum_\ell \left( ( d\phi)_\ell +2\pi w_\ell\right)^2
+{\lambda\over2} \sum_p [(dw)_p ]^2
\,,
\fe
with the additional Gauss constraints
\ie
    \exp\left( 2\pi i n_p \right)=1 \,, \qquad \exp{\left( i(\delta \tilde{A})_\ell - 2\pi i E_\ell -i(\phi \cup dw)_{\mathfrak{t}^{-1}(\ell)} \right)} = 1\,.
\fe
Putting all the constraints together, we find \eqref{eq:AgaugeGauss}.

\subsection{Gauging without Villain fields}\label{app:novillain}

Here, we discuss how to gauge a U(1) $r$-form symmetry, without a Villain field for the U(1) gauge fields.

We assume that the symmetry operator takes the form
\ie
   U = \exp\left( i\sum_{c_r} \chi_{c_r} j_{c_r}\right)\,, \qquad \text{for} \quad (d\chi)^{(r+1)}  = 0\,,
\fe
where the local charges $j_{c_r}$ are gauge invariant and commute with each other. Since the total charge is quantized, we assume that there exists an `improvement term' characterized by $\eta^{(r+1)}$ such that the local charges $(j+\delta \eta)_{c_{r}}$ are integer valued and commute with each other. When $d\chi=0$, we can rewrite the symmetry operator as
\ie
    U =  \exp\left(-i\sum_{c_r} \chi_{c_r} (j + \delta\eta )_{c_r} \right)\,.
\fe

\begin{enumerate}
    \item \textbf{Coupling to background gauge fields:} We couple the theory to U(1) background gauge field $A^{(r+1)}$, by conjugating the Hamiltonian $H$ and Gauss's law operators $\{G\}$ by the unitary operator
\ie
     U[\chi] = \exp\left( -i \sum_{c_r} \chi_{c_r} (j + \delta\eta )_{c_r} \right) \,,
\fe
To find 
\ie
    U[\chi] \,H \left(U[\chi]\right)^\dagger = H[d\chi] \,, \qquad U[\chi] \,G \left(U[\chi]\right)^\dagger = G[d\chi]\,,
\fe
for each Gauss's law operator $G$. Here, we have used the fact that the system has the global symmetry, and thus commutes with $U[\chi]$ for $d\chi = 0$,  to conclude that the right-hand side of the equations above only depends on $d\chi$.

\item \textbf{Making the gauge fields dynamical:} We add the conjugate variables $E^{(r+1)}$ satisfying $[A_{c_{r+1}},E_{c'_{r+1}}] = i\delta_{c_{r+1},c'_{r+1}}$. Setting $(d\chi)_{c_{r+1}} = A_{c_{r+1}}$ and adding kinetic terms for the gauge fields, we find the gauged system
\ie
    &H_{\text{gauged}} = \frac{1}{2 \gamma} \sum_{c_{r+1}} \left(E_{c_{r+1}} - (-1)^r \eta_{c_{r+1}} \right)^2 + \frac{\gamma}{2} \sum_{c_{r+2}} \cos((dA)_{c_{r+2}}) + H[A^{(r+1)}] \,, \\
    &G_{\text{gauged}} = G[A^{(r+1)}] \,,
\fe
and additional Gauss's law constraints
\ie
    \exp\left( 2\pi i E_{c_{r+1}} \right) = 1 \,, \qquad (\delta E)_{c_{r}} = (-1)^r \left(j + \delta\eta \right)_{c_r} \,.
\fe
The first constraint above commutes with the system, since the local charges $j_{c_{r}}+(\delta\eta)_{c_r}$ were assumed to be integer-valued. In the procedure above, it is crucial to first write the transformed Hamiltonian and Gauss's law constraints in terms of $d\chi$ and then set $d\chi = A$. Since the local charges $j +\delta \eta$ are not gauge-invariant, the naive electric field $E_{c_{r+1}}$ is not gauge invariant, and the correct gauge-invariant electric field is given by $E - (-1)^r \eta$. 

There is another presentation of the gauged model with the standard kinetic term and gauge-invariant $E_{c_{r+1}}$. We conjugate the system with
\ie
    \mathcal{V}=\exp{\left(i(-1)^{r+1}\sum_{c_{r+1}} A_{c_{r+1}} \eta_{c_{r+1}} \right)} \,,
\fe
to find
\ie
    &H_{\text{gauged}}' = \frac{1}{2 \gamma} \sum_{c_{r+1}} \left(E_{c_{r+1}} \right)^2 + \frac{\gamma}{2} \sum_{c_{r+2}} \cos((dA)_{c_{r+2}}) + \mathcal{V} H[A^{(r+1)}] \mathcal{V}^\dagger \,, \\
    &G_{\text{gauged}}' = \mathcal{V} G[A^{(r+1)}] \mathcal{V}^\dagger \,,
\fe
and additional Gauss's constraints 
\ie
    \exp\Big( 2\pi i \left(E_{c_{r+1}}+(-1)^r \eta_{c_{r+1}}\right) \Big) = 1 \,, \qquad (\delta E)_{c_{r}} = (-1)^r j_{c_{r}} \,.
\fe

\end{enumerate}

\subsubsection*{Gauging U(1)$_\mathrm{V}$ without Villain fields}

We demonstrate the idea by gauging the 0-form symmetry U(1)$_\mathrm{V}$ of the compact boson Hamiltonian \eqref{Hamiltonian} as in Section \ref{sec:anomaly}. The conserved gauge-invariant and quantized charge is
\ie
    Q_{\mathrm{V}} = \sum_s p_s = \sum_s \left( p_s -\frac{(\delta b)_s}{2\pi} \right) \,.
\fe
This corresponds to setting $r=0$, $j_s = p_s$, and $\eta_\ell = -\frac{b}{2\pi}$ in the steps above.

In the first presentation of the gauge theory, we conjugate the system by $e^{-i \sum_s \chi_s (p - \delta b/2\pi)_s }$ and then set $d\chi = A$, to find
\ie
    H_{\text{gauged}} &= \frac{1}{2 \gamma} \sum_{\ell} \left(E_\ell + \frac{b_\ell}{2\pi} \right)^2 + \frac{\gamma}{2} \sum_{p} \cos((dA)_p)  \\
    &+{1\over 2\beta}\sum_s p_s^2
+{\beta\over2}
\sum_\ell \left( ( d\phi)_\ell +2\pi w_\ell \right)^2
+{\lambda\over2} \sum_p [(dw)_p ]^2
\,.
\fe
with Gauss's constraints
\ie
    \exp\left(2\pi i w_\ell\right)&= \exp\left(-i A_\ell\right) \,, \qquad& \exp\left( 2\pi i p_s - i(\delta b)_s \right) &=1 \,,\\
    \exp\left( 2\pi i E_\ell \right) &= 1 \,, \qquad& (\delta E)_s - p_s + \frac{(\delta b)_s}{2\pi} &= 0 \,.
\fe
As mentioned above, it is important to first perform the transformation $e^{-i \sum_s \chi_s (p - \delta b/2\pi)_s }$ and then replace $d\chi$ by $A$. In particular, under the conjugation by $e^{-i \sum_s \chi_s (p - \delta b/2\pi)_s }$ we have $w \mapsto w+d\chi/2\pi$ and $dw \mapsto dw$. Then, by setting $d\chi = A$, we find $w \mapsto w+A_\ell$ and $dw \mapsto dw$, instead of $dw \mapsto dw + dA/2\pi$.

In the second presentation, we conjugate the model above with $e^{i \sum_{\ell} A_\ell b_\ell / 2\pi}$ to find
\ie
    H_{\text{gauged}}' &= \frac{1}{2 \gamma} \sum_{\ell} E_\ell  ^2 + \frac{\gamma}{2} \sum_{p} \cos((dA)_p)  \\
    &+{1\over 2\beta}\sum_s p_s^2 +{\beta\over2} \sum_\ell \left( ( d\phi)_\ell +2\pi w_\ell - A_\ell \right)^2 +{\lambda\over2} \sum_p \left(dw - \frac{dA}{2\pi}\right)_p ^2
\,.
\fe
with Gauss's constraints
\ie
    \exp\left(2\pi i w_\ell\right)&=1 \,, \qquad& \exp\left( 2\pi i p_s - i(\delta b)_s \right) &=1 \,,\\
    \exp\left( 2\pi i E_\ell - i b_\ell \right) &= 1 \,, \qquad& (\delta E)_s - p_s  &= 0 \,.
\fe
This reproduces the gauged Hamiltonian \eqref{gaugedH} and Gauss's law \eqref{ZVgauss} in Section \ref{sec:anomaly}. 

\subsubsection*{2-group background gauge fields}

Finally, we couple the model described by \eqref{eq:AgaugeHamiltonian} and \eqref{eq:AgaugeGauss} to background gauge fields $\mathcal{A}^{(1)}$ and $\mathcal{B}^{(2)}$ for the U$(1)_{\text{V}'}^{(0)}$ 0-form symmetry and U$(1)_m^{(1)}$ 1-form symmetry generated by \eqref{eq:AgaugeQV} and \eqref{eq:AgaugeQm}, respectively.

Here we choose the local charges to be quantized rather than gauge-invariant, so the gauge fields do not couple to the Hamiltonian and only modify the Gauss law constraints in \eqref{eq:AgaugeGauss}. This lets us study the 2-group structure purely kinematically, independent of the Hamiltonian. Also, we assume that the flatness for the Villain gauge field $n^{(2)}$, i.e., $dn=0$, which ensures that the 1-form symmetry is topological.

Following the steps described above, we conjugate the system with the unitary operators
\ie
     \exp\left( i \sum_s \alpha_s \left( p_s - { \left(\delta b\right)_s\over2\pi}+ (w\cup n)_{\mathfrak{t}(s)} \right)\right) \quad \text{and} \quad \exp\left( i \sum_\ell \chi_\ell n_{\mathfrak{t}(\ell)}  \right) = \exp\left( i \int \chi \cup n  \right)\,.
\fe
and then set $(d\alpha)_\ell = \mathcal{A}_\ell$ and $(d \chi)_p = -\mathcal{B}_p$. These unitary operators indeed commute with the Hamiltonian, and modify Gauss's laws in \eqref{eq:AgaugeGauss} to \eqref{eq:2groupGauss1}.

\section{ABJ anomaly in QED}\label{app:QED}

Here we review the ABJ anomaly for the axial symmetry in QED with one massless electron $\Psi$, following the discussion in \cite{Choi:2022jqy}. 

We start with the internal global symmetry of a free, massless Dirac fermion $\Psi$, which is equivalent to two Weyl fermions. 
The $\text{U(1)}_\text{V}\times \text{U(1)}_\text{A}\over \mathbb{Z}_2$ global symmetry acts on $\Psi$ as
\ie
&\text{U(1)}_\text{V}:~~~\Psi \mapsto e^{i \alpha} \Psi\,,~~~~\text{U(1)}_\text{A}:~~~\Psi \mapsto e^{i \gamma_5\alpha} \Psi\,.
\fe
The $\mathbb{Z}_2$ quotient arises because a rotation by $\pi$ in either U(1)$_\text{V}$ or U(1)$_\text{A}$ acts identically as fermion parity $(-1)^F$. 
In addition to the $\text{U(1)}_\text{A}\text{-}\text{U(1)}_\text{V}\text{-}\text{U(1)}_\text{V}$ mixed 't Hooft anomaly, U(1)$_\text{A}$ also has a self anomaly of the form $\text{U(1)}_\text{A}\text{-}\text{U(1)}_\text{A}\text{-}\text{U(1)}_\text{A}$, and a mixed gravitational anomaly $\text{U(1)}_\text{A}\text{-}R\text{-}R$.
In contrast, the global symmetry group of our lattice model is the direct product U(1)$_\text{V}\times \text{U(1)}_\text{A}$, and it only has the $\text{U(1)}_\text{A}\text{-}\text{U(1)}_\text{V}\text{-}\text{U(1)}_\text{V}$ anomaly, with no self anomaly or gravitational anomaly. 
Our lattice model has the same symmetry and anomaly as the Yukawa field theory in 
Section \ref{sec:Yukawa}.

Next, we gauge U(1)$_\text{V}$. 
The axial current and charge are
\ie\label{j}
&\hat j^\text{A}_\mu =  \frac 12\bar \Psi\gamma_5 \gamma_\mu \Psi\,,~~~~~
\widehat Q_\text{A} = \int_{M_3} d^3x\, \hat j^\text{A}_0\,.
\fe
While $\widehat Q_\text{A}$ is a U(1)$_\text{V}$ gauge-invariant local operator, it is not conserved for general spatial manifolds $M_3$ because of the anomalous conservation equation 
\ie
d\star \hat j^\text{A} = {1\over 8\pi^2} dA_\text{V} \wedge dA_\text{V}\,~~~(\text{QED}).
\fe
One can define another axial current and charge as
\ie\label{jhat}
\star j^\text{A} = \star \hat j^\text{A} -{1\over 8\pi^2} A_\text{V}\wedge dA_\text{V}\,,~~~~
 Q_\text{A} = \int_{M_3} d^3x\, j_0^\text{A}\,.
\fe
The new current obeys $\partial^\mu j_\mu^\text{A}=0$ and therefore the new axial charge $ Q_\text{A}$ is conserved. 
However, $ Q_\text{A}$ is not gauge-invariant. 
To summarize, the ABJ anomaly in the continuum QED is manifested by the fact that we do not have a conserved and gauge-invariant axial charge.

\bibliographystyle{JHEP}

\bibliography{ref}

\end{document}